\documentclass[secnumarabic,amssymb, amsmath,nobibnotes, aps, prd, showkeys, nofootinbib, longbibliography, twocolumn]{revtex4-2}

\usepackage{braket,slashed,bm}
\usepackage{multirow}

\usepackage[T1]{fontenc} 
\usepackage{cancel}
\usepackage{mathrsfs}

\usepackage{tikz}
\usetikzlibrary{arrows,decorations.pathmorphing,backgrounds,positioning,fit,petri,automata,shadows,calendar,mindmap,
decorations.markings,calc}
\usepackage{xcolor}
\usepackage[normalem]{ulem}
\usepackage[hidelinks]{hyperref}
\usepackage{orcidlink}
\usepackage{placeins}



\def\mcp{\mathcal P}

\begin{document}
\title{Probing lepton-flavor-violating four-lepton operators at a muon collider}



\author{Sukanta Dutta\,\orcidlink{0000-0002-9925-2085}}
\affiliation{SGTB Khalsa College, University of Delhi, Delhi, India 110007.}
\email{sukanta.dutta@sgtbkhalsa.du.ac.in}

\author{Purnath Unnikrishnan\,\orcidlink{0009-0008-4737-8840}}
\thanks{Contact author: }
\email{purnathuk@gmail.com}

\author{Yashasvi\,\orcidlink{0009-0005-8741-8550}}
\thanks{Contact author: }
\email{yamansangwan16@gmail.com}

\affiliation{Department of Physics \& Astrophysics, University of Delhi, Delhi, India 110007.}
\affiliation{SGTB Khalsa College, University of Delhi, Delhi, India 110007.}

\begin{abstract}
We investigate charged lepton-flavor violation (LFV) induced by dimension-six four-lepton operators within the Standard Model effective field theory(SMEFT) at a proposed high-energy muon collider.  We study the processes $\mu^{+}\mu^{-}\to e^{\pm}\tau^{\mp}$, $\mu^{+}\mu^{-}\to e^{\pm}\mu^{\mp}$, and $\mu^{+}\mu^{-}\to \mu^{\pm}\tau^{\mp}$ at $\sqrt{s}=3$, $10$, and $14$~TeV, incorporating beam polarization and hadronic $\tau$ reconstruction. Using an optimal-observable analysis of the angular distributions, we perform a global fit to the relevant set of four-lepton operators.   Projected sensitivities reach $C/\Lambda^{2}\sim(0.5-1.2) \times10^{-4}\,\mathrm{TeV}^{-2}$, depending on the flavor and chiral structure of the operator, exceeding current limits by up to an order of magnitude. A combined analysis of multiple center-of-mass energies and beam polarizations significantly improves the resolution of correlations among the Wilson coefficients. These results highlight the strong sensitivity of a future multi-TeV muon collider to charged LFV four-lepton interactions, establishing it as a powerful probe of the SMEFT parameter space.\\
\\
DOI: 10.1103/k249-jg2x
\end{abstract}

\maketitle
\tableofcontents
\section{Introduction}
\label{sec:intro}

Lepton-flavor violation in the charged lepton sector constitutes one of the most sensitive probes of physics beyond the Standard Model (SM). Neutrino oscillations provide clear evidence of lepton-flavor violation(LFV) in the neutral lepton sector, whereas charged LFV processes have not yet been observed. In the SM minimally extended by neutrino masses, such processes arise only through loop amplitudes suppressed by $(\Delta m_\nu^2/M_W^2)^2$~\cite{Cheng:1976uq,Petcov:1976ff}, leading to branching ratios far below any realistic experimental sensitivity. Consequently, any observation of charged LFV would provide a clean and unambiguous signal of new physics.

A wide class of beyond-SM scenarios predict potentially observable LFV effects. In particular, simple and well-motivated extensions such as heavy neutrino frameworks~\cite{Ilakovac:1994kj,Alonso:2012ji}, multi-Higgs models with generic Yukawa structures~\cite{Branco:2011iw,Pich:2009sp}, and \(Z^\prime\) constructions with flavor nonuniversal couplings~\cite{Langacker:2008yv} naturally induce LFV. More elaborate setups, including supersymmetric frameworks~\cite{Hisano:1995cp,Barbieri:1995tw}, grand unified theories~\cite{Barbieri:1994pv}, leptoquark models~\cite{Davidson:1993qk}, and extra-dimensional constructions~\cite{Agashe:2006iy}, also predict potentially observable signals.
A model-independent description of such phenomena is provided by the Standard Model effective field theory(SMEFT), in which heavy new physics is encoded through higher-dimensional operators consistent with SM symmetries~\cite{Buchmuller:1985jz,Grzadkowski:2010es}.

Experimentally, increasingly stringent bounds have been established across multiple channels. The MEG experiment constrains $\mu\to e\gamma$ at the level of $4.2\times10^{-13}$~\cite{MEG:2016leq}, while SINDRUM sets $\mathcal{B}(\mu\to eee) < 10^{-12}$~\cite{SINDRUM:1987nra}. Searches for LFV $\tau$ decays by Belle and BABAR probe branching fractions at the $10^{-8}$ level~\cite{Belle-II:2022cgf,Hayasaka:2007vc,BaBar:2010huw,Hayasaka:2010np}. In addition, $\mu$-$e$ conversion experiments currently reach sensitivities of $4.3\times10^{-12}$~\cite{SINDRUMII:2006dvw}, with upcoming facilities such as Mu2e and COMET expected to improve these limits by several orders of magnitude~\cite{Mu2e:2014fns,COMET:2018auw}. Complementary constraints also arise from LFV $B$-meson decays at LHCb~\cite{LHCb:2020khb,LHCb:2019gjj}.

Large electron positron collider constrained LFV $Z$ decays at the $10^{-5}$ level~\cite{Abreu:1996mj,Achard:2001qv}, while future $e^+e^-$ facilities (ILC, CLIC, FCC-ee, CEPC) are projected to reach sensitivities down to $10^{-9}$ in several channels~\cite{Calibbi:2022osm,Banerjee:2016foh,Barik_2026,Calibbi:2021pyh}. These developments motivate a systematic investigation of LFV processes at high-energy lepton colliders.

Within SMEFT, LFV effects can arise from multiple operator classes. In this work, we focus on four-lepton operators, which provide the leading contact-interaction contributions to purely leptonic LFV processes at high energies. This choice is motivated by their direct sensitivity at lepton colliders, their energy-enhanced behavior, and their minimal dependence on assumptions about additional operator sectors.

Most SMEFT analyses adopt a top-down approach, specifying Wilson coefficients at a high new-physics scale and evolving them downward. Such an approach can also relate collider observables to low-energy constraints. In contrast, we follow a bottom-up strategy in which experimentally allowed Wilson coefficients defined near the electroweak scale are evolved upward to the characteristic energy of a multi-TeV muon collider using renormalization-group equations~\cite{Jenkins:2013zja,Jenkins:2013wua,Alonso:2014csa}. This choice provides a direct, data-driven framework to propagate existing bounds to collider scales without imposing assumptions on the ultraviolet flavor structure.

A high-energy muon collider offers a clean probe of LFV interactions induced by four-lepton operators, with strong sensitivity to such structures at multi-TeV energies~\cite{Casarsa:2022,Han:2021kes,Buttazzo:2020eyl}. Future \(e^+e^-\) facilities such as FCC-ee, with \(\mathcal{O}(10^{12})\) \(Z\) bosons (\(\sim 20\,\text{ab}^{-1}\)), can test LFV decays \(Z \to \ell_i \ell_j\) down to branching ratios \( \sim 10^{-9}\text{--}10^{-10}\)~\cite{FCC:2018evy,Calibbi:2021pyh,Kamenik:2023hvi}. Complementary studies of LFV scattering at lepton colliders show sensitivity to four-fermion operators through nonresonant channels~\cite{Altmannshofer:2023tsa,Jahedi:2024lfv}, while a multi-TeV muon collider extends this reach through the energy growth of contact interactions.

LFV processes at high-energy muon colliders have also been studied in specific ultraviolet models ~\cite{AlAli:2021let, Homiller:2022iax} and in effective field theory(EFT) studies exploring LFV four-fermion operators ~\cite{Homhomiller2026} and more general flavor-violating interactions \cite{Glioti:2025zpn}.

In this work, we perform a gauge-invariant SMEFT analysis in the Warsaw basis, identifying the complete set of dimension-six four-lepton operators that contribute at tree level to the processes under consideration and using their chirality structure. We systematically account for operator redundancies, including Fierz relations, and perform a global sensitivity analysis using polarized $\mu^+\mu^-$ collisions at $\sqrt{s}=3,\,10,$ and $14$~TeV. The extraction of Wilson coefficients is carried out using an optimal-observable framework applied to differential kinematic distributions, enabling statistically efficient constraints and a detailed characterization of correlations in parameter space.

The paper is organzed as follows. In Sec.~\ref{sec:effLag}, we introduce the effective Lagrangian for four-lepton operators,   while Sec.\ref{sec:lowE}  discusses the corresponding theoretical and experimental constraints. Sec.~\ref{sec:RGE} presents the renormalization-group evolution of the Wilson coefficients. In Sec.~\ref{sec:Simulation}, we detail the collider simulation, together with signal and SM background event selections, and  Sec.\ref{sec:oot}  is devoted to the optimal observable analysis. Finally, our results and conclusions are summarized in Sec.~\ref{sec:sum}.

\section{Effective Lagrangian}
\label{sec:effLag}
In this work, we focus on LFV transitions accessible at a muon collider via the processes
\begin{equation}
\mu^+\mu^- \to e^\pm\tau^\mp,\qquad
\mu^+\mu^- \to \mu^\pm\tau^\mp,\qquad
\mu^+\mu^- \to e^\pm\mu^\mp,\label{LFVprocesses}
\end{equation}
which are forbidden in the SM.

Within the SMEFT, charged LFV in purely leptonic processes first appears at leading order through dimension-six four-fermion operators, neglecting higher-dimensional contributions. The SMEFT Lagrangian for six dimensions can be written as

\begin{equation}
\mathcal{L}
=
\mathcal{L}_{\mathrm{SM}}
+
\sum_i \frac{C_i}{\Lambda^2} \, \mathcal{O}_i
\end{equation}
Here, \(C_i/\Lambda^2\) denote the Wilson coefficients, where the \(C_i\) are dimensionless and \(\mathcal{O}_i\) are the dimension-six operators.  At dimension six, four-fermion operators relevant at lepton colliders, in the Warsaw basis, are given by Table \ref{tab:SMEFT-4ell} \cite{Buchmuller:1985jz,Grzadkowski:2010es}.

\begin{table}[!ht]
\centering
\renewcommand{\arraystretch}{1.1} 
\setlength{\tabcolsep}{10pt}      
\caption{Dimension-six four-fermion operators in the SMEFT relevant for charged LFV.}
\label{tab:SMEFT-4ell}
\begin{tabular}{lcc}
\hline
\hline
Operator & Wilson coefficient & Operator structure \\
\hline
$\mathcal{O}_{\ell\ell}^{prst}$ 
& $C_{\ell\ell}^{prst}/{\Lambda^2}$ 
& $(\bar{\ell}_p \gamma_\mu \ell_r)(\bar{\ell}_s \gamma^\mu \ell_t)$ \\
$\mathcal{O}_{ee}^{prst}$ 
& ${C_{ee}^{prst}}/{\Lambda^2}$ 
& $(\bar{e}_p \gamma_\mu e_r)(\bar{e}_s \gamma^\mu e_t)$ \\
$\mathcal{O}_{\ell e}^{prst}$ 
& ${C_{\ell e}^{prst}}/{\Lambda^2}$ 
& $(\bar{\ell}_p \gamma_\mu \ell_r)(\bar{e}_s \gamma^\mu e_t)$ \\
\hline
\hline
\end{tabular}
\end{table}
Here, $\ell_p$ denotes the left-handed SU$(2)_L$ lepton doublet, $e_p$ the right-handed charged-lepton singlet, and $p,r,s,t$ are flavor indices. Other classes of operators at dimension six also contribute to the LFV process, such as the Dipole operators or the Higgs-mediated operators. Because of the presence of $s$-channel suppression in the case of dipole and Higgs operators, their contribution to the LFV cross section, at the same order of Wilson coefficient, is smaller compared to the four-fermion operator\cite{Jahedi:2024lfv}. Hence, throughout this work, we consider only purely leptonic four-fermion operators given in Table \ref{tab:SMEFT-4ell}, and Wilson coefficients are assumed to be real to avoid additional $CP$-violating phases.

Since the initial state at a muon collider is fixed, only operators with $(p,r)=(\mu^\pm,\mu^\mp)$ contribute at tree level. The LFV final states correspond to $f_1^\pm f_2^\mp$ with $f_1\neq f_2$, where $f_1 f_2 \in \{e\tau,\;\mu\tau,\;e\mu\}$. So, the contributing four-fermion operators can be identified by the final-state leptons alone. Accordingly, for simplicity, the Wilson coefficients can be written as
$$C^{prst}_{\ell\ell}\equiv C^{f_1f_2}_{\ell\ell}, \quad C^{prst}_{\ell e}\equiv C^{f_1f_2}_{\ell e}, \quad C^{prst}_{ee}\equiv C^{f_1f_2}_{ee}$$
where \(f_1\) and \(f_2\) denote the outgoing leptons.
\par Using standard Fierz rearrangements, the operators in the Warsaw basis can be expressed in terms of the commonly used low-energy vector and scalar contact-interaction operators \cite{Davidson:2016edt}.
The corresponding operator basis is
\begin{align*}
\frac{(C_{\mathrm{LR}}^{S})^{f_1 f_2}}{\Lambda^2}\,&(\bar{\mu} P_L \mu)(\bar{f}_1 P_R f_2),\;\\
\frac{(C_{\mathrm{RL}}^{S})^{f_1 f_2}}{\Lambda^2}\,&(\bar{\mu} P_R \mu)(\bar{f}_1 P_L f_2),\;\\
\frac{(C_{\mathrm{LL}}^{V})^{f_1 f_2}}{\Lambda^2}\,&(\bar{\mu} \gamma^\sigma P_L \mu)(\bar{f}_1 \gamma_\sigma P_L f_2),\;\\
\frac{(C_{\mathrm{RR}}^{V})^{f_1 f_2}}{\Lambda^2}\,&(\bar{\mu} \gamma^\sigma P_R \mu)(\bar{f}_1 \gamma_\sigma P_R f_2),\;\\
\frac{(C_{\mathrm{LR}}^{V})^{f_1 f_2}}{\Lambda^2}\,&(\bar{\mu} \gamma^\sigma P_L \mu)(\bar{f}_1 \gamma_\sigma P_R f_2),\;\\
\frac{(C_{\mathrm{RL}}^{V})^{f_1 f_2}}{\Lambda^2}\,&(\bar{\mu} \gamma^\sigma P_R \mu)(\bar{f}_1 \gamma_\sigma P_L f_2)
\end{align*}
where the Wilson coefficient is related through 
\begin{align}
\left(\frac{C_{\ell e}}{\Lambda^2}\right)^{st} &= \left(\frac{C^{V}_{LR}}{\Lambda^2}\right)^{st} = \left(\frac{C^{V}_{RL}}{\Lambda^2}\right)^{st},\nonumber \\
\left(\frac{C_{\ell e}}{\Lambda^2}\right)^{st}&= -\frac{1}{2}\left(\frac{C^{S}_{LR}}{\Lambda^2}\right)^{st}=-\frac{1}{2}\left(\frac{C^{S}_{RL}}{\Lambda^2}\right)^{st},\nonumber \\
\left(\frac{C_{\ell\ell}}{\Lambda^2}\right)^{st} &= \frac{1}{4}\left(\frac{C^{V}_{LL}}{\Lambda^2}\right)^{st},\nonumber \\
\left(\frac{C_{ee}}{\Lambda^2}\right)^{st} &= \frac{1}{4}\left(\frac{C^{V}_{RR}}{\Lambda^2}\right)^{st}.
\label{smeft_relation}
\end{align}
 At high-energy colliders, where lepton masses can be neglected, chirality and helicity coincide, which allows us to write the amplitudes in the helicity basis. The full set of helicity amplitudes is given by
\begin{align}
\mathcal{M}^{\mathrm{EFT}}_{\lambda_{\mu}\lambda_{\bar{\mu}}\to\lambda_l\lambda_{\bar{l'}}}
&= \frac{s}{\Lambda^2} \Big[ C^{LR}_{V}(1-\cos\theta)\,\,\delta_{\left[--;\,++\right]}\nonumber\\ 
&+ C^{RL}_{V}(1-\cos\theta)\,\,\delta_{\left[++;\,--\right]}\nonumber\\ 
&+ C^{LR}_{S}\,\,\delta_{\left[-+;\,+-\right]}\nonumber\\
&+C^{RL}_{S}\,\,\delta_{\left[+-;\,-+\right]}\nonumber\\
&-C^{LL}_{V}(1+\cos\theta)\,\,\delta_{\left[--;\,--\right]}\nonumber\\
&-C^{RR}_{V}(1+\cos\theta)\,\,\delta_{\left[++;\,++\right]}
\Big],
\label{eq:del}
\end{align}
where \(\delta_{\left[--;\,++\right]}=\delta_{\lambda_{\mu,-1}}\delta_{\lambda_{\bar{\mu},-1}}\delta_{\lambda_{\ell,+1}}\delta_{\lambda_{\bar{\ell},+1}}\) projects onto a specific helicity configuration of the four leptons. Since different operators contribute to distinct helicity configurations, interference terms between them vanish in the cross section. The differential cross section, therefore, can be written as 
\begin{equation}
\frac{d\sigma_{\text{EFT}}}{d\Omega}
=
\frac{1}{64\pi^2 s}
\sum_{\lambda_{\mu},\,\lambda_{\bar{\mu}}\,,\lambda_l,\,\lambda_{\bar{l'}}}
\left|
\mathcal{M}^{\text{EFT}}_{\lambda_{\mu}\lambda_{\bar{\mu}}\to\lambda_l\lambda_{\bar{l'}}}(\sqrt{s})
\right|^2.
\end{equation}

Since in the SM, the LFV amplitude vanishes identically, there is no interference between the SM and EFT amplitudes\footnote{However,  with neutrino masses, these amplitudes are nonzero but arise at loop level only and are proportional to neutrino masses, leaving them highly suppressed. }. Moreover, as mentioned above, due to the helicity structure of the EFT amplitudes, no interference between dimension-six EFT amplitudes survives, leaving the amplitude only quadratic [\(\mathcal{O}(1/\Lambda^4)\)] in Wilson Coefficients. Although dimension-eight operators can, in principle, contribute through interference with the SM amplitude at \(\mathcal{O}(1/\Lambda^4)\), in the present case this vanishes as the SM contributions vanish. The leading interference between dimension-six and dimension-eight amplitudes is of the order \(\mathcal{O}(1/\Lambda^6)\) and is therefore further suppressed.

\begin{table*}[ht]
\centering
\renewcommand{\arraystretch}{1.35}
\caption{ Experimental upper bounds on the relevant LFV branching ratios together with the corresponding constraints on the Warsaw-basis four-lepton Wilson coefficients.}
\label{tab:LFVbounds}
\begin{tabular}{l c c  c c c}
\hline\hline
Process
& Upper bound on branching ratio
& Wilson coefficients
& Extracted bounds on $C/\Lambda^2$ (TeV$^{-2}$)

\\
\hline
$\mu^+\to e^+\gamma$
&
$<4.2\times10^{-13}$
&
$C_{eB}$ (dipole)
&
$\lesssim 6\times10^{-5}$

\\
\hline
$\tau\to\mu\mu\mu$
&
$<2.1\times10^{-8}$
&
$\begin{array}{c}
 C_{\ell\ell}^{\mu\tau}, C_{ee}^{\mu\tau}    \\ 
 C_{\ell e}^{\mu\tau}   
\end{array}$
&
$\begin{array}{c}
   \lesssim 2.8\times10^{-3}    \\
   \lesssim 5.7\times10^{-3}     
\end{array}$

\\
\hline
$\tau\to\mu\mu e$
&
$<2.7\times10^{-8}$
&
$\begin{array}{c}
 C_{\ell\ell}^{e\tau}, C_{ee}^{e\tau}    \\ 
 C_{\ell e}^{e\tau}    
\end{array}$
&
$\begin{array}{c}
   \lesssim 3.26\times10^{-3}   \\
   \lesssim 6.51\times10^{-3}     
\end{array}$

\\
\hline\hline
\end{tabular}
\end{table*}

\section{Experimental and Unitarity Constraints}
\label{sec:lowE}

The dimension-six four-lepton operators introduced in
Sec.~\ref{sec:effLag} are constrained by low-energy searches
for LFV. Since these operators induce
LFV processes at tree level, the most stringent bounds arise from
three-body charged-lepton decays of the form
$\ell_p \to \ell_r \ell_s \ell_t$, followed by coherent
$\mu$-$e$ conversion in nuclei. Additionally,
radiative decays such as $\ell_p \to \ell_r \gamma$ mainly probe dipole
operators and affect the four-lepton operators only through loop-induced
mixing.

\par In this work, for the two processes involving a \(\tau\) lepton in the final state, the bounds on the Wilson coefficients can be extracted directly from the upper bounds on the flavor-violating decays of \(\tau\) leptons. These channels have been extensively investigated in several experiments, with the most stringent bounds obtained at low-energy B-factory experiments such as Belle \cite{Hayasaka:2010np} and BABAR \cite{BaBar:2010huw},

\begin{equation}
\mathcal{B}(\tau \to \mu\mu\mu) < 2.1 \times 10^{-8},
\qquad
\mathcal{B}(\tau \to \mu\mu e) < 2.7 \times 10^{-8},
\end{equation}
both quoted at the 90\% confidence level~\cite{Hayasaka:2010np}.
These limits were normalized to well-measured flavor-conserving leptonic \(\tau\) decay rates,     thereby eliminating common normalization factors and allowing the branching ratios to be expressed directly in terms of the SMEFT Wilson coefficients.

\begin{align}
\frac{\mathcal{B}(\tau \to \mu \mu \mu)}
{\mathcal{B}(\tau \to \mu \nu_\tau \bar{\nu}_\mu)}
&\simeq 
v^4
\!\left[
\left|\frac{2 C_{\ell\ell}^{\tau\mu}}{\Lambda^2}\right|^2
+\left|\frac{2 C_{ee}^{\tau\mu}}{\Lambda^2}\right|^2
+\left|\frac{C_{\ell e}^{\tau\mu}}{\Lambda^2}\right|^2
\right],
 \nonumber \\
\frac{\mathcal{B}(\tau \to \mu \mu e)}
{\mathcal{B}(\tau \to e \nu_\tau \bar{\nu}_e)}
&\simeq 
v^4
\!\left[
\left|\frac{2 C_{\ell\ell}^{\tau e}}{\Lambda^2}\right|^2
+\left|\frac{2 C_{ee}^{\tau e}}{\Lambda^2}\right|^2
+\left|\frac{C_{\ell e}^{\tau e}}{\Lambda^2}\right|^2
\right],
\label{deriv-bound}
\end{align}
where the numerical factor of 2 arises from identical-particle combinatorics, and $v\simeq246$~GeV denotes the Higgs vacuum expectation value. The branching ratios for the flavor-conserving decays are taken from the PDG \cite{PDG:2024cfk}.

\par  The situation differs for the process with $\mu^\pm e^\mp$ as final state, where no direct tree-level constraint exists for the corresponding purely leptonic Wilson coefficients. The relevant low-energy probes for charged LFV in this case are \(\mu\to e\gamma\) and coherent  \(\mu\to e\) conversion in nuclei. The latter constrains semileptonic operators and therefore does not constrain the purely leptonic operators directly. The bounds on the semileptonic Wilson coefficients are of \(\mathcal{O}(10^{-8})\) \cite{Ardu:2022pzk}, which, when translated to purely leptonic four-fermion operators through operator mixing, becomes even weaker.
Similarly, the decay $\mu\to e\gamma$ receives no tree-level contribution from the four-lepton operators and vanishes at one loop when all the diagrams are taken into account\cite{Jahedi:2024lfv}. As a result, constraints on these operators arise only through higher loops or through operator mixing, which are weaker. Consequently, a future multi-TeV muon collider offers an unparalleled environment to directly probe the Wilson coefficients for contact interactions contributing to this process.

All existing upper bounds on the branching ratios and the derived bounds on the Wilson coefficients for the Warsaw-basis four-lepton operators considered in this work are
summarized in Table ~\ref{tab:LFVbounds}.

At high energies, the amplitudes from dimension-six four-lepton operators grow with the center-of-mass energy, which may eventually lead to a breakdown of perturbative unitarity. An estimate of the perturbative consistency of the EFT description can be obtained from the partial-wave unitarity condition \(|\mathrm{Re}(a_0)| \le 1/2,\)
where the $s$-wave partial amplitude is defined as
\begin{equation}
a_0
=
\frac{1}{32\pi}
\int_{-1}^{1}
d\cos\theta\,
\mathcal{M}(s,\theta).
\end{equation}

The \(s\)-wave provides the most stringent bounds, as \(p\)-wave amplitudes have angular dependence, making them subleading. This leads to an approximate upper bound on the effective couplings of the form
\begin{equation}
\left|\frac{C}{\Lambda^2}\right|
\lesssim
\left(\frac{8\pi}{s}\right),
\end{equation}
up to numerical factors depending on the operator normalization and helicity structure. In the present analysis, the Wilson coefficients are chosen such that the explored parameter space remains within the perturbatively reliable regime of the EFT description for the center-of-mass energies considered.


\section{Renormalization-group evolution of LFV four-lepton operators}
\label{sec:RGE}
We study the renormalization-group equation(RGE) of dimension-six four-lepton operators in the SMEFT that induce charged LFV. The renormalization-group equations governing the SMEFT Wilson coefficients are
\begin{equation}
\dot{C}_i\equiv16\pi^2\mu \frac{d C_i}{d\mu_{RS}}
=\Gamma_{ij}\,C_j,
\end{equation}
where \(\mu_{RS}\) is the renormalization scale. The renormalization-group evolution of the gauge couplings is given by
\begin{equation}
\mu_{RS} \frac{d g_i}{d\mu_{RS}} = \frac{b_i}{16\pi^2} g_i^3, \qquad (i=1,2,3),
\end{equation}
with the SM coefficients $b_1 = \tfrac{41}{6}$,  $b_2 = -\tfrac{19}{6}$ and $b_{3}=-7$. This running is consistently included in the evolution of the SMEFT Wilson coefficients. 

The one-loop RGEs for the relevant four-lepton operator classes, derived from the latest version of \cite{Alonso:2014csa}, are given below.

\begin{figure*}[!ht]
    \centering
    \includegraphics[width=0.31\textwidth]{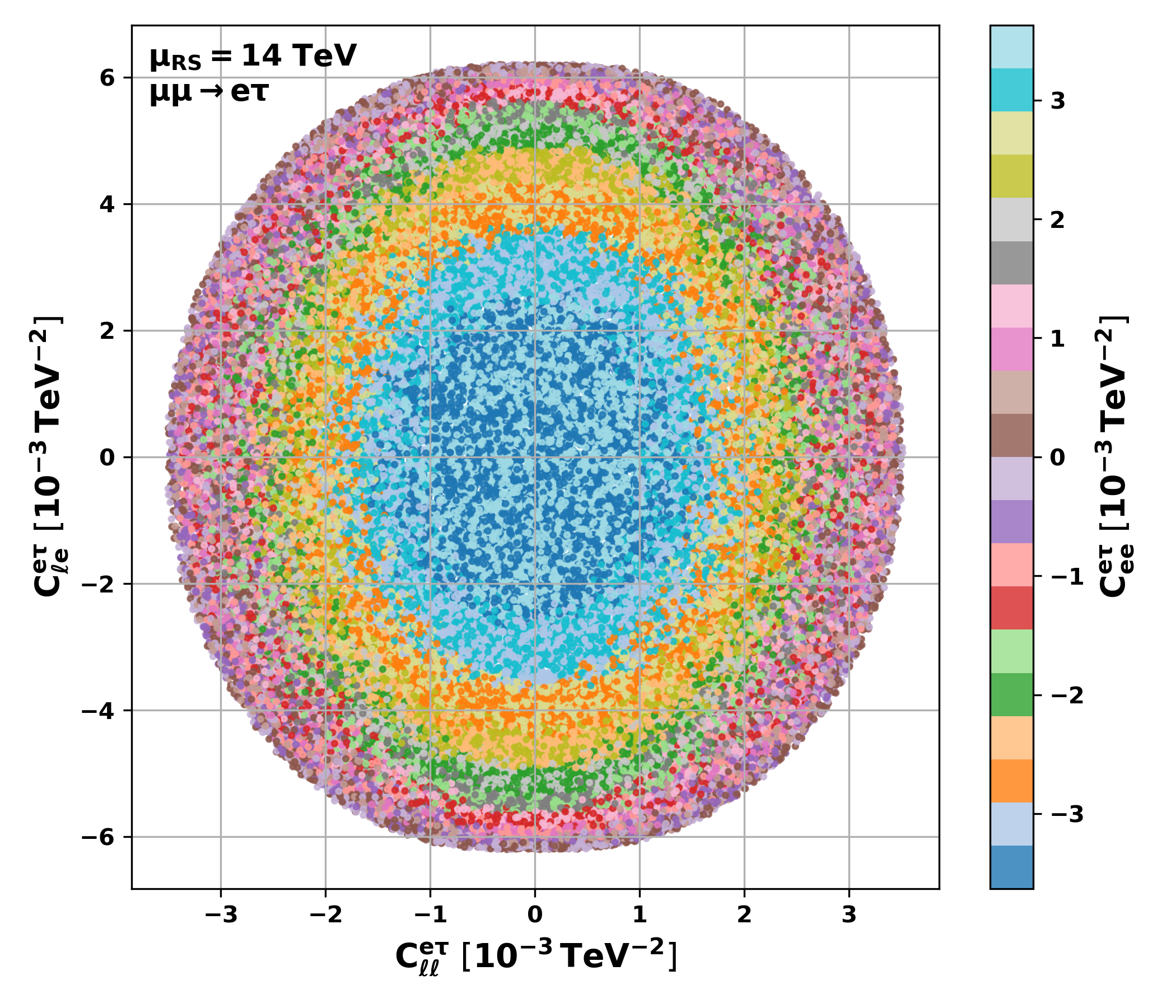}
    \includegraphics[width=0.31\textwidth]{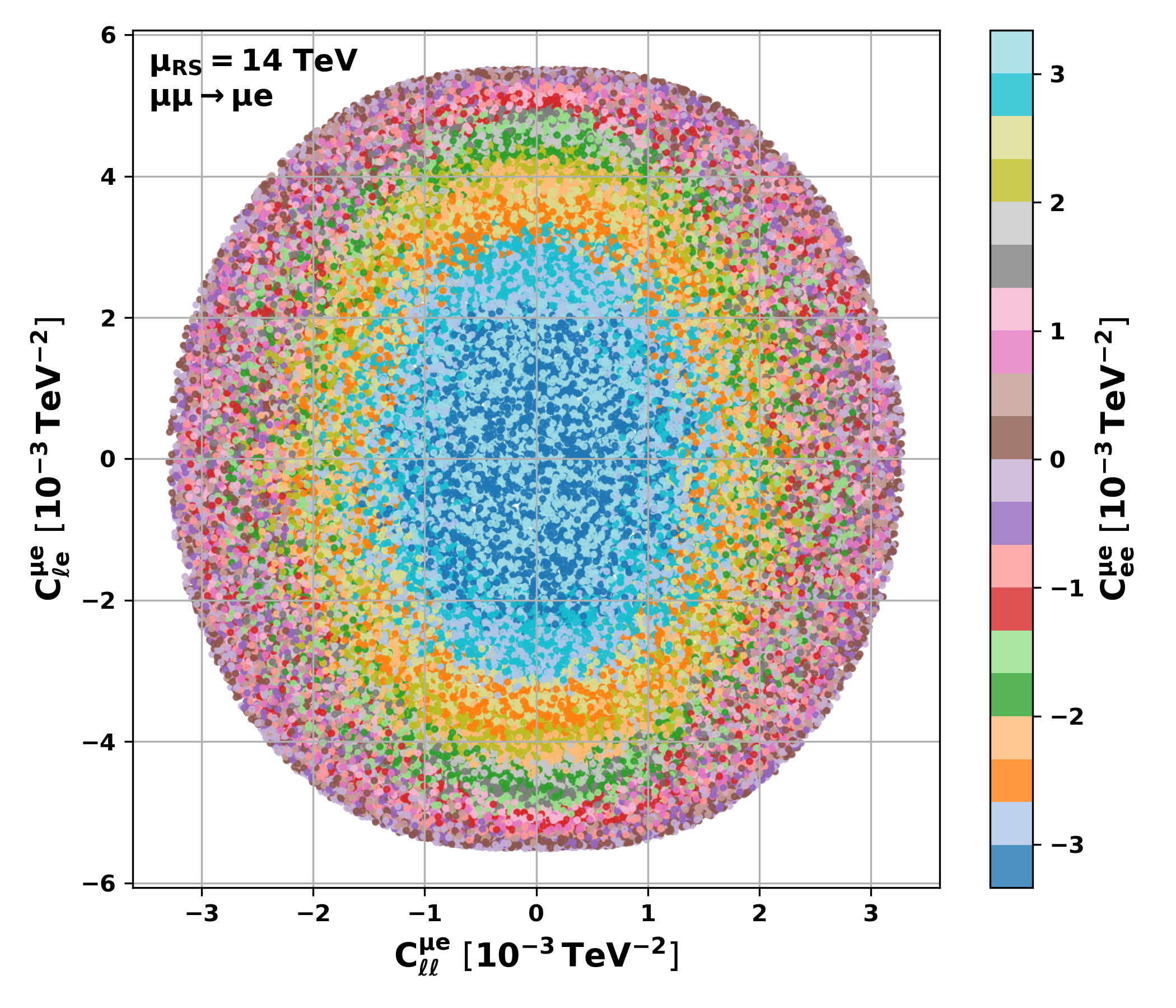}
    \includegraphics[width=0.31\textwidth]{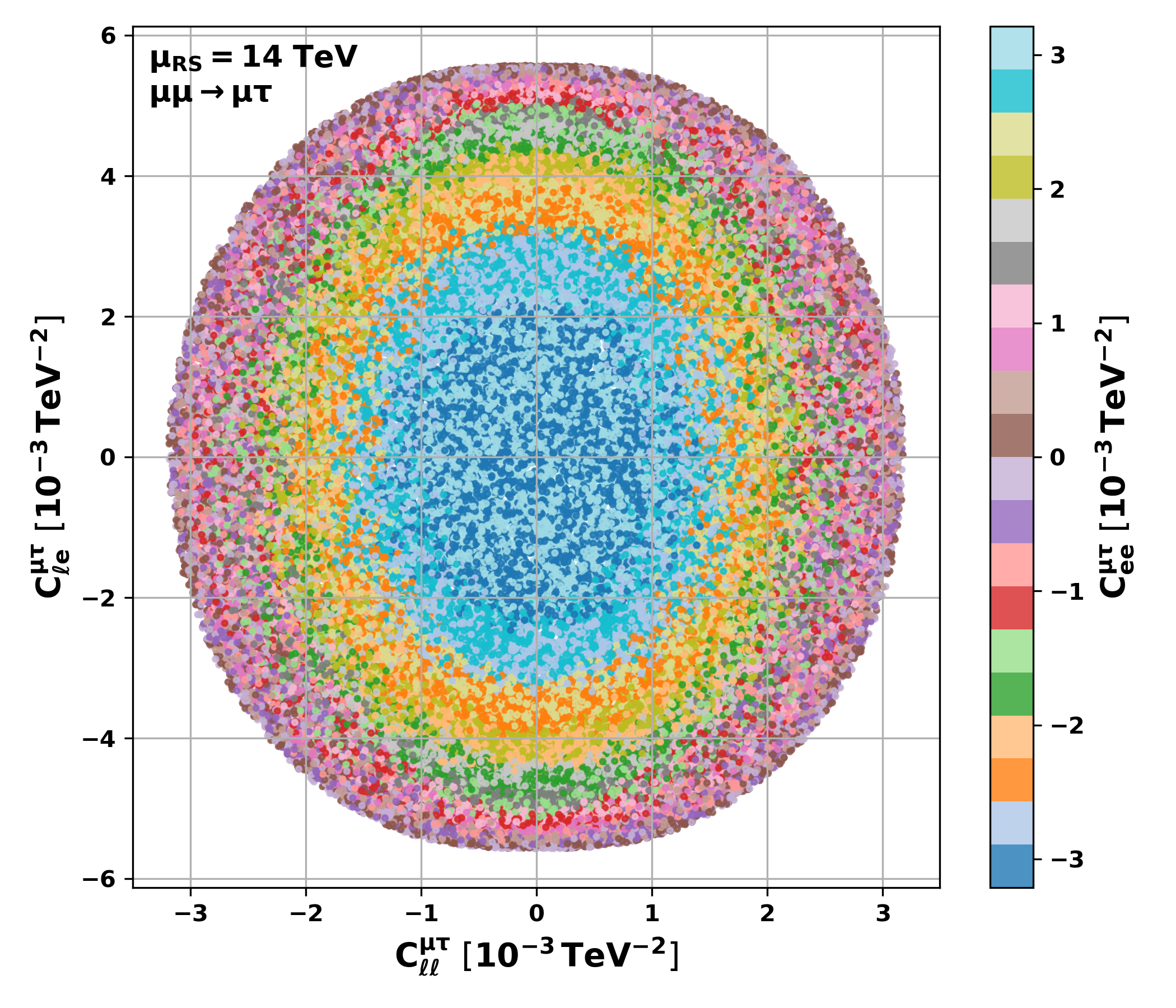}
    \caption{Evolution of the allowed Wilson-coefficient parameter space from the low-energy scale $\mu_{RS}=m_\tau$, where the constraints are extracted from LFV $\tau$ decays \eqref{deriv-bound}, to the collider scale $\mu_{RS}=14~\mathrm{TeV}$. The points shown correspond to the RG-evolved coefficients in the $(C_{\ell\ell},\,C_{\ell e})$ plane, with $C_{ee}$ indicated by the colour scale.}
 \label{rgerunning}
\end{figure*}

For the operator \(O_{\ell\ell}^{prst} =(\bar \ell_p \gamma_\mu \ell_r)(\bar \ell_s \gamma^\mu \ell_t)\), the RGE reads
\begin{align}
\dot C_{\substack{ \ell\ell \\ prst}}& =
    \frac{8}{3} g_1^2 y_\ell^2 C_{\substack{\ell\ell\\prww}} \delta_{st}
+   \frac{8}{3} g_1^2 y_\ell^2 C_{\substack{\ell\ell\\stww}} \delta_{pr} \nonumber\\
&\quad
+   \frac{4}{3} g_1^2 y_\ell^2 C_{\substack{\ell\ell\\pwwr}} \delta_{st}
+   \frac{4}{3} g_1^2 y_\ell^2 C_{\substack{\ell\ell\\swwt}} \delta_{pr} \nonumber\\
&\quad 
-   \frac{2}{6} g_2^2 C_{\substack{\ell\ell\\pwwr}} \delta_{st}
-   \frac{2}{6} g_2^2 C_{\substack{\ell\ell\\swwt}} \delta_{pr} \nonumber\\
&\quad
+   \frac{2}{3} g_2^2 C_{\substack{\ell\ell\\swwr}} \delta_{pt}
+   \frac{2}{3} g_2^2 C_{\substack{\ell\ell\\pwwt}} \delta_{rs}\nonumber\\
&\quad 
+   \frac{2}{3} g_1^2 y_e y_\ell C_{\substack{\ell e\\prww}} \delta_{st}
+   \frac{2}{3} g_1^2 y_e y_\ell C_{\substack{\ell e\\stww}} \delta_{pr}\nonumber\\
&\quad
+   6 g_2^2 C_{\substack{\ell\ell\\ptsr}}
-   3 \left(g_2^2 - 4 y_\ell^2 g_1^2 \right) C_{\substack{\ell\ell\\prst}}
\label{eq:RGE_ll}
\end{align}

\noindent For the operator class \(
O_{\substack{ ee \\ prst}} = (\bar e_p \gamma_\mu e_r) (\bar e_s \gamma^\mu e_t)
\), the RGE is
\begin{align}
\dot C_{\substack{ee\\ prst}} &=
        \frac23 g_1^2 y_e y_\ell  C_{ \substack {\ell e \\ w w s t } } \delta_{p r} 
+       \frac23 g_1^2 y_e y_\ell  C_{ \substack {\ell e \\ w w p r } } \delta_{s t}\nonumber\\
&\quad
+       \frac23 g_1^2 y_e y_\ell  C_{ \substack {\ell e \\ w w s r } } \delta_{p t} 
+       \frac23 g_1^2 y_e y_\ell  C_{ \substack {\ell e \\ w w p t } } \delta_{s r}\nonumber \\
&\quad
+       \frac23 g_1^2 y_e^2 C_{ \substack {ee \\ s t w w } } \delta_{p r} 
+       \frac23 g_1^2 y_e^2 C_{ \substack {ee \\ p r w w } } \delta_{s t}\nonumber\\
&\quad
+       \frac23 g_1^2 y_e^2 C_{ \substack {ee \\ s r w w } } \delta_{p t} 
+       \frac23 g_1^2 y_e^2 C_{ \substack {ee \\ p t w w } } \delta_{s r}\nonumber\\
&\quad
+       \frac23 g_1^2 y_e^2 C_{ \substack {ee \\ s w w t } } \delta_{p r} 
+       \frac23 g_1^2 y_e^2 C_{ \substack {ee \\ p w w r } } \delta_{s t}\nonumber\\
&\quad 
+       \frac23 g_1^2 y_e^2 C_{ \substack {ee \\ s w w r } } \delta_{p t} 
+       \frac23 g_1^2 y_e^2 C_{ \substack {ee \\ p w w t } } \delta_{s r}\nonumber\\
&\quad 
+            12 g_1^2 y_e^2 C_{ \substack {ee \\ p r s t } } .
\label{eq:RGE_ee}
\end{align}

For \(O_{\ell e}^{prst} =(\bar \ell_p \gamma_\mu \ell_r)\,(\bar e_s \gamma^\mu e_t)\,,\)
\begin{align}
\dot C_{\substack{le \\ prst}} &=    
        \frac{16}{3} g_1^2 y_e y_\ell C_{ \substack {\ell\ell \\ p r w w } } \delta_{s t} 
+       \frac83 g_1^2 y_e y_\ell C_{ \substack {\ell\ell \\ p w w r } } \delta_{s t}\nonumber\\
&\quad
+       \frac83 g_1^2 y_e y_\ell C_{ \substack {ee \\ s t w w } } \delta_{p r} 
+       \frac83 g_1^2 y_e y_\ell C_{ \substack {ee \\ s w w t } } \delta_{p r}   \nonumber \\
&
+       \frac43 g_1^2 y_e^2      C_{ \substack {\ell e \\ p r w w } } \delta_{s t} 
+       \frac83 g_1^2 y_\ell^2   C_{ \substack {\ell e \\ w w s t } } \delta_{p r} \nonumber\\
&\quad
-            12 g_1^2 y_e y_\ell C_{ \substack { \ell e \\ p r s t } } 
\label{eq:RGE_le}
\end{align}
Here, \(y_\ell=-\frac{1}{2}, \,y_e=-1
\) denote the SM hypercharges of the left-handed lepton doublet $\ell$ and right-handed charged-lepton singlet $e$, respectively. 

For the LFV processes \(\mu^+\mu^- \to f_1^\pm f_2^\mp\) where \(f_1 \neq f_2\), the set of Wilson coefficients we analyze are, 
\begin{equation}
\vec C=
\left(
C_{\ell\ell}^{\mu\mu f_1f_2},
\,\,
C_{\ell e}^{\mu\mu f_1f_2},
\,\,
C_{ee}^{\mu\mu f_1f_2}
\right)^T
\end{equation}
Since the initial-state current is flavor conserving, we have
\(p=r=\mu\), \(s=f_1\), \(t=f_2\), for \(f_1\neq f_2\), which implies
\begin{align}
\delta_{pr}=\delta_{\mu\mu}=1&, \qquad \delta_{st}=\delta_{f_1f_2}=0,\nonumber \\
\delta_{pt} = \delta_{\mu f_2}&, \qquad \delta_{rs} = \delta_{\mu f_1}
\end{align}

Incorporating these relations, the anomalous-dimension matrix \(\Gamma_{ij}\) for our analysis, including all possible nonvanishing permutations, can be written as

\begin{align}
\mathbf{\Gamma}_{\mu\mu f_1 f_2} =
\underbrace{
\begin{pmatrix}
\big(A g_1^2 + Bg_2^2\big) & \frac13g_1^2 & 0 \\
0 & -\frac{16}{3}g_1^2 & Dg_1
^2\\ 
0 & Eg_1^2 & Fg_1^2
\end{pmatrix}}_{\text{gauge}}\nonumber\\
\qquad +\underbrace{\begin{pmatrix}
2\,y_\mu^2 & 0 & 0\\
0 & 2\,y_\mu^2+y_{f_1}^2+y_{f_2}^2 & 0\\
0 & 0 & y_{f_1}^2+y_{f_2}^2
\end{pmatrix}}_{\text{Yukawa part}},
\end{align}
where, 
\begin{align*}
A &=\frac{11}{3}+\frac{1}{3}\mcp_{f_1\mu\mu f_2}, \\ 
B &= -\frac{2}{6}\mcp_{f_1\mu\mu f_2}+ \frac23\mcp_{f_1f_2\mu\mu}\delta_{\mu f_2}+\frac23\delta_{\mu f_1}+6\mcp_{\mu f_!f_2\mu}-3,\\ 
D &=\frac43\left(1+\mcp_{f_1\mu\mu f_2}\right),\\
E&=\frac13+\frac13\mcp_{\mu f_2 f_1 \mu}\left(\delta_{\mu f_1}+\delta_{\mu f_2}\right) \,\, \text{and} \\ 
F &= \frac{38}{3}+\frac23\mcp_{\mu f_2 f_1 \mu}\left(\delta_{\mu f_2}+\delta_{\mu f_1}+1\right)+\frac23\left(\delta_{\mu f_2}+\delta_{\mu f_1}\right)
\end{align*}
and  \(\mcp\) is a permutation operator such that \(\mcp_{psrt}C_{prst}  \to C_{psrt}\). At the one-loop level, Yukawa contributions are comparable to those of two-loop gauge contributions, and hence can be neglected. Consequently, the renormalization-group evolution is overwhelmingly dominated by the gauge sector.

Figure~\ref{rgerunning} shows the renormalization-group-evolved values of the experimentally allowed Wilson coefficients at $\mu_{\rm RS}=\sqrt{s}=14~\mathrm{TeV}$, obtained by evolving the low-energy constraints extracted from LFV $\tau$ decays at $\mu_{\rm RS}=m_\tau$. The initial points are generated on the ellipsoidal surface defined by the low-energy constraints and subsequently evolved using the one-loop SMEFT RGEs. Since the evolution includes operator mixing, individual Wilson coefficients need not vary monotonically; depending on the direction in coefficient space, they may either increase or decrease slightly. The figure therefore represents the RG evolution of the allowed SMEFT parameter space rather than the evolution of any individual Wilson coefficient. The shift remains small, indicating that the characteristic scale of the allowed coefficient space is largely preserved between the $\tau$ mass and multi-TeV energies. 


\section{Event Generation and Significance Estimation}
\label{sec:Simulation}
 Having established the SMEFT parameter space at the collider scale through renormalization-group evolution, we now investigate the sensitivity of a future muon collider to these Wilson coefficients using detector-level simulations. The analysis is based on Monte Carlo event samples for both the LFV signal processes and the relevant SM backgrounds capable of producing experimentally indistinguishable final-state signatures. We consider the three benchmark LFV channels.
\[
\mu^+\mu^- \to e^\pm\tau^\mp,\qquad
\mu^+\mu^- \to \mu^\pm\tau^\mp,\qquad
\mu^+\mu^- \to e^\pm\mu^\mp.
\]
Here, unlike electrons and muons, which leave clean and directly identifiable signatures in detectors, \(\tau\) leptons decay promptly and cannot be observed as stable charged tracks. Hence, \(\tau\) leptons must be inferred from their decay products. Approximately \(65\%\) of \(\tau\) leptons decay through hadronic channels, producing narrow, low-multiplicity jets, while the remaining (\(\sim35\%\)) decay leptonically into electrons or muons accompanied by neutrinos. At first sight, leptonic \(\tau\) decays may appear advantageous because they produce relatively clean final states containing isolated charged leptons. However, SM processes such as \(\mu^+\mu^- \to W^+W^-\) can produce identical visible final states together with missing energy from neutrinos, leading to kinematic distributions that are difficult to distinguish from the signal.

Hence, the present study focuses exclusively on hadronically decaying \(\tau\) leptons. In addition, this choice benefits from the larger hadronic branching fraction, resulting in a higher signal yield than purely leptonic \(\tau\) decay channels. The hadronic \(\tau\) reconstruction is implemented using jet-reconstruction algorithms, \(\tau\)-identification criteria, and a cut-based analysis framework designed to maximize the signal sensitivity while maintaining efficient suppression of the SM backgrounds.
\begin{figure}[!ht]
  \centering
  \includegraphics[width=0.4\textwidth]{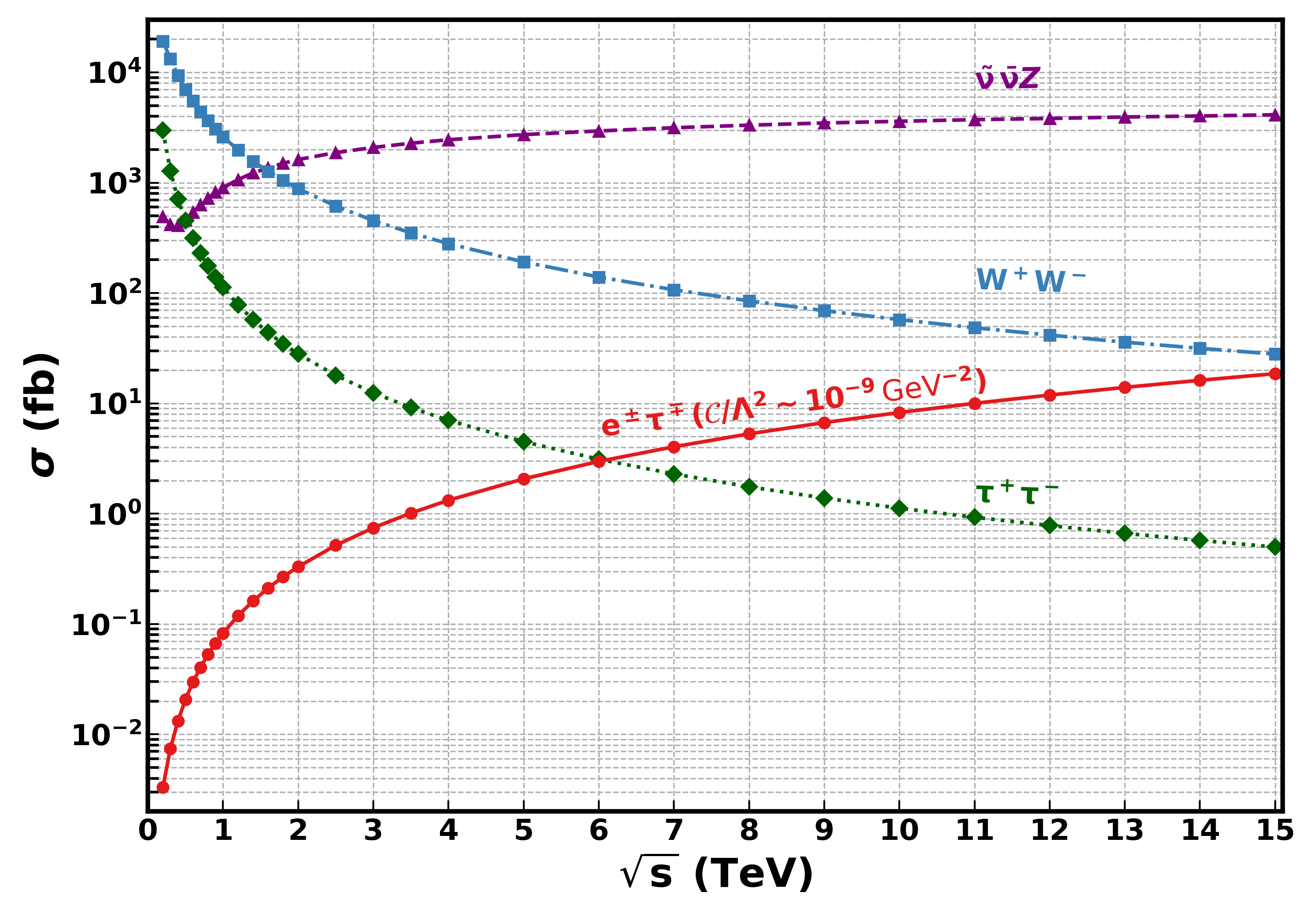}
  \caption{Leading-order cross sections as a function of the center-of-mass energy for the LFV signal process $\mu^+\mu^-\to e^\pm\tau^\mp$ 
  and the dominant SM background processes. The signal is evaluated for  \(C^{e\tau}/\Lambda^2=1.0\times 10^{-3}\) TeV\(^{-2}\),  with one Wilson coefficient switched on at a time while all others are set to zero. Since all four-lepton operators exhibit identical energy dependence, only a single representative signal curve is shown.}
  \label{etauxsecvariation}
\end{figure}

\begin{table*}[!ht]
    \footnotesize
    \centering
    \renewcommand{\arraystretch}{1.4}
    \caption{Leading-order cross sections (in femtobarn) for LFV signal processes with \(e^\pm\tau^\mp\), \(\mu^\pm\tau^\mp\), and \(e^\pm\mu^\mp\) final states, together with the dominant SM backgrounds, at \(\sqrt{s}=3\), 10, and 14~TeV for different muon beam polarizations. The signal rates are shown 
    before particle decays and detector effects. Columns under each energy correspond to \(P_{\mu^-}=0\%\), \(-80\%\), \(-30\%\), \(+30\%\), and \(+80\%\), respectively.}
    \label{tab:xsec}
    \begin{tabular}{p{3.5cm}*{5}{p{.8cm}}*{5}{p{.8cm}}*{5}{p{.8cm}}}
        \hline
        \hline
        \multicolumn{16}{c}{\textbf{Process: $\mu^+ \mu^- \to e^\pm \tau^\mp$ }} \\
        \hline
        
        & \multicolumn{5}{c}{\textbf{3~TeV}} 
        & \multicolumn{5}{c}{\textbf{10~TeV}} 
        & \multicolumn{5}{c}{\textbf{14~TeV}} \\
        \cline{2-16}
         \textbf{Couplings }$ (10^{-3}~\mathrm{TeV}^{-2})$& 0\% & -80\% & -30\% & +30\% & +80\% & 0\% & -80\% & -30\% & +30\% & +80\% & 0\% & -80\% & -30\% & +30\% & +80\% \\
        \hline
         $ C_{\ell\ell}^{f_1f_2}/\Lambda^2 $  & 0.74 & 1.3 & 0.965 & 0.521 & 0.15 & 8.26 & 14.87 & 10.74 & 5.78 & 1.65 & 16.2 & 29.15 & 21.04 & 11.33 & 3.24 \\
         $ C_{\ell e}^{f_1f_2}/\Lambda^2 $  & 0.37 & 0.37 & 0.37 & 0.37 & 0.37 & 4.13 & 4.13 & 4.13 & 4.13 & 4.13 & 8.09 & 8.09 & 8.09 & 8.09 & 8.09 \\
         $ C_{ee}^{f_1f_2}/\Lambda^2 $  & 0.74 & 0.15 & 0.52 & 0.96 & 1.3 & 8.26 & 1.65 & 5.79 & 10.74 & 14.87 & 16.2 & 3.24 & 11.33 & 21.05 & 29.16 \\
        Total BG & 2560 & 4592 & 3322 & 1798 & 529 & 3663 & 6592 & 4761 & 2564 & 734 & 4068 & 7322 & 5287 & 2848 & 814.5 \\
        \hline
    \end{tabular}
   
    \begin{tabular}{p{3.5cm}*{5}{p{.8cm}}*{5}{p{.8cm}}*{5}{p{.8cm}}}
        \hline
        \multicolumn{16}{c}{\textbf{Process: $\mu^+ \mu^- \to \mu^\pm \tau^\mp$ }} \\
        \hline
        & \multicolumn{5}{c}{\textbf{3~TeV}} 
        & \multicolumn{5}{c}{\textbf{10~TeV}} 
        & \multicolumn{5}{c}{\textbf{14~TeV}} \\
        \cline{2-16}
         \textbf{Couplings }$ (10^{-3}~\mathrm{TeV}^{-2})$& 0\% & -80\% & -30\% & +30\% & +80\% & 0\% & -80\% & -30\% & +30\% & +80\% & 0\% & -80\% & -30\% & +30\% & +80\% \\
        \hline
         $ C_{\ell\ell}^{f_1f_2}/\Lambda^2 $  & 2.98 & 5.35 & 3.87 & 2.08 & 0.59 & 33.05 & 59.49 & 42.97 & 23.14 & 6.61 & 64.78 & 116.6 & 84.22 & 45.35 & 12.96 \\
         $ C_{\ell e}^{f_1f_2}/\Lambda^2 $  & 1.49 & 1.49 & 1.49 & 1.49 & 1.49 & 16.53 & 16.53 & 16.53 & 16.53 & 16.53 & 32.39 & 32.39 & 32.39 & 32.39 & 32.39 \\
         $ C_{ee}^{f_1f_2}/\Lambda^2 $  & 2.98 & 0.59 & 2.08 & 3.87 & 5.35 & 33.05 & 6.61 & 23.14 & 42.97 & 59.49 & 64.78 & 12.96 & 45.35 & 84.22 & 116.6 \\
        Total BG & 2560 & 4592 & 3322 & 1798 & 529 & 3663 & 6592 & 4761 & 2564 & 734 & 4068 & 7322 & 5287 & 2848 & 814.5 \\
        \hline
    \end{tabular}

    \begin{tabular}{p{3.5cm}*{5}{p{.8cm}}*{5}{p{.8cm}}*{5}{p{.8cm}}}
        \hline
        \multicolumn{16}{c}{\textbf{Process: $\mu^+ \mu^- \to e^\pm \mu^\mp$ }} \\
        \hline
        & \multicolumn{5}{c}{\textbf{3~TeV}} 
        & \multicolumn{5}{c}{\textbf{10~TeV}} 
        & \multicolumn{5}{c}{\textbf{14~TeV}} \\
        \cline{2-16}
        \textbf{Couplings }$ (10^{-3}~\mathrm{TeV}^{-2})$& 0\% & -80\% & -30\% & +30\% & +80\% & 0\% & -80\% & -30\% & +30\% & +80\% & 0\% & -80\% & -30\% & +30\% & +80\% \\
        \hline
        $ C_{\ell\ell}^{f_1f_2}/\Lambda^2 $  & 2.98 & 5.35 & 3.87 & 2.08 & 0.59 & 33.05 & 59.49 & 42.97 & 23.14 & 6.61 & 64.78 & 116.6 & 84.22 & 45.35 & 12.96 \\
         $ C_{\ell e}^{f_1f_2}/\Lambda^2 $  & 1.49 & 1.49 & 1.49 & 1.49 & 1.49 & 16.53 & 16.53 & 16.53 & 16.53 & 16.53 & 32.39 & 32.39 & 32.39 & 32.39 & 32.39 \\
         $ C_{ee}^{f_1f_2}/\Lambda^2 $  & 2.98 & 0.59 & 2.08 & 3.87 & 5.35 & 33.05 & 6.61 & 23.14 & 42.97 & 59.49 & 64.78 & 12.96 & 45.35 & 84.22 & 116.6 \\
        Total BG & 7568 & 9797 & 8405 & 6732 & 5339 & 34121 & 7068 & 5225 & 3014 & 1172 & 4301 & 7565 & 5525 & 3077 & 1038 \\
        \hline
        \hline
    \end{tabular}
\end{table*}

The dominant SM background processes considered in this analysis are
\[\mu^+\mu^- \to W^+W^- \qquad \mu^+\mu^- \to \tau^+\tau^- \qquad \mu^+\mu^- \to Z\nu\bar{\nu}\]
For the specific \(e^\pm\mu^\mp\) final state, an additional background arises from \(\mu^+\mu^-\to \mu^+\mu^-\). These background processes have been extensively studied in collider searches for charged LFV, including analyses of LFV $Z$ boson decays by the ATLAS Collaboration~\cite{ATLAS:2018sky}. Their inclusion is therefore important for constructing realistic event-selection criteria.

In addition to the conventional SM backgrounds discussed above, muon colliders are subject to beam-induced backgrounds (BIBs), which arise from the decay of beam muons along the accelerator beamline. These decays generate large fluxes of soft electrons, photons, low-energy hadrons, and neutrinos that propagate into the detector volume, predominantly in the forward direction and with time delays relative to particles originating from the primary hard-scattering interaction~\cite{Mokhov:2011zzd}. The principal effect of BIB is the generation of diffuse low-\(p_T\) activity in the detector, particularly within the forward regions. Importantly, BIB contributions rarely produce hard and isolated objects in the central detector volume.  Qualitative studies of beam-induced backgrounds have been carried out within the Muon Accelerator Program and subsequent muon-collider detector studies~\cite{Delahaye:2019omf,Black:2022cth}.
The hard kinematic cuts considered in our analysis strongly suppress beam-induced backgrounds, which therefore do not contribute significantly in the kinematic regions of our interest.

The experimental setup considered in this study follows future muon collider proposals operating at multiple~TeV center-of-mass energies corresponding to different stages of the proposed muon collider program \cite{Delahaye:2019omf,Long:2020wfp,Han:2020uid,Aime:2022flm}. Signal and SM background events are simulated by implementing the LFV model in
\texttt{FeynRules}~\cite{Alloul:2013bka} followed by event generation with \texttt{MadGraph5\_aMC@NLO} \cite{Alwall:2011uj}, and parton showering and hadronization are performed using \texttt{Pythia8} \cite{Sjostrand:2014zea}. Finally, detector-level simulation is carried out with \texttt{Delphes3} \cite{deFavereau:2013fsa} using the \texttt{delphes\_card\_MuonColliderDet.tcl} \cite{Lucchesi:2024zenodo} \cite{Delphes:MuonColliderCard}, developed based on the proposed muon collider detector designs. At this stage, the reconstruction of hadronic \(\tau\) decays is carried out using the Valencia algorithm in \texttt{FastJet} \cite{Cacciari:2011ma} considering inclusive jet reconstruction with the jet radius $\Delta R \leq 0.2$ followed by jet association and \(\tau\)-tagging.

Figure~\ref{etauxsecvariation} compares the energy dependence of the LFV signal evaluated for a benchmark  Wilson coefficient and the dominant SM background cross sections.  As expected, the dominant SM background cross sections, arising primarily from the $W^+W^-$ and $\tau^+\tau^-$ production channels, decrease with increasing center-of-mass energy owing to \(s\)-channel suppression. In contrast, the LFV signal, induced by dimension-six four-lepton contact interactions, increases with \(\sqrt{s}\), thereby improving the signal-to-background ratio at higher collision energies. Table~\ref{tab:xsec} summarizes the corresponding leading-order signal and SM background cross sections for all center-of-mass energies and beam-polarization configurations. We observe that the cross section of $\mu\mu \to e\tau$ is different from the other two signal processes ($\mu\mu \to \mu e$ and $\mu\mu \to \mu\tau$), which are identical to each other; this is because they have the same signal topology, where the presence of a final-state muon introduces additional permutation channels compared to \(\mu\mu\to e\tau\).

We have also simulated events considering different polarizations of the $\mu^-$ beam, keeping the $\mu^+$ beam unpolarized. In our study, the simulations are performed for $P_{\mu^-}=0\%,\ \pm30\%$ and $\pm80\%$.

To define exclusive LFV final states for signal and SM background, we impose channel-specific particle selection requirements before any analysis,
\begin{align}
\text{Cut}_0^{(e\tau)} &: (N_e,N_{\tau_h},N_\mu)=(1,1,0), \nonumber\\
\text{Cut}_0^{(\mu\tau)} &: (0,1,1), \\
\text{Cut}_0^{(e\mu)} &: (1,0,1). \nonumber
\end{align}


\begin{figure*}[ht]
     \centering
     \includegraphics[width=0.29\textwidth]{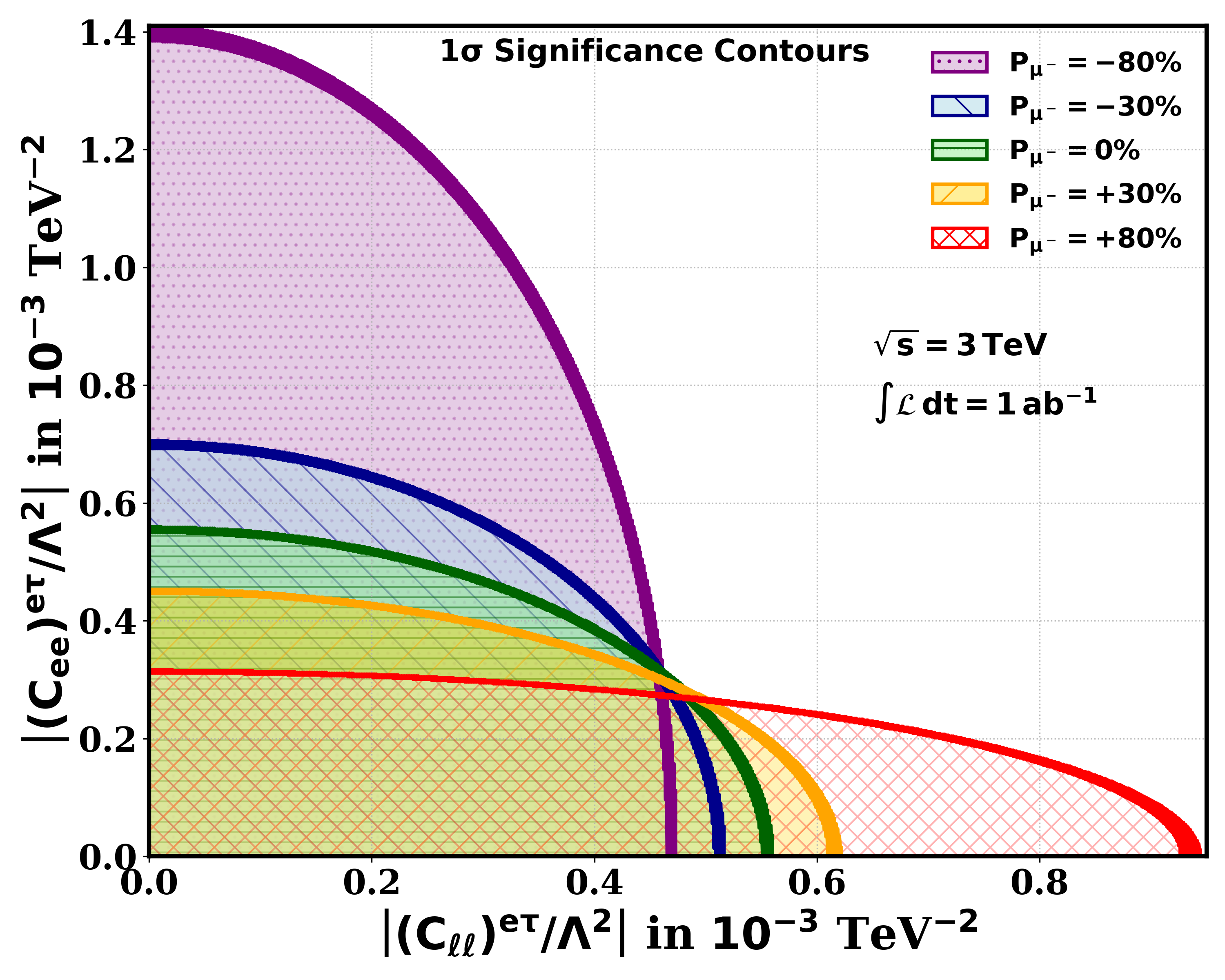}
     \includegraphics[width=0.29\textwidth]{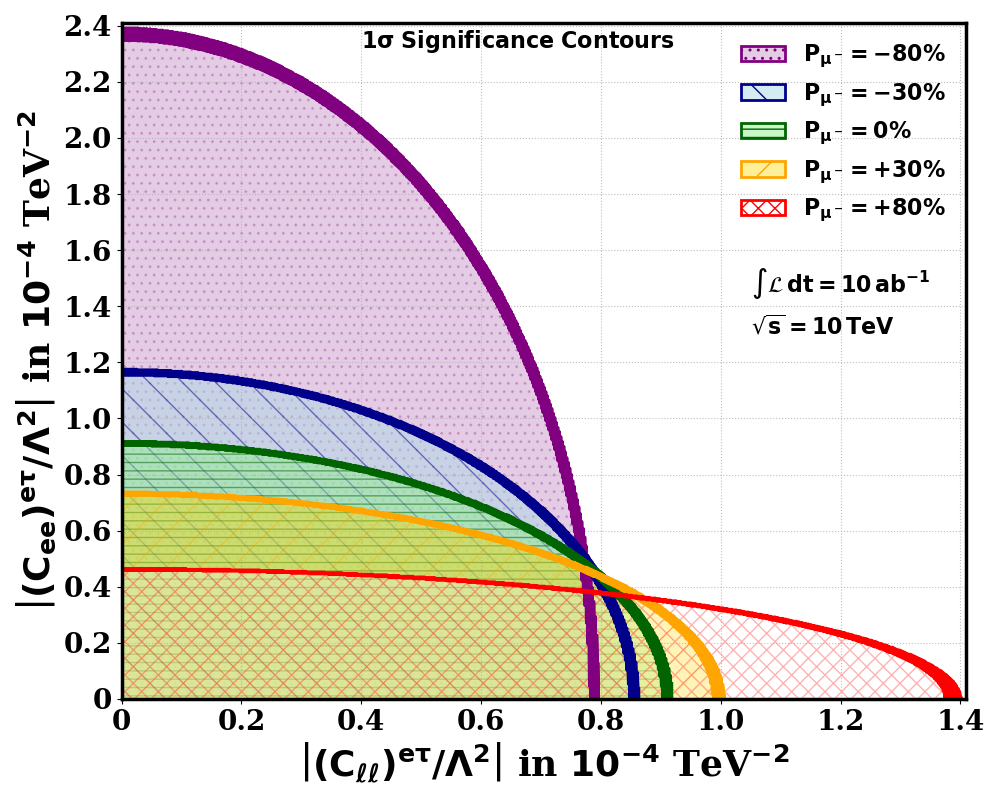}    
     \includegraphics[width=0.29\textwidth]{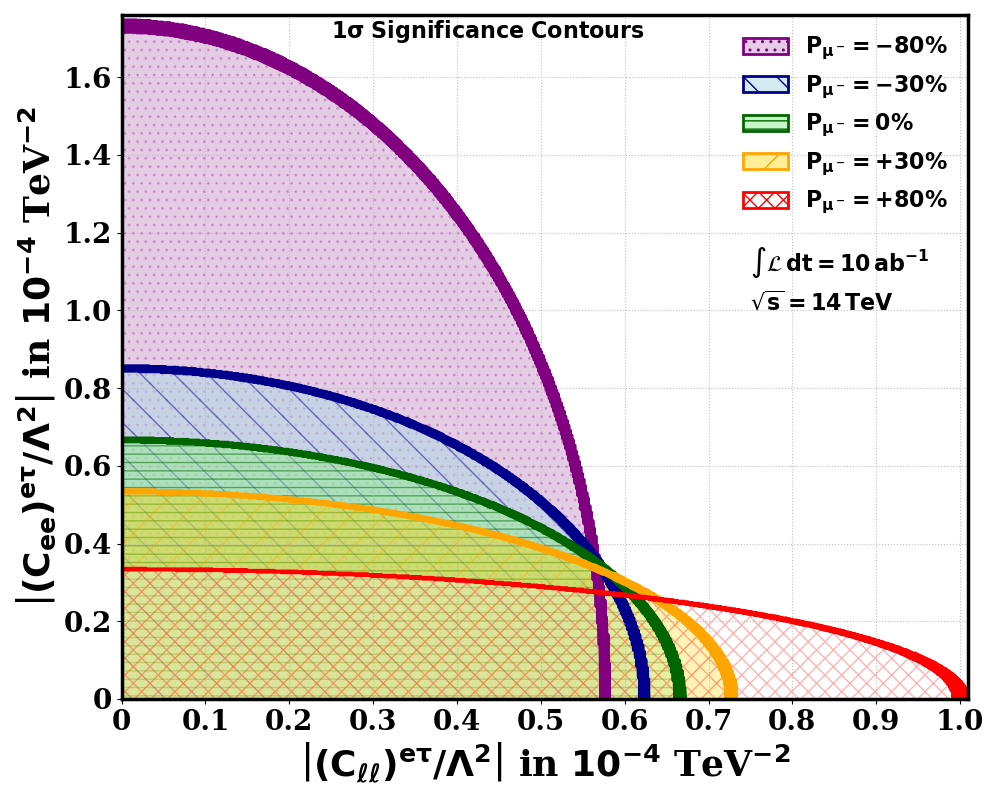} \\
     \includegraphics[width=0.29\textwidth]{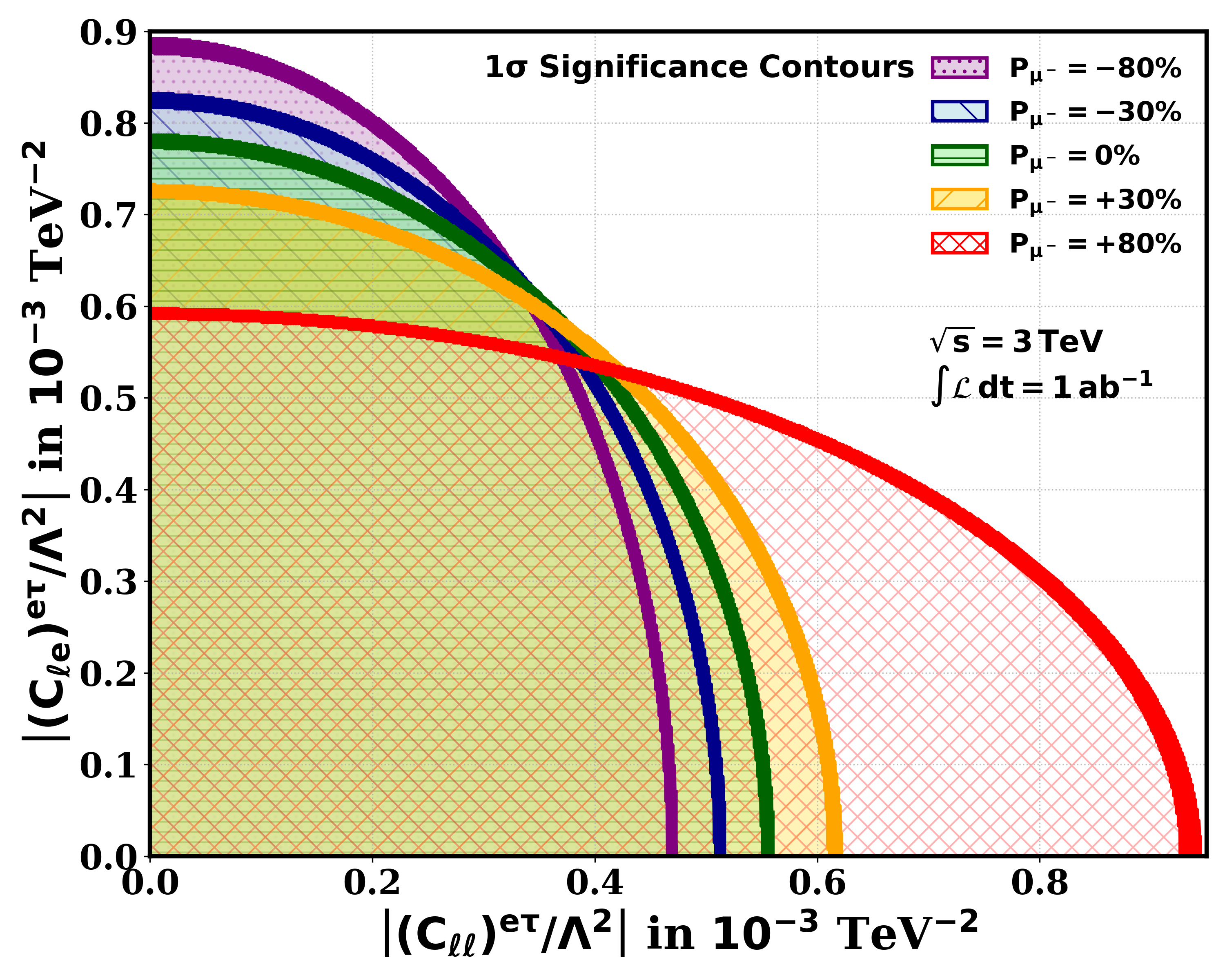} 
     \includegraphics[width=0.29\textwidth]{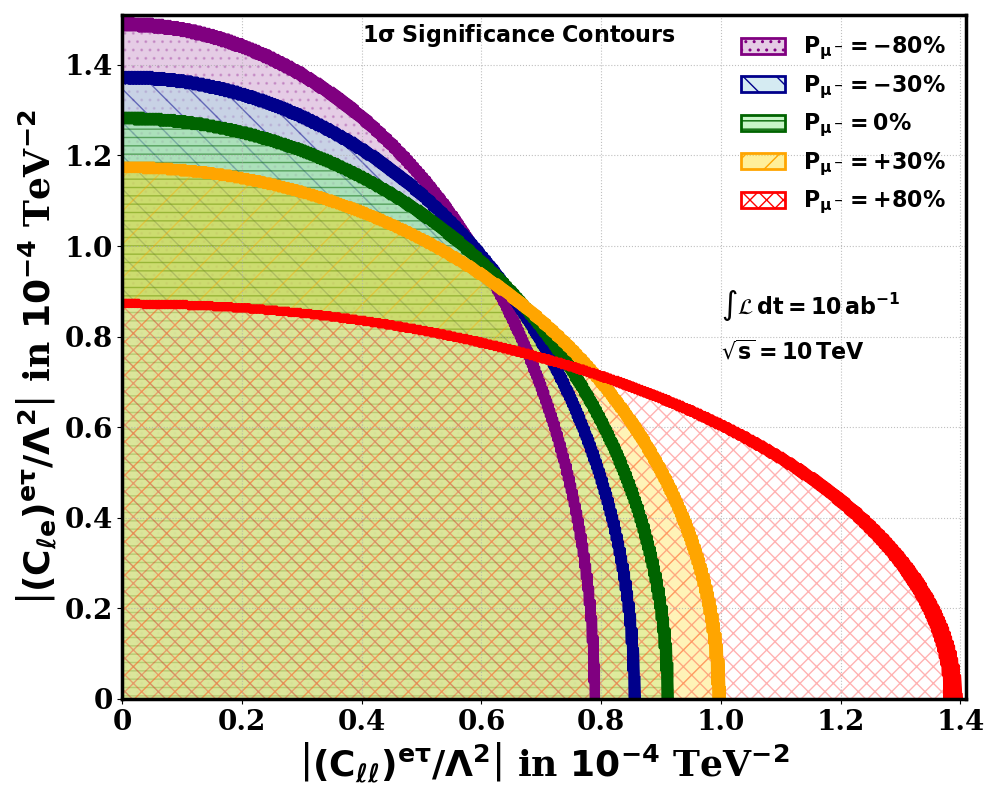} 
     \includegraphics[width=0.29\textwidth]{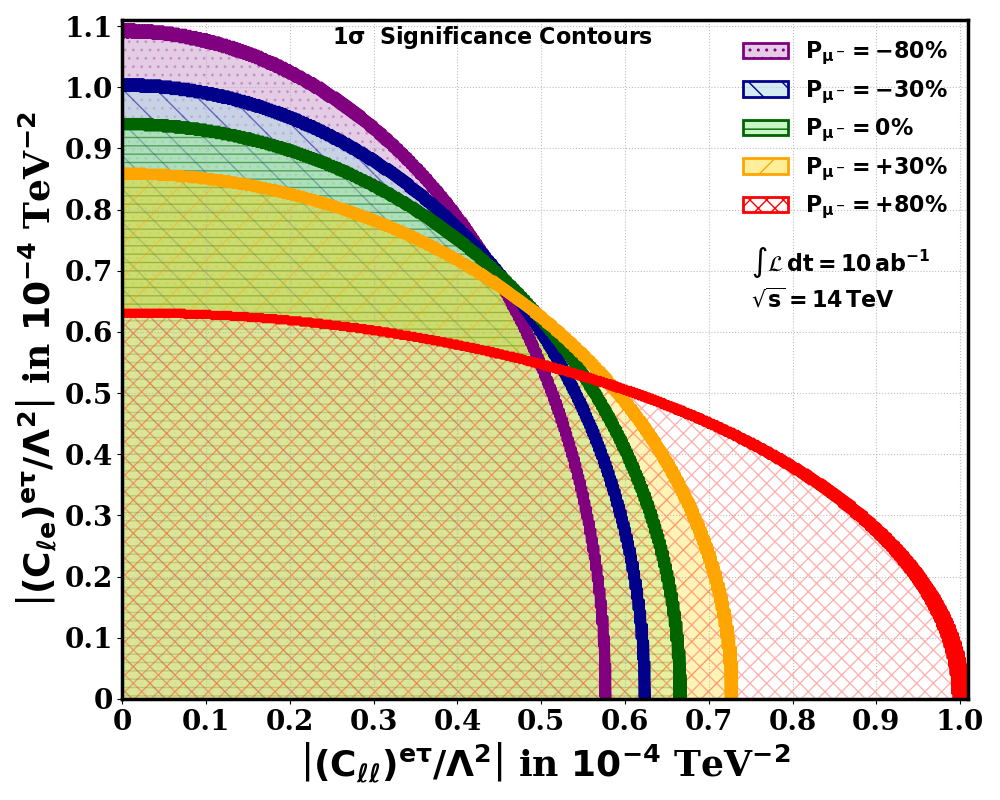} \\
     \includegraphics[width=0.29\textwidth]{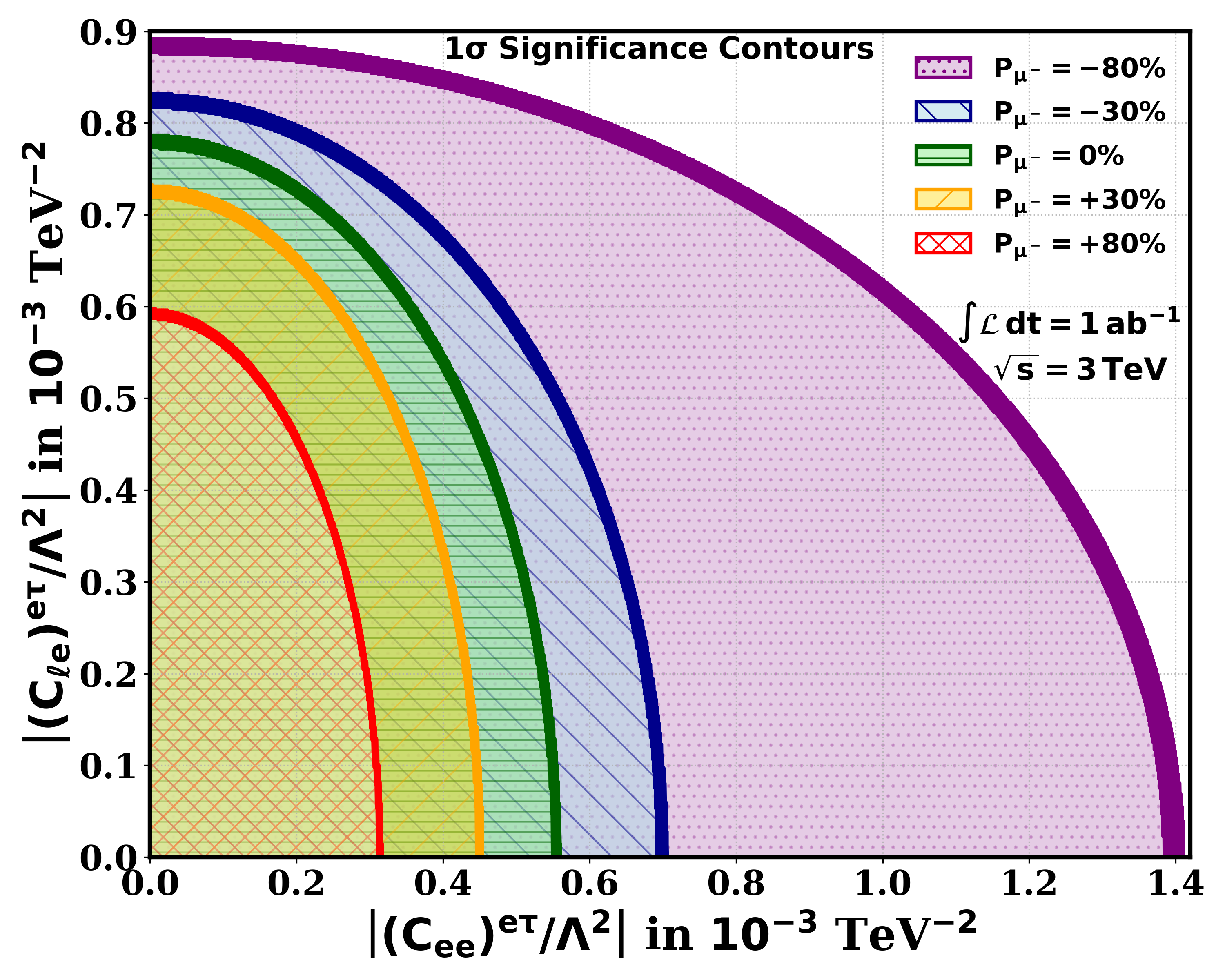} 
     \includegraphics[width=0.29\textwidth]{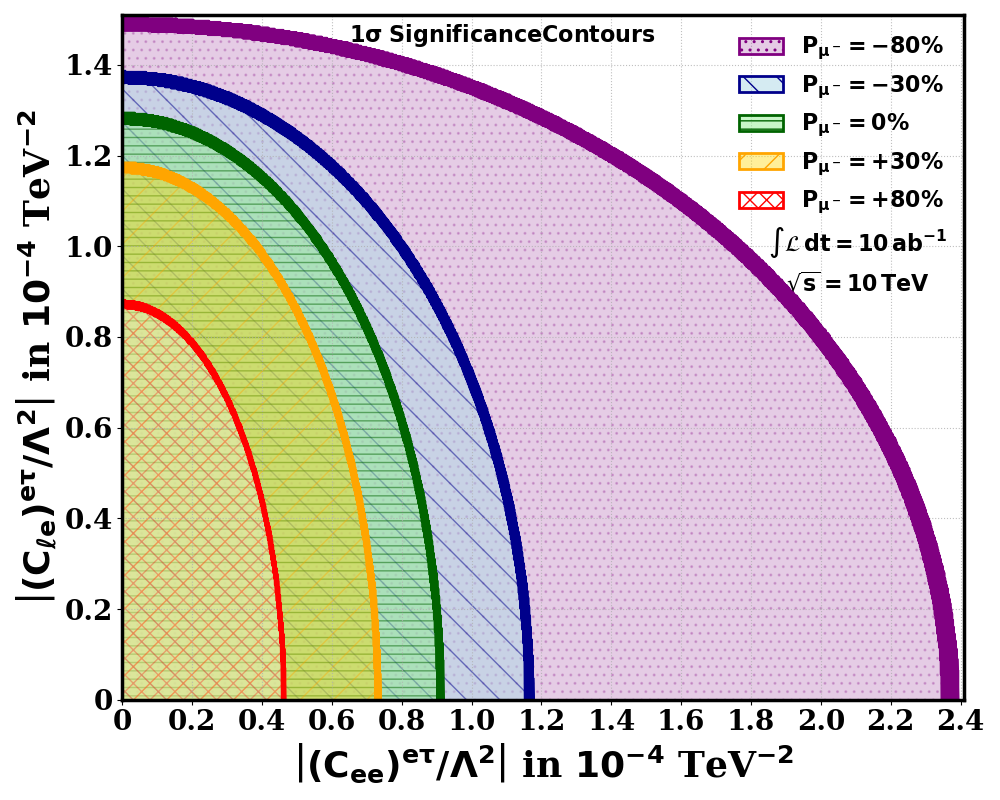}         
     \includegraphics[width=0.29\textwidth]{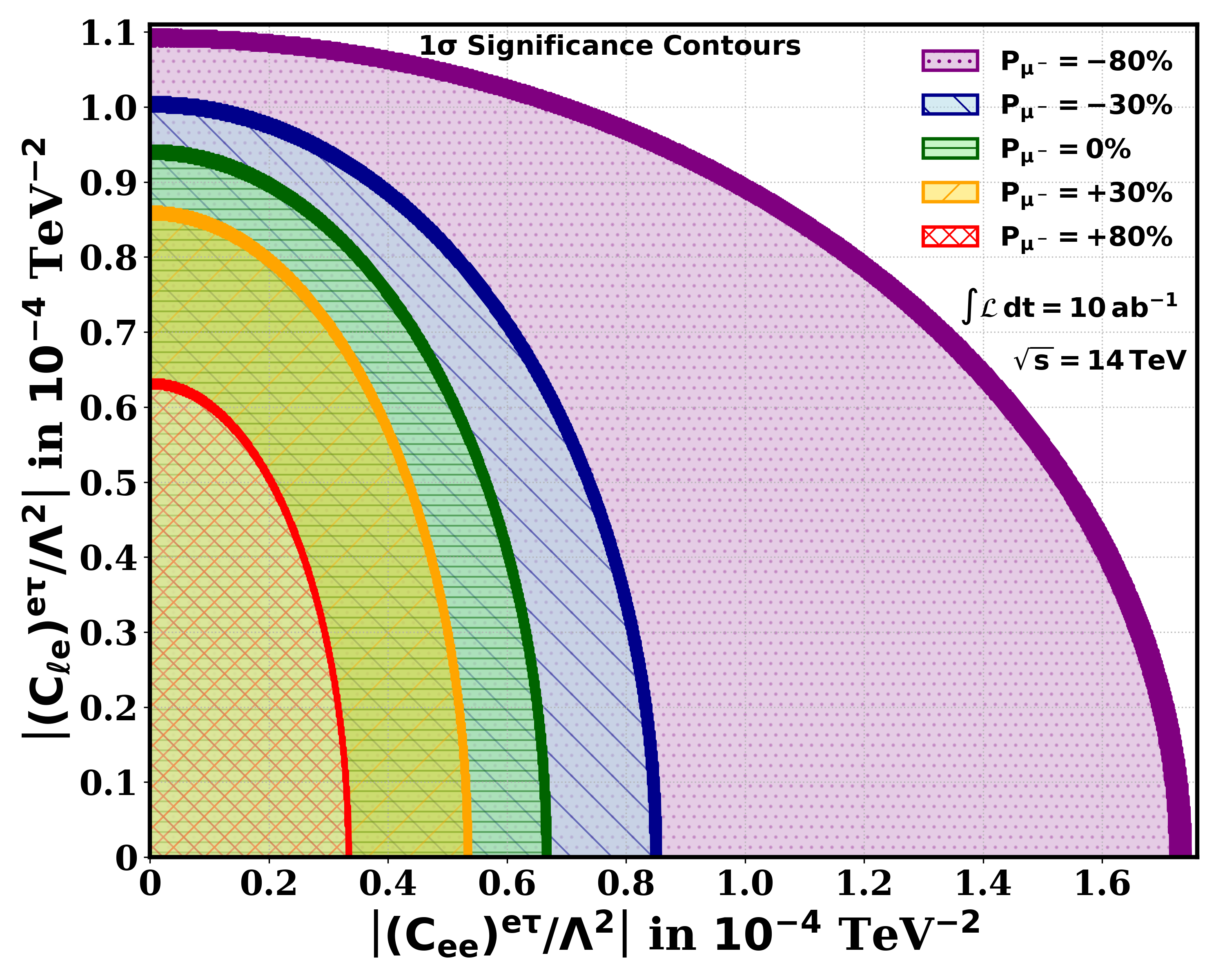}        
    \caption{Projected $1\sigma$ discovery contours for Wilson coefficients for the process \(\mu^\pm\mu^\mp \to e^\pm \tau^\mp\), with all the polarizations  for three benchmark energy center-of-mass energies. }
\label{signif}
 \end{figure*}
 
The impact of these selection requirements becomes evident once the hadronic $\tau$ decays and detector efficiencies discussed above are taken into account.  Table~\ref{xsec-cuts} presents the cross sections after the application of Cut$_0$. The observed reduction in the signal cross section is primarily due to the hadronic branching fraction of the $\tau$ lepton ($\sim65\%$), followed by the $\tau$-tagging efficiency of $80\%$ for $p_T>10$ GeV and the electron-identification efficiency ranging from $58$ to $98 \%$ across different $p_T$ and $\eta$ bins as defined in the muon collider Delphes card. For the $e^\pm\mu^\mp$ channel, which does not contain $\tau$ leptons, the reduction is correspondingly much smaller and arises predominantly from the finite electron-identification efficiency.

Subsequent selections, labeled Cut$_1^{(f_1f_2)}$, are applied cumulatively
on top of Cut$_0^{(f_1f_2)}$ and impose progressively tighter kinematic requirements on the leading lepton. Their precise definitions are provided in the following sections, and their results are tabulated.

\begin{figure*}[t]
    \centering
    \includegraphics[width=0.30\textwidth]{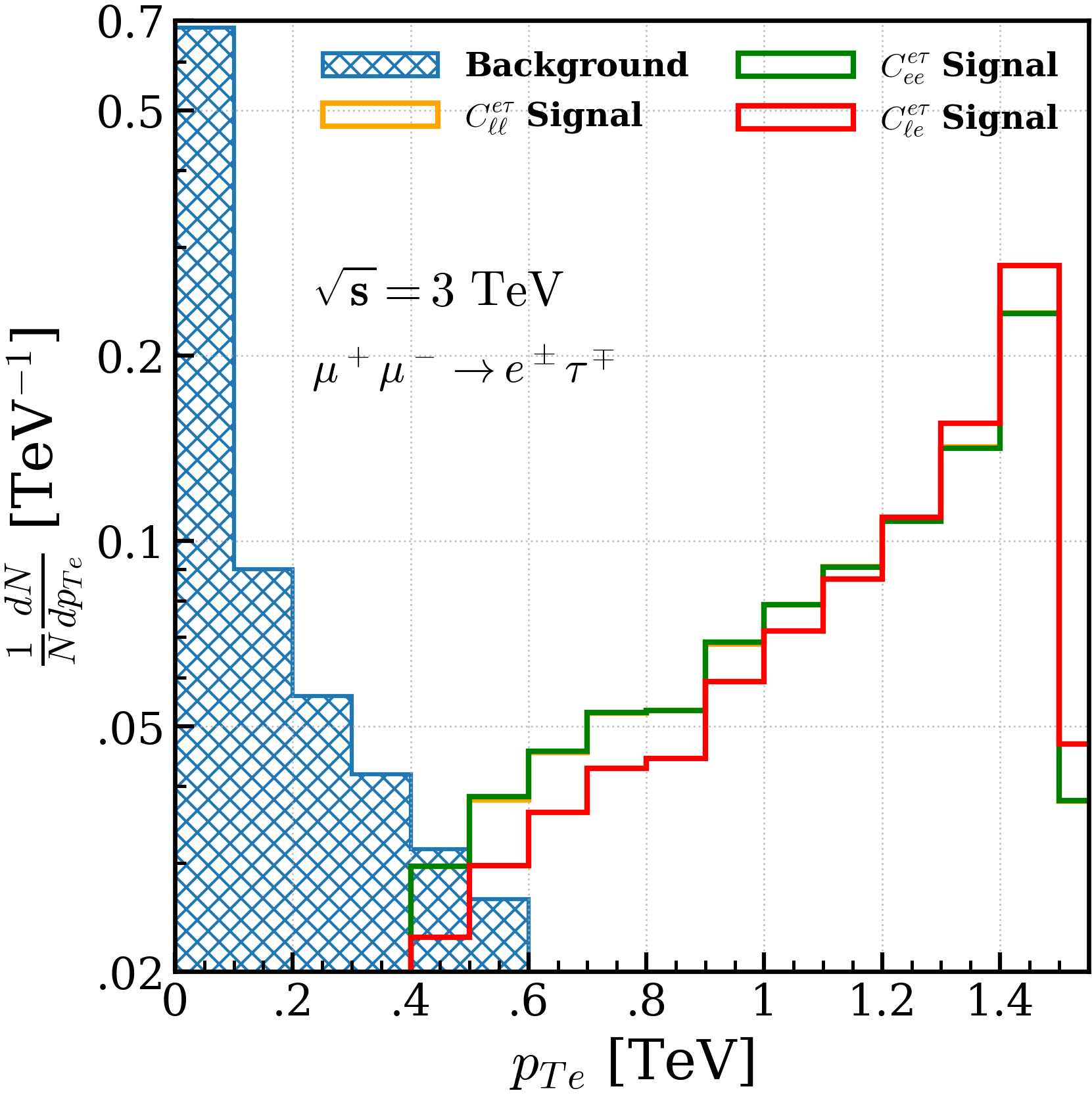}
    \includegraphics[width=0.30\textwidth]{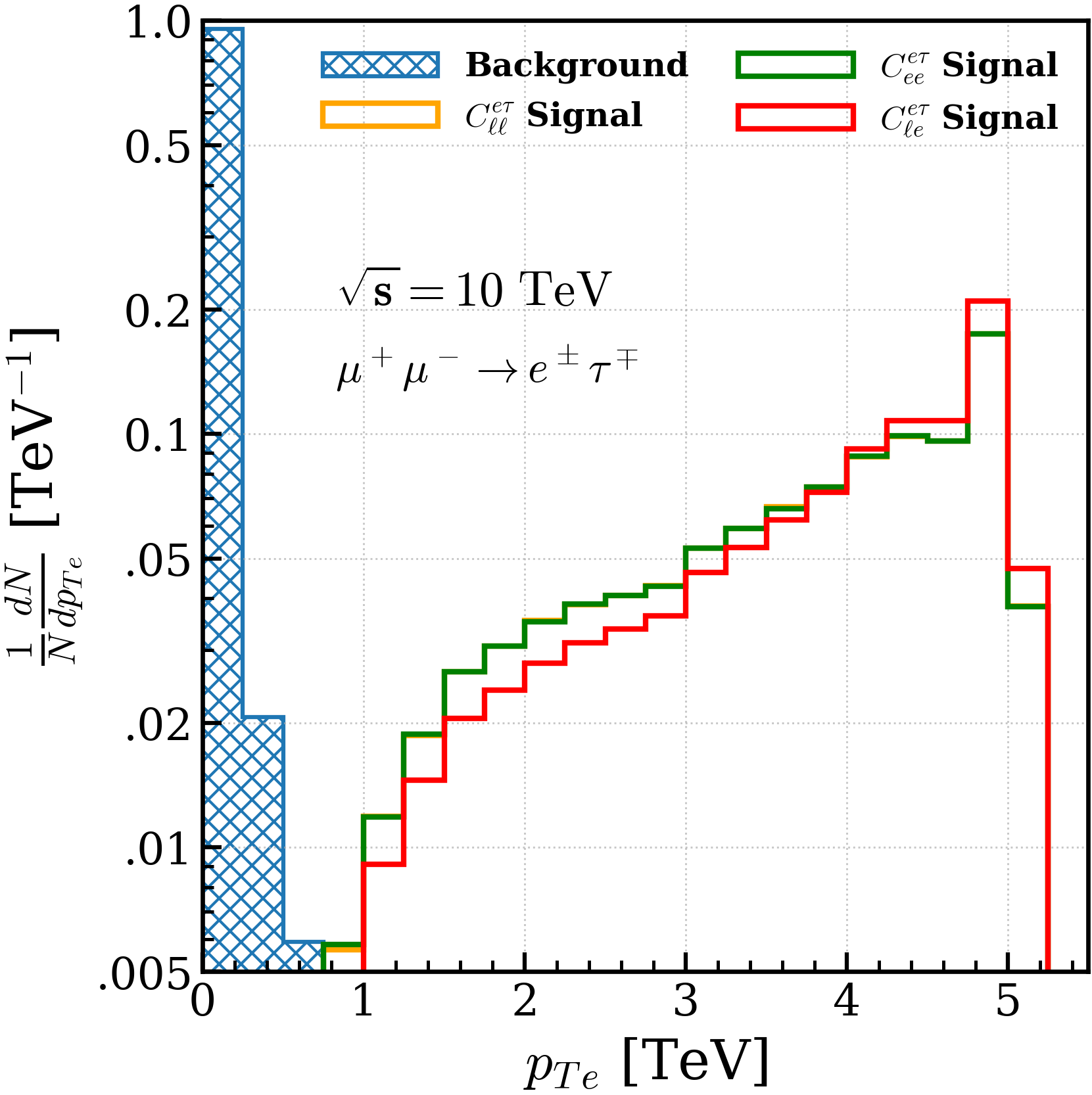}
    \includegraphics[width=0.30\textwidth]{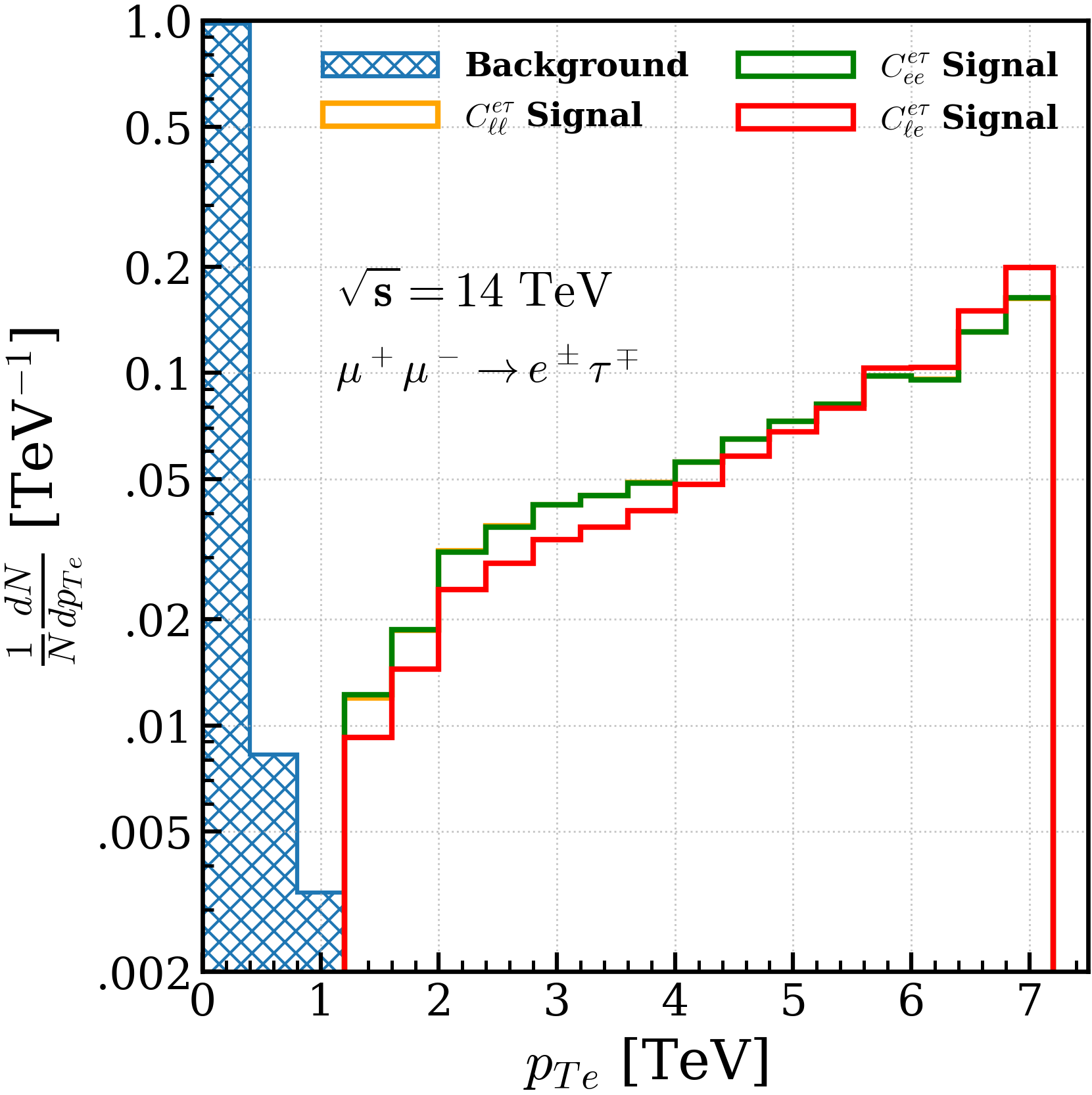}  
    \caption{Distributions of the transverse momentum $p_T$ of the leading electron in LFV process \(\mu^+\mu^- \to e^\pm\tau^\mp\) at $\sqrt{s}=3,\,10$, and $14$~TeV. The signal arises from dimension-six LFV four-lepton operators in the SMEFT, while the background corresponds to purely SM contributions. Here the bin widths are $100$ GeV, $250$ GeV and $400$ GeV for  \(\sqrt{s}=3,\,10\), and \(14\)~TeV respectively.}
    \label{PtKin}
\end{figure*}
\subsection{Signal significance estimation}
\label{sec:significance}
Having established the event selection together with the corresponding signal and SM background rates, we now quantify the discovery potential of the proposed muon collider using the profile-likelihood formalism. The expected statistical significance is evaluated using the asymptotic likelihood formalism~\cite{Cowan:2010js},
\begin{align}
Z &= \Bigg[2 \Bigg((N_S+N_B) \ln \left(\frac{(N_S+N_B)(N_B+\Delta_b^2)}{N_B^2+ (N_S+N_B)\Delta_b^2}\right) \nonumber\\ 
&-\frac{N_B^2}{\Delta_b^2} \ln \left (1+\frac{\Delta_b^2 N_S}{N_B(N_B+\Delta_b^2)}\right)\Bigg)\Bigg]^{1/2},
\label{eq:significance}    
\end{align}
where  $ N_S=\sigma_S \mathcal{L}_{\text{int}} $  and  $ N_B=\sigma_B \mathcal{L}_{\text{int}} $  are the expected number of signal and SM background events, respectively, and  $ \Delta_b $  is the relative systematic uncertainty in the background. The systematic uncertainty is taken to be 1\%. Here ${\cal L}_{\text{int}}$ is the integrated luminosity. The uncertainty in the integrated luminosity is included in the quoted systematic uncertainty.
 
Each panel in Fig. \ref{signif} shows the allowed region in the space of two SMEFT Wilson coefficients while keeping the third fixed. The shaded regions enclosed by the contours correspond to the parameter space where a $1\sigma$ discovery would be achievable. 
The polarization dependence of the corresponding significance contours
for the $\mu^\pm\tau^\mp$ and $e^\pm\mu^\mp$ final states is identical to that for the $e^\pm\tau^\mp$ case, although the numerical sensitivities differ because of channel-dependent backgrounds and reconstruction efficiencies. 

\subsection{Kinematic distributions}
We first examine the kinematic properties of the final-state leptons in order to identify observables that maximize the sensitivity to LFV contact interactions and optimize background suppression. In particular, we focus on the transverse momentum $p_T$ and polar-angle distribution $\cos\theta$ of the reconstructed electron (muon in case of \(\mu^\pm\tau^\mp\) signal) in the final state. 

The resulting distributions for the process $\mu^+\mu^- \to e^\pm\tau^\mp$ are shown in Fig.~\ref{PtKin} for $\sqrt{s}=3,\,10$, and $14$~TeV.

As shown in Fig.~\ref{PtKin}, the LFV signal exhibits a markedly hard transverse-momentum spectrum, with the lepton distribution concentrated near the kinematic end point, \(p_T \sim \sqrt{s}/2\). In contrast, the SM backgrounds originate predominantly from electroweak gauge-boson decays and therefore populate significantly lower transverse-momentum regions.

Motivated by this clear kinematic separation, we impose the following hard transverse-momentum requirements on the leading electron:
\begin{align}
p_T^e > 0.8~\text{TeV} \text{ for } \sqrt{s}=3~\text{TeV}, \nonumber\\ 
p_T^e > 1.0~\text{TeV} \text{ for } \sqrt{s}=10~\text{TeV},\\
p_T^e > 1.2~\text{TeV} \text{ for } \sqrt{s}=14~\text{TeV}.\nonumber
\end{align}

These selections substantially reduce the SM background while preserving most of the signal contribution. A small number of reconstructed events extend beyond the nominal kinematic end point at \(p_T \simeq \sqrt{s}/2\). This feature originates from detector-level energy smearing implemented in the Delphes muon-collider card and reflects purely instrumental fluctuations near the end point region. Such effects also mildly modify the overall normalization and induce slight shape smearing in the distributions, but this does not affect the stark difference between signal and SM background in the high-\(p_T\) region.

\begin{table*}[!ht] 
    \centering
    \renewcommand{\arraystretch}{1.1}
    \setlength{\tabcolsep}{3.1pt} 
    \caption{Cross sections (in femtobarn) after sequential cuts for (a) $\sqrt{s}=3$~TeV with an integrated luminosity of $1~\mathrm{ab}^{-1}$, and (b) $\sqrt{s}=10$~TeV and $14$~TeV with an integrated luminosity of $10~\mathrm{ab}^{-1}$ each. Results are shown for the three effective operators and the corresponding background, assuming unpolarized ($0\%$), positively polarized ($+80\%,\, +30\%$), and negatively polarized ($-80\%,\, -30\%$) muon beams.}
    \label{xsec-cuts}   
    \resizebox{\textwidth}{!}{%
    \begin{tabular}{l*{6}{ccc}}
        \hline
        \hline
        \multicolumn{19}{c}{\textbf{Process: $\mu^+ \mu^- \to e^\pm \tau^\mp$ }} \\
        \hline
        \multirow{4}{*}{\textbf{Couplings}}
        & \multicolumn{6}{c}{$\sqrt{s}=3$~TeV}
        & \multicolumn{6}{c}{$\sqrt{s}=10$~TeV}
        & \multicolumn{6}{c}{$\sqrt{s}=14$~TeV} \\
        \cline{2-19}
       
        & \multicolumn{3}{c}{\makebox[3.0cm]{Cut$^{(e\tau)}_0: N_\mu=0$}} & \multicolumn{3}{c}{\makebox[3.0cm]{Cut$^{(e\tau)}_1$}}
        & \multicolumn{3}{c}{\makebox[3.0cm]{Cut$^{(e\tau)}_0: N_\mu=0$}} & \multicolumn{3}{c}{\makebox[3.0cm]{Cut$^{(e\tau)}_1$}}
        & \multicolumn{3}{c}{\makebox[3.0cm]{Cut$^{(e\tau)}_0: N_\mu=0$}} & \multicolumn{3}{c}{\makebox[3.0cm]{Cut$^{(e\tau)}_1$}} \\
       
        & \multicolumn{3}{c}{\makebox[3.0cm]{$N_e=1,N_{\tau_h}=1$}} & \multicolumn{3}{c}{\makebox[3.0cm]{$p_T^e > 0.8$~TeV}}
        & \multicolumn{3}{c}{\makebox[3.0cm]{$N_e=1,N_{\tau_h}=1$}} & \multicolumn{3}{c}{\makebox[3.0cm]{$p_T^e > 1$~TeV}}
        & \multicolumn{3}{c}{\makebox[3.0cm]{$N_e=1,N_{\tau_h}=1$}} & \multicolumn{3}{c}{\makebox[3.0cm]{$p_T^e > 1.2$~TeV}} \\
        \cline{2-19}
       
        $\begin{array}{c}
        (1\times10^{-3}\\ \mathrm{TeV}^{-2})\end{array}$
       
        & 0\% & $\begin{array}{c}+30\\-30\end{array}$ & $\begin{array}{c}+80\\-80\end{array}$
        & 0\% & $\begin{array}{c}+30\\-30\end{array}$ & $\begin{array}{c}+80\\-80\end{array}$
       
        & 0\% & $\begin{array}{c}+30\\-30\end{array}$ & $\begin{array}{c}+80\\-80\end{array}$
        & 0\% & $\begin{array}{c}+30\\-30\end{array}$ & $\begin{array}{c}+80\\-80\end{array}$
       
        & 0\% & $\begin{array}{c}+30\\-30\end{array}$ & $\begin{array}{c}+80\\-80\end{array}$
        & 0\% & $\begin{array}{c}+30\\-30\end{array}$ & $\begin{array}{c}+80\\-80\end{array}$ \\
        \hline
       
        $C_{\ell\ell}/\Lambda^2$
        & 0.34 & $\begin{array}{c}0.24\\0.45\end{array}$ & $\begin{array}{c}0.07\\0.62\end{array}$
        & 0.28 & $\begin{array}{c}0.20\\0.36\end{array}$ & $\begin{array}{c}0.06\\0.50\end{array}$
       
        & 3.87 & $\begin{array}{c}2.71\\5.03\end{array}$ & $\begin{array}{c}0.77\\6.96\end{array}$
        & 3.85 & $\begin{array}{c}2.69\\5.00\end{array}$ & $\begin{array}{c}0.77\\6.92\end{array}$
       
        & 7.36 & $\begin{array}{c}5.15\\9.56\end{array}$ & $\begin{array}{c}1.47\\13.24\end{array}$
        & 7.35 & $\begin{array}{c}5.14\\9.54\end{array}$ & $\begin{array}{c}1.47\\13.21\end{array}$ \\
        \hline
       
        $C_{\ell e}/\Lambda^2$
        & 0.17 & $\begin{array}{c}0.17\\0.17\end{array}$ & $\begin{array}{c}0.17\\0.17\end{array}$
        & 0.15 & $\begin{array}{c}0.15\\0.15\end{array}$ & $\begin{array}{c}0.15\\0.15\end{array}$
       
        & 1.95 & $\begin{array}{c}1.95\\1.95\end{array}$ & $\begin{array}{c}1.95\\1.95\end{array}$
        & 1.94 & $\begin{array}{c}1.95\\1.94\end{array}$ & $\begin{array}{c}1.94\\1.94\end{array}$
       
        & 3.70 & $\begin{array}{c}3.70\\3.69\end{array}$ & $\begin{array}{c}3.70\\3.70\end{array}$
        & 3.69 & $\begin{array}{c}3.69\\3.69\end{array}$ & $\begin{array}{c}3.69\\3.69\end{array}$ \\
        \hline
       
        $C_{ee}/\Lambda^2$
        & 0.34 & $\begin{array}{c}0.45\\0.24\end{array}$ & $\begin{array}{c}0.62\\0.07\end{array}$
        & 0.28 & $\begin{array}{c}0.37\\0.20\end{array}$ & $\begin{array}{c}0.50\\0.06\end{array}$
       
        & 3.86 & $\begin{array}{c}5.03\\2.71\end{array}$ & $\begin{array}{c}6.96\\0.77\end{array}$
        & 3.84 & $\begin{array}{c}5.00\\2.69\end{array}$ & $\begin{array}{c}6.91\\0.77\end{array}$
       
        & 7.35 & $\begin{array}{c}9.55\\5.14\end{array}$ & $\begin{array}{c}13.23\\1.47\end{array}$
        & 7.33 & $\begin{array}{c}9.53\\5.13\end{array}$ & $\begin{array}{c}13.20\\1.47\end{array}$ \\
        \hline
       
        \multicolumn{19}{c}{Background} \\
        \hline
       
        Total BG
        & 11.21 & $\begin{array}{c}8.37\\13.98\end{array}$ & $\begin{array}{c}3.70\\18.55\end{array}$
        & 0.43 & $\begin{array}{c}0.32\\0.46\end{array}$ & $\begin{array}{c}0.25\\0.55\end{array}$
       
        & 10.29 & $\begin{array}{c}7.24\\13.52\end{array}$ & $\begin{array}{c}2.20\\18.77\end{array}$
        & 0.15 & $\begin{array}{c}0.14\\0.18\end{array}$ & $\begin{array}{c}0.09\\0.25\end{array}$
       
        & 10.65 & $\begin{array}{c}7.42\\13.82\end{array}$ & $\begin{array}{c}2.17\\19.42\end{array}$
        & 0.09 & $\begin{array}{c}0.07\\0.11\end{array}$ & $\begin{array}{c}0.05\\0.13\end{array}$ \\
        \hline
       
    \end{tabular}%
    }
   
    \resizebox{\textwidth}{!}{%
    \begin{tabular}{l*{6}{ccc}}
        \hline
        \multicolumn{19}{c}{\textbf{Process: $\mu^+ \mu^- \to \mu^\mp \tau^\pm$ }} \\
        \hline
        \multirow{4}{*}{\textbf{Couplings}}
        & \multicolumn{6}{c}{$\sqrt{s}=3$~TeV}
        & \multicolumn{6}{c}{$\sqrt{s}=10$~TeV}
        & \multicolumn{6}{c}{$\sqrt{s}=14$~TeV} \\
        \cline{2-19}
       
        & \multicolumn{3}{c}{\makebox[3.0cm]{Cut$^{(\mu\tau)}_0: N_e=0$}} & \multicolumn{3}{c}{\makebox[3.0cm]{Cut$^{(\mu\tau)}_1$}}
        & \multicolumn{3}{c}{\makebox[3.0cm]{Cut$^{(\mu\tau)}_0: N_e=0$}} & \multicolumn{3}{c}{\makebox[3.0cm]{Cut$^{(\mu\tau)}_1$}}
        & \multicolumn{3}{c}{\makebox[3.0cm]{Cut$^{(\mu\tau)}_0: N_e=0$}} & \multicolumn{3}{c}{\makebox[3.0cm]{Cut$^{(\mu\tau)}_1$}} \\
       
        & \multicolumn{3}{c}{\makebox[3.0cm]{$N_\mu=1,N_{\tau_h}=1$}} & \multicolumn{3}{c}{\makebox[3.0cm]{$p_T^\mu > 0.8$~TeV}}
        & \multicolumn{3}{c}{\makebox[3.0cm]{$N_\mu=1,N_{\tau_h}=1$}} & \multicolumn{3}{c}{\makebox[3.0cm]{$p_T^\mu > 1$~TeV}}
        & \multicolumn{3}{c}{\makebox[3.0cm]{$N_\mu=1,N_{\tau_h}=1$}} & \multicolumn{3}{c}{\makebox[3.0cm]{$p_T^\mu > 1.2$~TeV}} \\
        \cline{2-19}
       
        $\begin{array}{c}
        (1\times10^{-3}\\ \mathrm{TeV}^{-2})\end{array}$
       
        & 0\% & $\begin{array}{c}+30\\-30\end{array}$ & $\begin{array}{c}+80\\-80\end{array}$
        & 0\% & $\begin{array}{c}+30\\-30\end{array}$ & $\begin{array}{c}+80\\-80\end{array}$
       
        & 0\% & $\begin{array}{c}+30\\-30\end{array}$ & $\begin{array}{c}+80\\-80\end{array}$
        & 0\% & $\begin{array}{c}+30\\-30\end{array}$ & $\begin{array}{c}+80\\-80\end{array}$
       
        & 0\% & $\begin{array}{c}+30\\-30\end{array}$ & $\begin{array}{c}+80\\-80\end{array}$
        & 0\% & $\begin{array}{c}+30\\-30\end{array}$ & $\begin{array}{c}+80\\-80\end{array}$ \\
        \hline
       
        $C_{\ell\ell}/\Lambda^2$
        & 1.51 & $\begin{array}{c}1.06\\1.96\end{array}$ & $\begin{array}{c}0.30\\2.72\end{array}$
        & 1.22 & $\begin{array}{c}0.85\\1.58\end{array}$ & $\begin{array}{c}0.24\\2.19\end{array}$
       
        & 17.24 & $\begin{array}{c}12.07\\22.43\end{array}$ & $\begin{array}{c}3.45\\31.08\end{array}$
        & 17.08 & $\begin{array}{c}11.95\\22.22\end{array}$ & $\begin{array}{c}3.42\\30.79\end{array}$
       
        & 34.38 & $\begin{array}{c}24.08\\44.73\end{array}$ & $\begin{array}{c}6.89\\61.93\end{array}$
        & 34.30 & $\begin{array}{c}24.02\\44.62\end{array}$ & $\begin{array}{c}6.87\\61.79\end{array}$ \\
        \hline
       
        $C_{\ell e}/\Lambda^2$
        & 0.76 & $\begin{array}{c}0.76\\0.76\end{array}$ & $\begin{array}{c}0.76\\0.76\end{array}$
        & 0.64 & $\begin{array}{c}0.64\\0.64\end{array}$ & $\begin{array}{c}0.64\\0.64\end{array}$
       
        & 8.69 & $\begin{array}{c}8.68\\8.69\end{array}$ & $\begin{array}{c}8.69\\8.69\end{array}$
        & 8.63 & $\begin{array}{c}8.62\\8.63\end{array}$ & $\begin{array}{c}8.63\\8.63\end{array}$
       
        & 17.32 & $\begin{array}{c}17.33\\17.31\end{array}$ & $\begin{array}{c}17.33\\17.32\end{array}$
        & 17.29 & $\begin{array}{c}17.30\\17.28\end{array}$ & $\begin{array}{c}17.30\\17.29\end{array}$ \\
        \hline
       
        $C_{ee}/\Lambda^2$
        & 1.52 & $\begin{array}{c}1.97\\1.06\end{array}$ & $\begin{array}{c}2.73\\0.30\end{array}$
        & 1.22 & $\begin{array}{c}1.58\\0.85\end{array}$ & $\begin{array}{c}2.19\\0.24\end{array}$
       
        & 17.24 & $\begin{array}{c}22.41\\12.08\end{array}$ & $\begin{array}{c}31.04\\3.45\end{array}$
        & 17.08 & $\begin{array}{c}22.20\\11.97\end{array}$ & $\begin{array}{c}30.75\\3.41\end{array}$
       
        & 34.35 & $\begin{array}{c}44.67\\24.07\end{array}$ & $\begin{array}{c}61.85\\6.87\end{array}$
        & 34.27 & $\begin{array}{c}44.57\\24.01\end{array}$ & $\begin{array}{c}61.71\\6.85\end{array}$ \\
        \hline
       
        \multicolumn{19}{c}{Background} \\
        \hline
       
        Total BG
        & 13.33 & $\begin{array}{c}9.96\\16.82\end{array}$ & $\begin{array}{c}4.29\\22.44\end{array}$
        & 0.43 & $\begin{array}{c}0.40\\0.50\end{array}$ & $\begin{array}{c}0.29\\0.60\end{array}$
       
        & 12.57 & $\begin{array}{c}9.05\\16.56\end{array}$ & $\begin{array}{c}2.68\\22.88\end{array}$
        & 0.18 & $\begin{array}{c}0.15\\0.22\end{array}$ & $\begin{array}{c}0.11\\0.26\end{array}$
       
        & 13.06 & $\begin{array}{c}9.17\\16.81\end{array}$ & $\begin{array}{c}2.66\\23.45\end{array}$
        & 0.10 & $\begin{array}{c}0.07\\0.10\end{array}$ & $\begin{array}{c}0.06\\0.16\end{array}$ \\
        \hline
       
    \end{tabular}%
    }
   
    \resizebox{\textwidth}{!}{%
    \begin{tabular}{l*{6}{ccc}}
        \hline
        \multicolumn{19}{c}{\textbf{Process: $\mu^+ \mu^- \to e^\pm \mu^\mp$ }} \\
        \hline
        \multirow{4}{*}{\textbf{Couplings}}
        & \multicolumn{6}{c}{$\sqrt{s}=3$~TeV}
        & \multicolumn{6}{c}{$\sqrt{s}=10$~TeV}
        & \multicolumn{6}{c}{$\sqrt{s}=14$~TeV} \\
        \cline{2-19}
       
        & \multicolumn{3}{c}{\makebox[3.0cm]{Cut$^{(e\mu)}_0: N_{\tau_h}=0$}} & \multicolumn{3}{c}{\makebox[3.0cm]{Cut$^{(\mu\tau)}_1$}}
        & \multicolumn{3}{c}{\makebox[3.0cm]{Cut$^{(e\mu)}_0: N_{\tau_h}=0$}} & \multicolumn{3}{c}{\makebox[3.0cm]{Cut$^{(\mu\tau)}_1$}}
        & \multicolumn{3}{c}{\makebox[3.0cm]{Cut$^{(e\mu)}_0: N_{\tau_h}=0$}} & \multicolumn{3}{c}{\makebox[3.0cm]{Cut$^{(\mu\tau)}_1$}} \\
       
        & \multicolumn{3}{c}{\makebox[3.0cm]{$N_e=1,N_\mu=1$}} & \multicolumn{3}{c}{\makebox[3.0cm]{$p_T^e > 0.8$~TeV}}
        & \multicolumn{3}{c}{\makebox[3.0cm]{$N_e=1,N_\mu=1$}} & \multicolumn{3}{c}{\makebox[3.0cm]{$p_T^e > 1$~TeV}}
        & \multicolumn{3}{c}{\makebox[3.0cm]{$N_e=1,N_\mu=1$}} & \multicolumn{3}{c}{\makebox[3.0cm]{$p_T^e > 1.2$~TeV}} \\
        \cline{2-19}
       
        $\begin{array}{c}
        (1\times10^{-3}\\ \mathrm{TeV}^{-2})\end{array}$
       
        & 0\% & $\begin{array}{c}+30\\-30\end{array}$ & $\begin{array}{c}+80\\-80\end{array}$
        & 0\% & $\begin{array}{c}+30\\-30\end{array}$ & $\begin{array}{c}+80\\-80\end{array}$
       
        & 0\% & $\begin{array}{c}+30\\-30\end{array}$ & $\begin{array}{c}+80\\-80\end{array}$
        & 0\% & $\begin{array}{c}+30\\-30\end{array}$ & $\begin{array}{c}+80\\-80\end{array}$
       
        & 0\% & $\begin{array}{c}+30\\-30\end{array}$ & $\begin{array}{c}+80\\-80\end{array}$
        & 0\% & $\begin{array}{c}+30\\-30\end{array}$ & $\begin{array}{c}+80\\-80\end{array}$ \\
        \hline
       
        $C_{\ell\ell}/\Lambda^2$
        & 2.60 & $\begin{array}{c}1.82\\3.37\end{array}$ & $\begin{array}{c}0.52\\4.67\end{array}$
        & 2.11 & $\begin{array}{c}1.48\\2.75\end{array}$ & $\begin{array}{c}0.42\\3.80\end{array}$
       
        & 28.22 & $\begin{array}{c}19.76\\36.72\end{array}$ & $\begin{array}{c}5.65\\50.81\end{array}$
        & 28.07 & $\begin{array}{c}19.65\\36.52\end{array}$ & $\begin{array}{c}5.61\\50.53\end{array}$
       
        & 52.73 & $\begin{array}{c}36.90\\68.56\end{array}$ & $\begin{array}{c}10.55\\94.88\end{array}$
        & 52.67 & $\begin{array}{c}36.86\\68.48\end{array}$ & $\begin{array}{c}10.54\\94.77\end{array}$ \\
        \hline
       
        $C_{\ell e}/\Lambda^2$
        & 1.31 & $\begin{array}{c}1.31\\1.31\end{array}$ & $\begin{array}{c}1.31\\1.31\end{array}$
        & 1.12 & $\begin{array}{c}1.12\\1.12\end{array}$ & $\begin{array}{c}1.12\\1.12\end{array}$
       
        & 14.25 & $\begin{array}{c}14.24\\14.24\end{array}$ & $\begin{array}{c}14.24\\14.24\end{array}$
        & 14.19 & $\begin{array}{c}14.18\\14.18\end{array}$ & $\begin{array}{c}14.18\\14.18\end{array}$
       
        & 26.49 & $\begin{array}{c}26.49\\26.48\end{array}$ & $\begin{array}{c}26.49\\26.49\end{array}$
        & 26.47 & $\begin{array}{c}26.46\\26.46\end{array}$ & $\begin{array}{c}26.47\\26.47\end{array}$ \\
        \hline
       
        $C_{ee}/\Lambda^2$
        & 2.60 & $\begin{array}{c}3.37\\1.82\end{array}$ & $\begin{array}{c}4.67\\0.52\end{array}$
        & 2.11 & $\begin{array}{c}2.74\\1.48\end{array}$ & $\begin{array}{c}3.80\\0.42\end{array}$
       
        & 28.23 & $\begin{array}{c}36.71\\19.76\end{array}$ & $\begin{array}{c}50.81\\5.65\end{array}$
        & 28.08 & $\begin{array}{c}36.50\\19.65\end{array}$ & $\begin{array}{c}50.53\\5.61\end{array}$
       
        & 52.73 & $\begin{array}{c}68.54\\36.90\end{array}$ & $\begin{array}{c}94.89\\10.55\end{array}$
        & 52.67 & $\begin{array}{c}68.47\\36.86\end{array}$ & $\begin{array}{c}94.79\\10.54\end{array}$ \\
        \hline
       
        \multicolumn{19}{c}{Background} \\
        \hline
       
        Total BG
        & 8.57 & $\begin{array}{c}6.03\\10.59\end{array}$ & $\begin{array}{c}2.17\\14.77\end{array}$
        & 0.38 & $\begin{array}{c}0.27\\0.42\end{array}$ & $\begin{array}{c}0.12\\0.59\end{array}$
       
        & 6.37 & $\begin{array}{c}4.52\\8.23\end{array}$ & $\begin{array}{c}1.30\\11.54\end{array}$
        & 0.13 & $\begin{array}{c}0.10\\0.16\end{array}$ & $\begin{array}{c}0.04\\0.24\end{array}$
       
        & 6.43 & $\begin{array}{c}4.41\\8.34\end{array}$ & $\begin{array}{c}1.32\\11.56\end{array}$
        & 0.07 & $\begin{array}{c}0.04\\0.09\end{array}$ & $\begin{array}{c}0.02\\0.14\end{array}$ \\
        \hline
        \hline
       
    \end{tabular}%
    }
\end{table*}
The corresponding signal and SM background cross sections after the baseline selection \(\mathrm{Cut}_0^{(f_1f_2)}\) and after the additional hard transverse-momentum requirement \(\mathrm{Cut}_1^{(f_1f_2)}\) are summarized in Table~\ref{xsec-cuts}. The results demonstrate that the hard-\(p_T\) selection provides the dominant kinematic discrimination between LFV signals and SM backgrounds across all collider energies considered in this study. The hard transverse-momentum requirement suppresses the SM backgrounds by more than an order of magnitude at all center-of-mass energies while retaining most of the LFV signal, thereby providing the principal source of signal-background discrimination in the cut-based analysis.

Although the distributions displayed in Fig.~\ref{PtKin} correspond explicitly to the \(e^\pm\tau^\mp\) final state, similar distributions are obtained for the \(\mu^\pm\tau^\mp\) and \(e^\pm\mu^\mp\) channels. We therefore employ an analogous kinematic selection strategy for all LFV final states in the subsequent optimal-observable analysis. In the \(\mu^\pm\tau^\mp\) channel, the transverse-momentum selection is applied to the prompt muon.


\section{Optimal-Observable Technique}
\label{sec:oot}
The method of optimal observables has been widely used in precision studies of anomalous gauge-boson couplings~\cite{Diehl:1993br}, Higgs couplings at lepton colliders~\cite{Gunion:1996vv,Dutta:2008bh}, and more recently in SMEFT studies of lepton-flavor violation (LFV)~\cite{Bhattacharya:2023lfv,Jahedi:2024lfv}.

In this section, we apply the OOT to quantify the sensitivity of a muon collider to LFV four-lepton operators using the \(\cos\theta\) distribution of the prompt charged lepton in the LFV final state \(f_1f_2\), where \(\theta\) denotes its polar angle with respect to the beam axis. For fixed center-of-mass energy \(\sqrt{s}\) and muon-beam polarization \(P_{\mu^-}\), the differential event yield is
\begin{align}
\mathcal{L}_{\rm int}&\frac{d\sigma_{\rm total}(\sqrt{s},P_{\mu^-})}{d(\cos\theta)} =
\mathcal{L}_{\rm int}
\frac{d\sigma_{\rm SM}(\sqrt{s},P_{\mu^-})}{d(\cos\theta)}\nonumber\\
&+ \mathcal{L}_{\rm int}
\sum_{\alpha,\,\beta}
C_\alpha^{f_1f_2}\, C_\beta^{f_1f_2}\
\frac{d\sigma_{\alpha\beta}(\sqrt{s},P_{\mu^-})}{d(\cos\theta)} ,
\end{align}
where \(\alpha,\beta\in \left\{\ell\ell,\ell e,ee\right\}\). The term \(\sigma_{\rm SM}\) denotes the flavor-conserving SM background, while \(\sigma_{\alpha\beta}\) contains the pure EFT contributions, including interference among dimension-six operators. Since LFV amplitudes are absent in the SM, there is no linear SM-EFT interference.

For a binned analysis in \(\cos\theta\), the expected number of events in the \(k\)th bin is
\begin{equation}
\mathcal{N}^{k}=\mathcal{N}^{k}_{\rm SM}
+ \sum_{\alpha,\,\beta}
C_\alpha^{f_1f_2}\, C_\beta^{f_1f_2}
\left(\mathcal{N}_{\alpha\beta}^{f_1f_2}\right)^k,
\end{equation}
where \( \left(\mathcal{N}_{\alpha\beta}^{f_1f_2}\right)^k=\mathcal{L}_{\rm int}\,
\sigma_{\alpha\beta}^{k}\).

\par As discussed earlier, the delta functions in Eq.~\eqref{eq:del} ensure that the cross terms between different Wilson coefficients vanish. As a result, the inverse covariance matrix can be written as
\begin{equation}
\left(V^{-1}\right)_{\alpha\beta}^{f_1f_2}=\sum_k
\frac{
\left(\mathcal{N}_{\alpha\alpha}^{f_1f_2}\right)^k
\left(\mathcal{N}_{\beta\beta}^{f_1f_2}\right)^k
}
{\left(\Delta\mathcal{N}^{k}\right)^2},
\end{equation}
where
\begin{equation}
\Delta\mathcal{N}^{k}=\sqrt{
(\delta \mathcal{N}_{\rm stat}^{k})^2
+
(\delta \mathcal{N}_{\rm sys}^{k})^2
}.
\end{equation}
\par We assume Poisson statistics, such that the statistical uncertainty is given by
\(\delta \mathcal{N}_{\rm stat}^{k} = \sqrt{\mathcal{N}_k}\). The systematic uncertainty is parametrized as
\(\delta \mathcal{N}_{\rm sys}^{k} = \epsilon\,\mathcal{N}_k\), where we take \(\epsilon = 1\%\).

The sensitivity obtained with the optimal-observable method depends on the binning of the angular distribution. After verifying the stability of the extracted limits under variations of the bin size, we adopt a bin width of  \(\Delta\cos\theta=0.1\), which provides an optimal compromise between statistical fluctuations and the ability to resolve differences in the angular distributions induced by the various operator structures.

The corresponding test statistic  is 
\begin{equation}
\left(\chi^2\right)_{\alpha\beta}^{f_1f_2}(\sqrt{s},P_{\mu^-})=\left(C_\alpha^{f_1f_2}\right)^2
(V_{\alpha\beta})^{-1}
\left(C_\beta^{f_1f_2}\right)^2 ,
\end{equation}
\par
\begin{figure*}[!ht]
    \centering
    \includegraphics[width=0.32\textwidth]{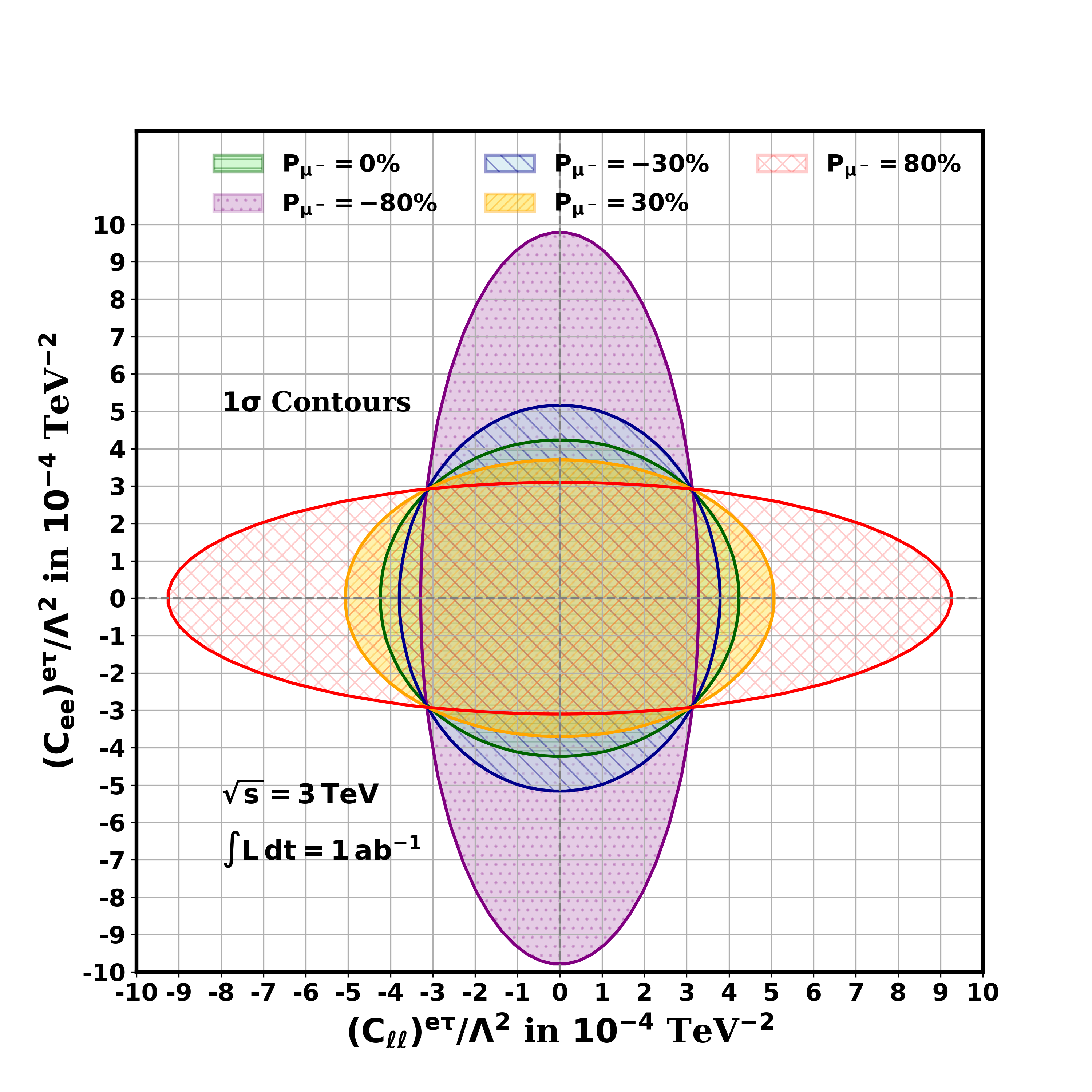}
    \includegraphics[width=0.32\textwidth]{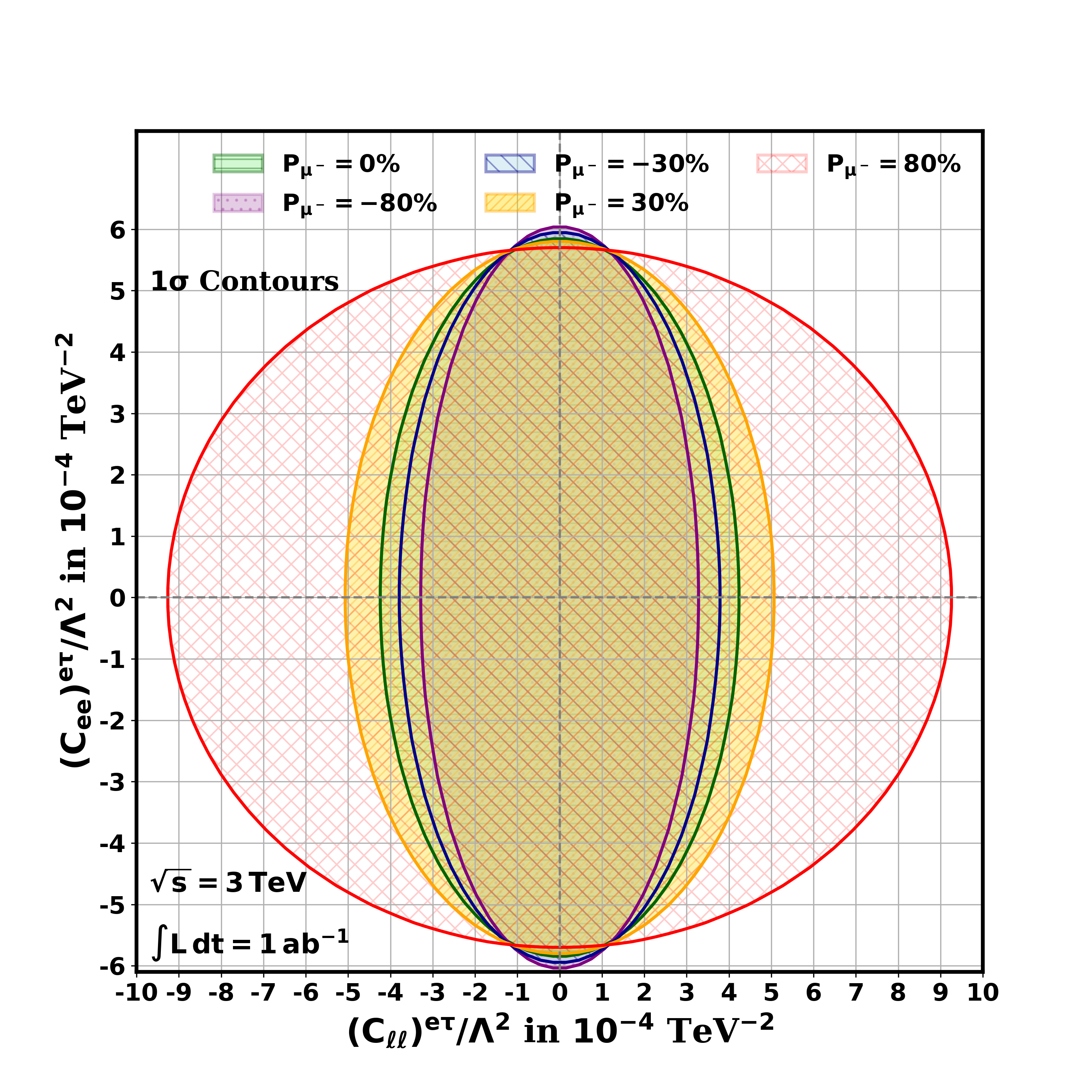}
    \includegraphics[width=0.32\textwidth]{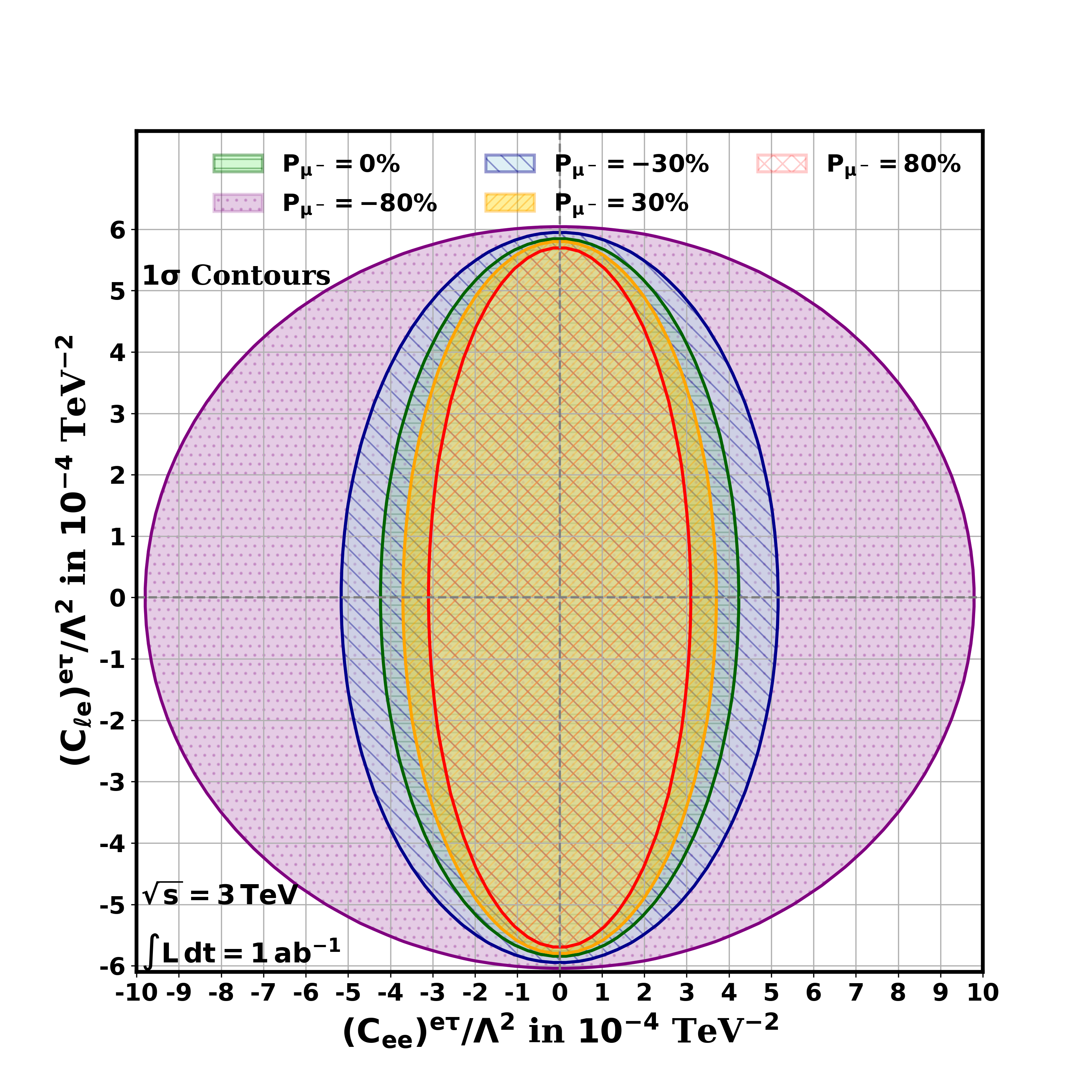}
    \includegraphics[width=0.32\textwidth]{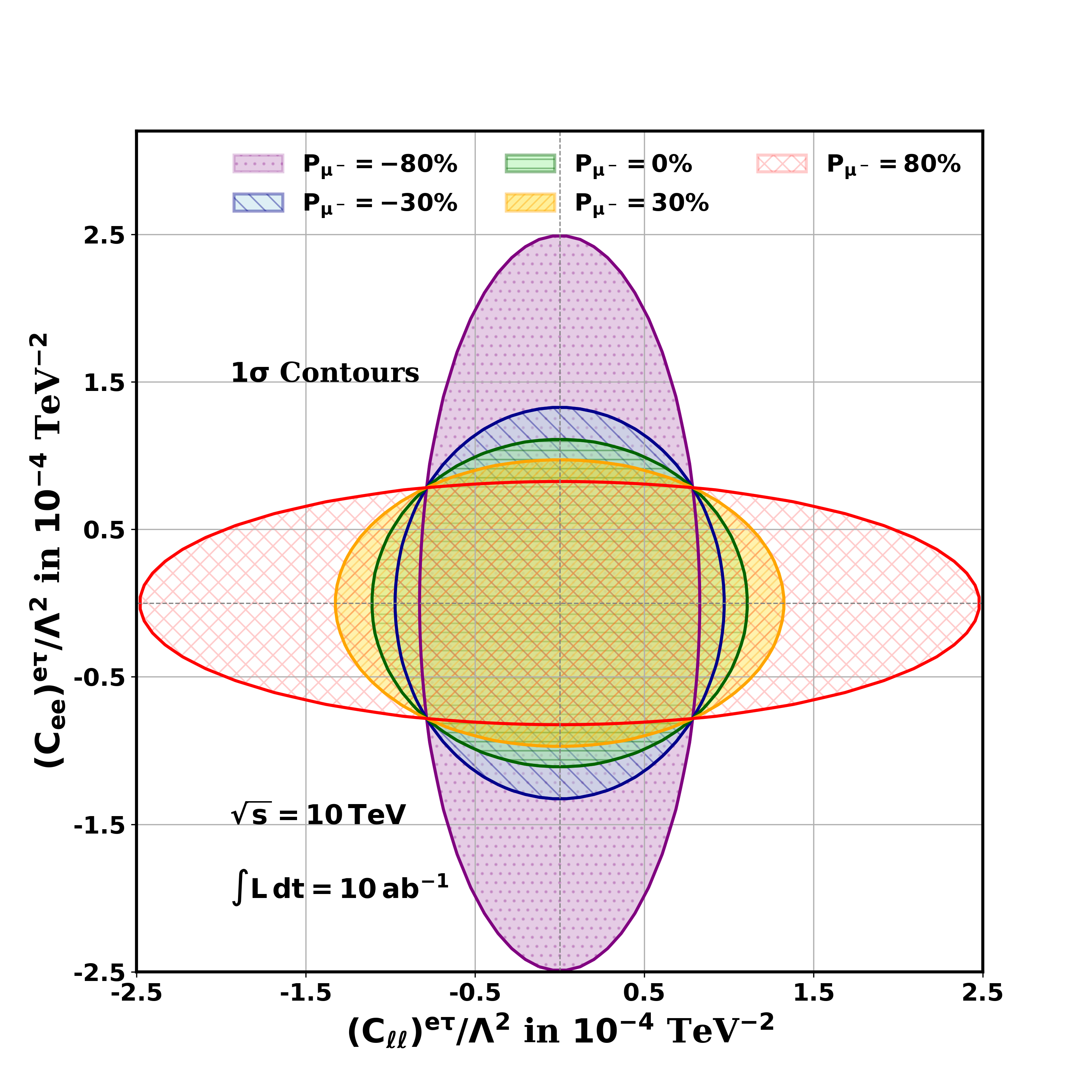}
    \includegraphics[width=0.32\textwidth]{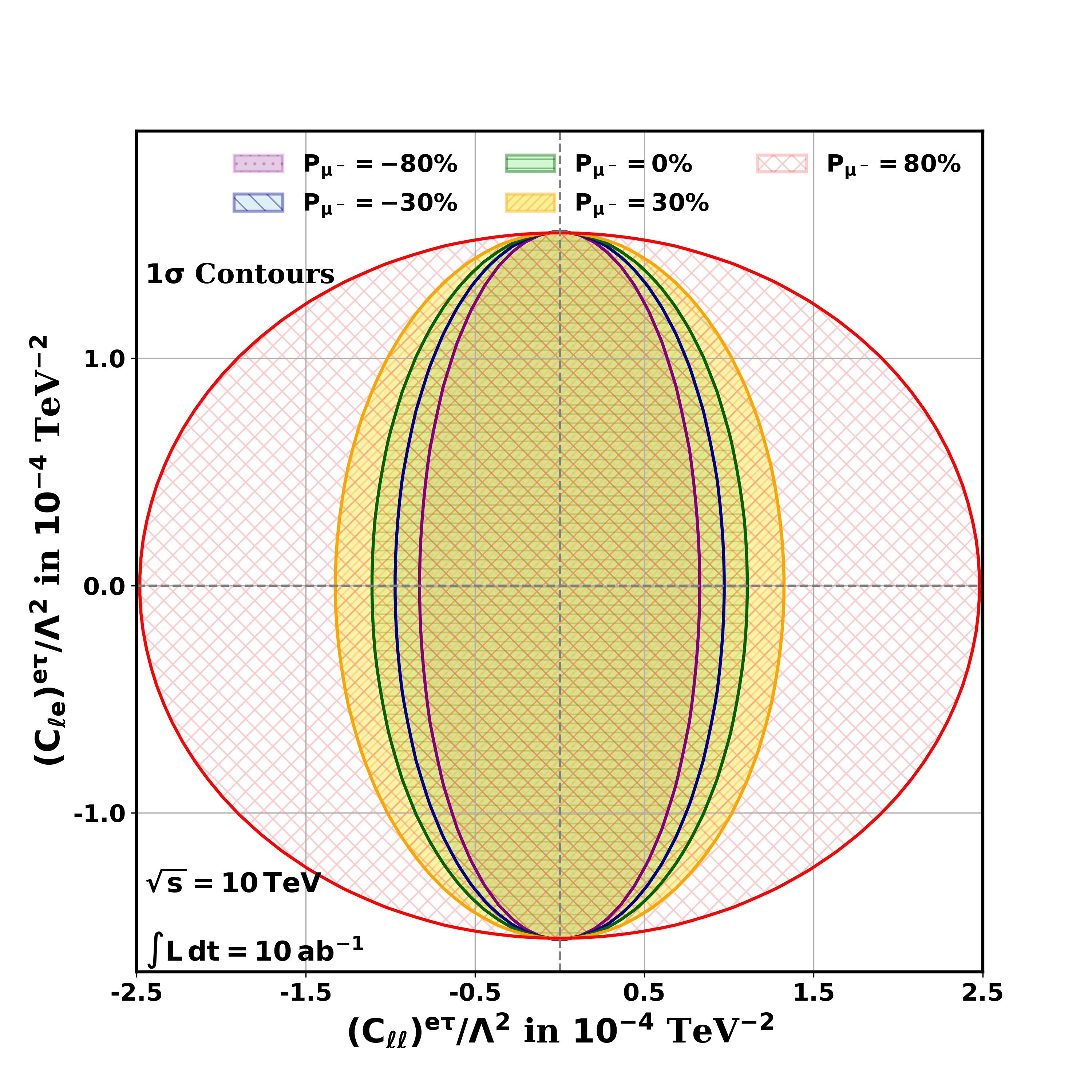}
    \includegraphics[width=0.32\textwidth]{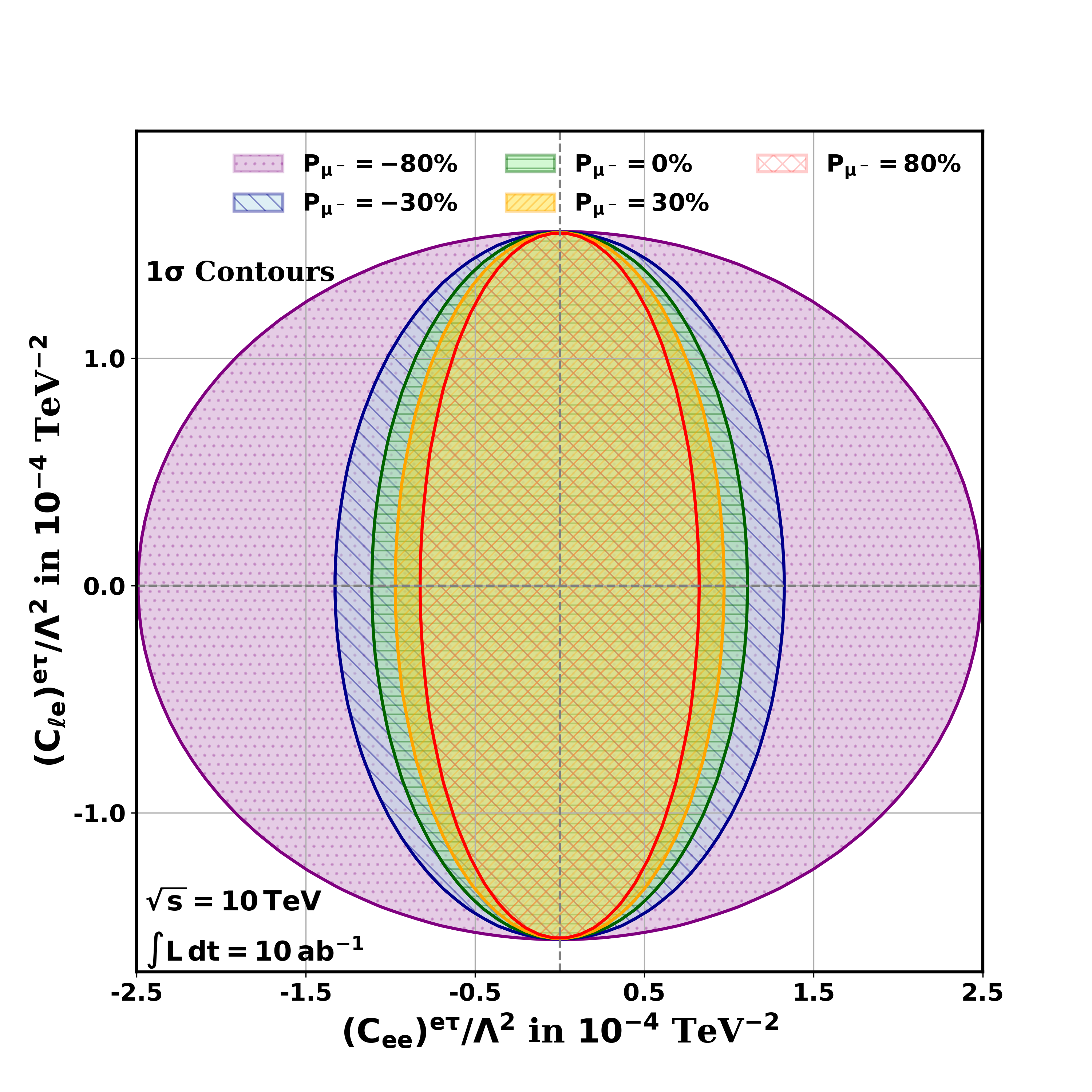}
    \includegraphics[width=0.32\textwidth]{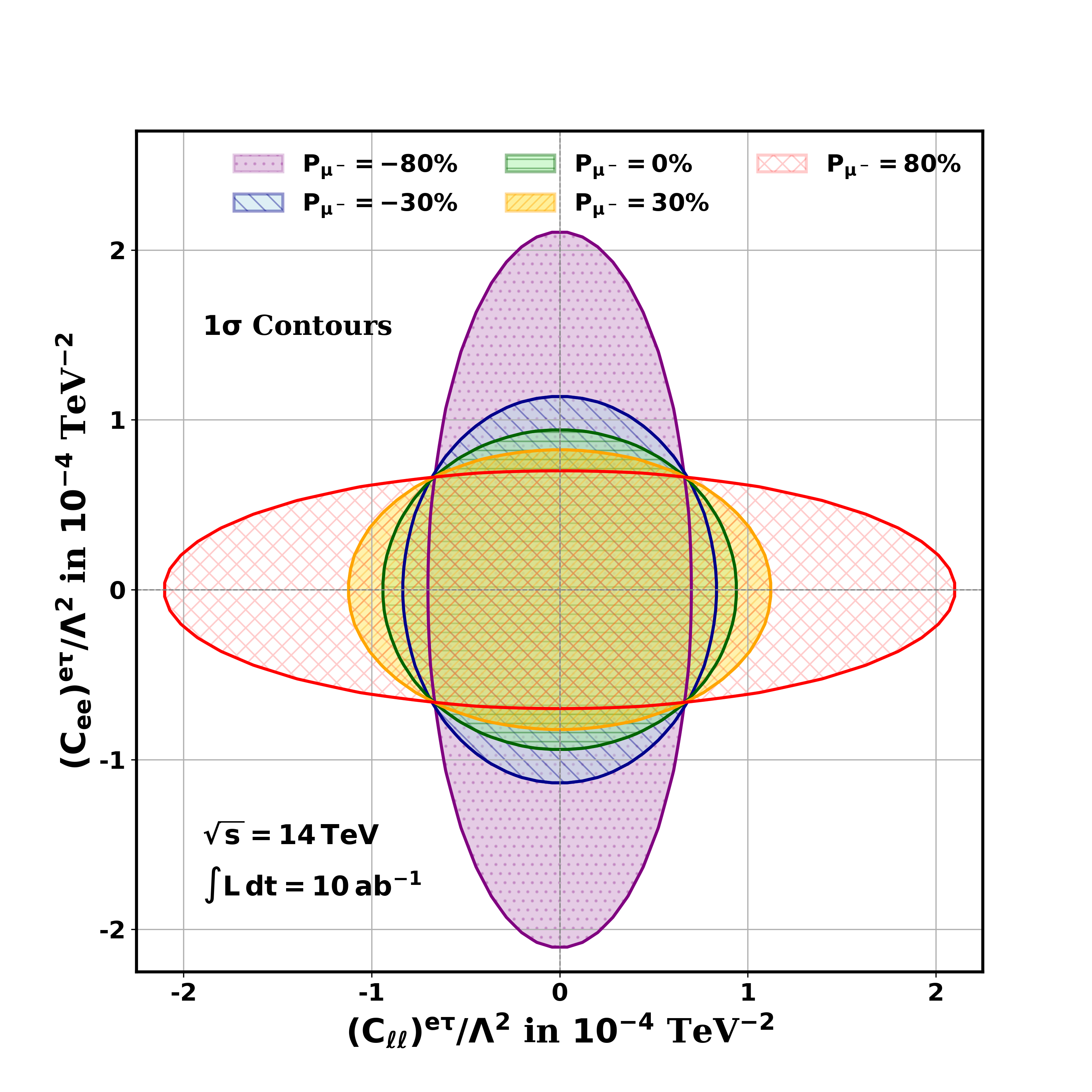}
    \includegraphics[width=0.32\textwidth]{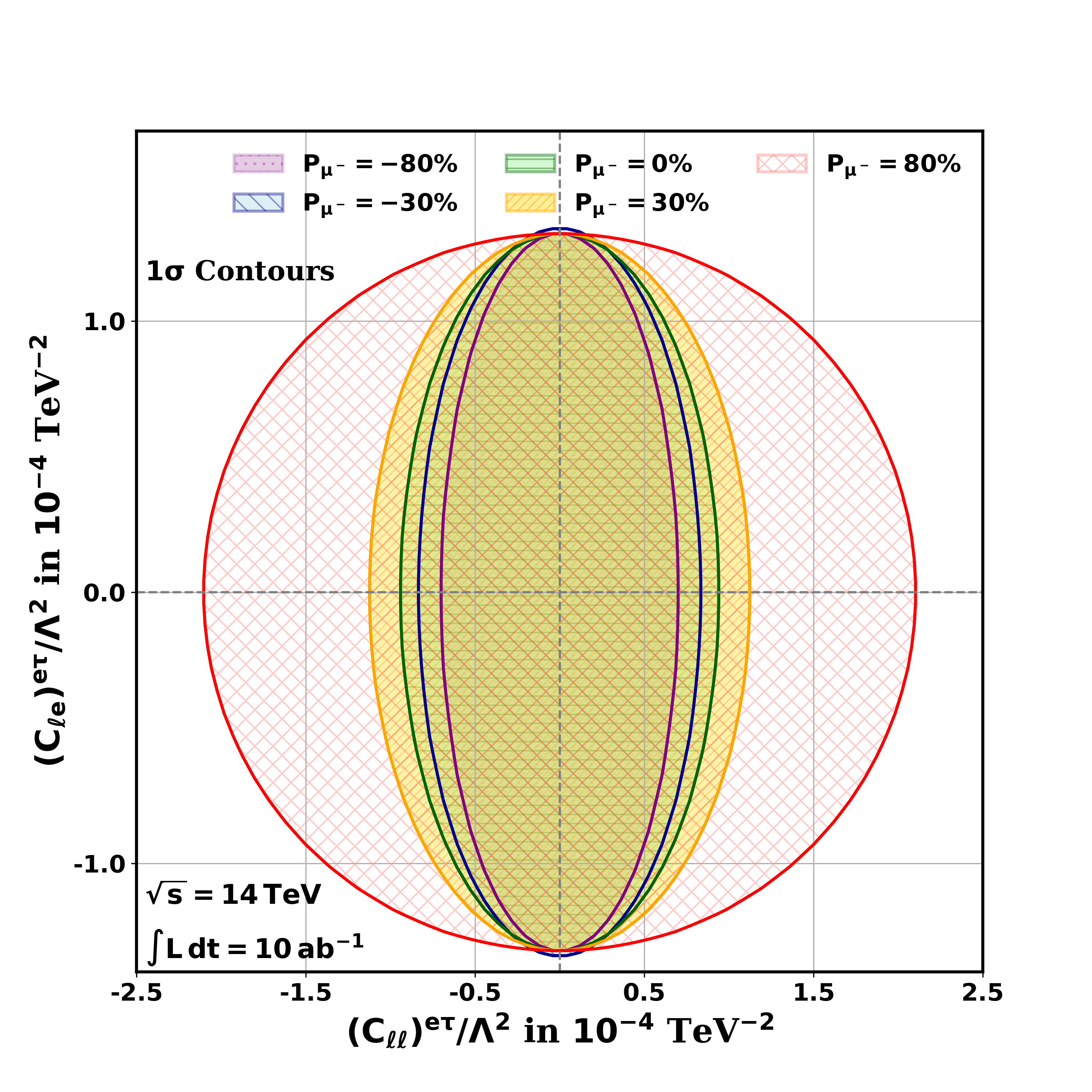}
    \includegraphics[width=0.32\textwidth]{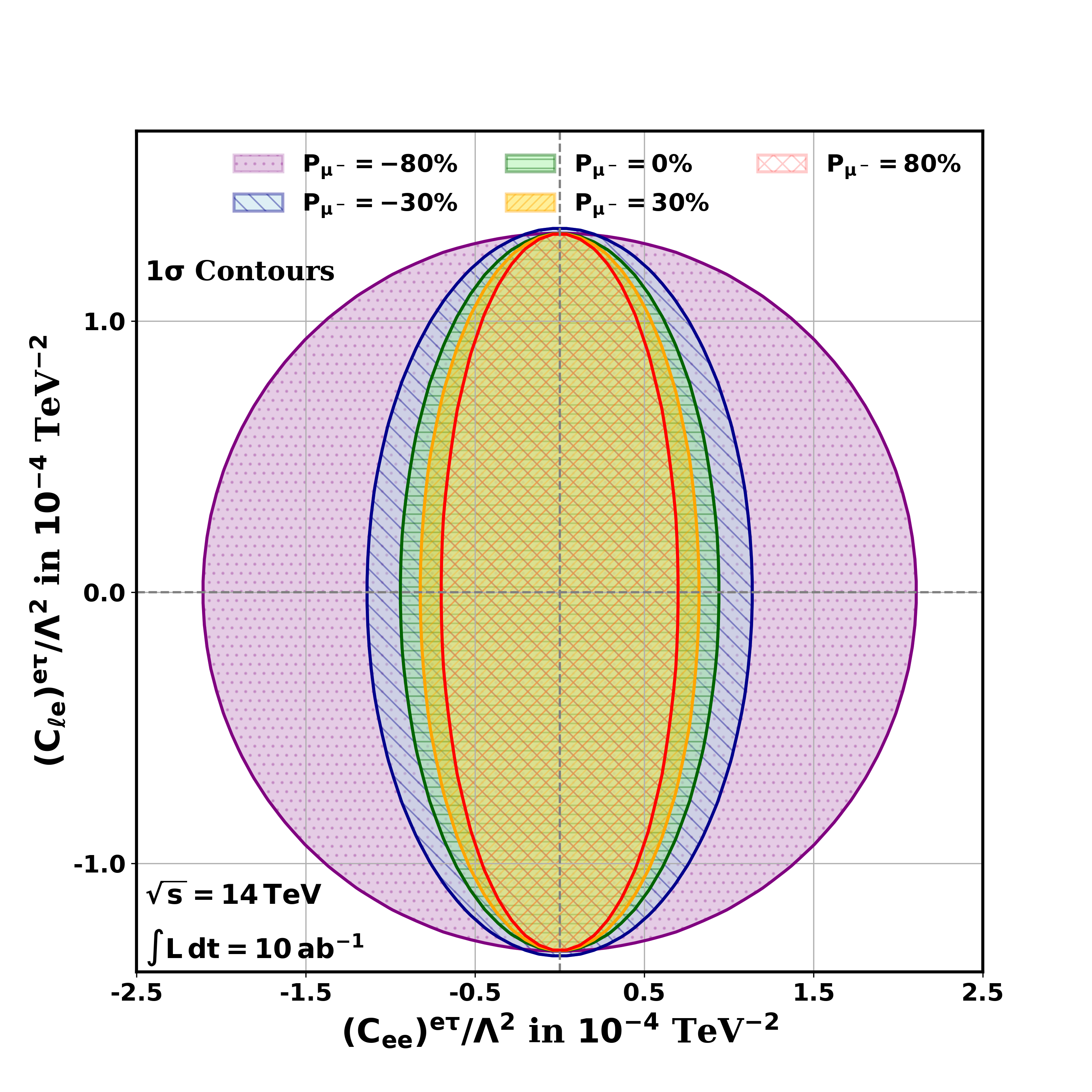}      
    \caption{Shaded \(1\sigma\)(\(\Delta\chi^2=2.30\)) C.L. region in the \((C_{\ell\ell}^{e\tau}/\Lambda^2,; C_{ee}^{e\tau}/\Lambda^2)\), \((C_{\ell\ell}^{e\tau}/\Lambda^2,; C_{\ell e}^{e\tau}/\Lambda^2)\), and \((C_{ee}^{e\tau}/\Lambda^2,; C_{\ell e}^{e\tau}/\Lambda^2)\) planes for the process \(\mu^+\mu^- \to e^\pm\tau^\mp\) for a given beam-polarization, shown for three center-of-mass energies with their corresponding integrated luminosities.}
    \label{etauConts}
\end{figure*}

 The total \(\left(\chi^2\right)_{\alpha\beta}^{f_1f_2}\) is obtained by summing the binwise contributions over all \(\cos\theta\) bins for each beam-polarization and center-of-mass-energy configuration. Figure~\ref{etauConts} shows the projected \(1\sigma\) \((\left(\Delta\chi^2\right)_{\alpha\beta}^{f_1f_2}=2.30)\) contours in the (\(C_{\ell\ell}^{e\tau}/\Lambda^2\,-\, C_{ee}^{e\tau}/\Lambda^2\)), (\(C_{\ell\ell}^{e\tau}/\Lambda^2\,-\, C_{\ell e}^{e\tau}/\Lambda^2\)), and (\(C_{ee}^{e\tau}/\Lambda^2\,-\, C_{\ell e}^{e\tau}/\Lambda^2\)) planes for the LFV process \(\mu^+\mu^- \to e^\pm\tau^\mp\). The corresponding \(1\sigma\) contours for the remaining LFV channels, \(\mu^+\mu^-\to \mu^\pm\tau^\mp\) and  \(\mu^+\mu^-\to e^\pm\mu^\mp\)  are provided in Appendix~A and shown in Figs.~\ref{mtauConts} and \ref{emuConts}, respectively.

The chirality structure of the operators plays a key role in determining the sensitivity. The strongest constraints on the left-chiral operator \(O_{\ell\ell}\) are obtained with left-polarized muon beams \((P_{\mu^-}=-80\%)\), while \(O_{ee}\) is most effectively constrained by right-polarized beams. In contrast, the mixed-chirality operator \(O_{\ell e}\) is largely insensitive to beam polarization due to its simultaneous coupling to left- and right-handed leptons. The orientation and eccentricity of the allowed ellipses are governed by operator correlations encoded in the covariance matrix, with stronger correlations producing more elongated contours.

For a fixed center-of-mass energy, the sensitivities differ among LFV channels due to the varying reconstruction efficiencies of the final-state leptons. Channels containing \(\tau\) leptons generally yield weaker constraints than those involving only electrons and muons because of the lower \(\tau\)-tagging and background-rejection efficiencies.

The projected sensitivities improve the current experimental limits on several LFV Wilson coefficients by up to an order of magnitude, depending on the operator and final state. The allowed parameter space contracts with increasing center-of-mass energy because the pure EFT contribution grows approximately as \((s/\Lambda^2)^2\), enhancing both the signal-to-background ratio and the sensitivity to LFV operators.

A higher integrated luminosity leads to stronger projected constraints, since the $\chi^2$ statistic scales linearly with the integrated luminosity. For a fixed value of $\chi^2$, the coupling strength therefore scales as
\begin{equation}
    C_{\alpha} \propto \mathcal{L}^{-1/4},
\end{equation}
implying that the sensitivity to the coupling improves with the fourth root of the integrated luminosity.

Finally, the sensitivity is dominated by quadratic EFT terms proportional to \((C_\alpha/\Lambda^2)^2\). In the absence of SM-EFT interference, positive and negative values of the Wilson coefficients yield identical \(\chi^2\) values, leading to contours that are symmetric under \(C_\alpha \rightarrow -C_\alpha\).

\subsection{Covariance analysis at a fixed center-of-mass energy}
\label{sec:covariance}
While the projected confidence contours quantify the sensitivity to individual Wilson coefficients, the covariance matrix provides a more complete description of the fit by determining both their projected uncertainties and mutual correlations. For a given center-of-mass energy, beam polarization, and LFV final state \(f_1f_2\), the elements of correlation matrix (\(\rho_{\alpha\beta}\)) and the projected \(1\sigma\) uncertainty of each squared Wilson coefficient (\(\varepsilon_{C^2_\alpha}\)) are respectively defined as
\begin{equation}
\rho_{\alpha\beta}^{f_1f_2}=\frac{V_{\alpha\beta}^{f_1f_2}}{ \sqrt{V_{\alpha\alpha}^{f_1f_2} V_{\beta\beta}^{f_1f_2}}}\,;  
\varepsilon_{\alpha}^{f_1f_2} \equiv \varepsilon_{(C_\alpha^{f_1f_2})^2}=\sqrt{V_{\alpha\alpha}^{f_1f_2}}.   
\end{equation}

To demonstrate the covariance structure, we consider the benchmark process
\(\mu^+\mu^-\rightarrow e^\pm\tau^\mp\)
at
\(\sqrt{s}=3~{\rm TeV}\)
with an unpolarized muon beam. The projected \(1\sigma\) uncertainties and corresponding correlation matrix are
\begin{equation}
\begin{aligned}
\varepsilon_{\ell\ell}^{e\tau}&=\pm51.66,\\
\varepsilon_{ee}^{e\tau}&=\pm50.75,\\
\varepsilon_{\ell e}^{e\tau}&=\pm4.604,
\end{aligned}
\qquad
\begin{pmatrix}
1 & -0.99 & -0.34\\
-0.99 & 1 & 0.29\\
-0.34 & 0.29 & 1
\end{pmatrix}.
\end{equation}

The large off-diagonal elements indicate that the squared Wilson coefficients are strongly correlated and therefore cannot be extracted independently. Instead, the measured angular distributions constrain specific combinations of operators.

To identify these combinations, we diagonalize the covariance matrix. Its eigenvectors define the principal directions of the fit, while the corresponding eigenvalues represent the variances along those directions. The principal direction associated with the largest eigenvalue is the least-constrained direction, corresponding to the combination of squared Wilson coefficients that produces the smallest change in the measured angular distributions.

For the unpolarized beam configuration, the least-constrained direction is
\begin{equation}
-0.71
\left(C_{\ell\ell}^{e\tau}\right)^2
+
0.70
\left(C_{ee}^{e\tau}\right)^2
+
0.02
\left(C_{\ell e}^{e\tau}\right)^2
=
\pm72.4,
\end{equation}
with a condition number, defined here as the square root of the ratio of the largest to the smallest of the eigenvalues of the covariance matrix,
\begin{equation}
\kappa=
\sqrt{\frac{\lambda_{\rm max}}
{\lambda_{\rm min}}}
=
392.7,
\end{equation}
Such a large condition number indicates that certain combinations of squared Wilson coefficients are much more weakly constrained than others.

The least-constrained direction depends strongly on the beam polarization. For positively polarized beams
(\(P_{\mu^-}=+80\%\) and \(+30\%\)),
the dominant eigenvectors are approximately
\[
(-0.99,\,0.11,\,0.001),
\qquad
(-0.88,\,0.47,\,-0.008),
\]
whereas for negatively polarized beams
(\(P_{\mu^-}=-80\%\) and \(-30\%\))
they become
\[
(-0.11,\,0.99,\,-0.001),
\qquad
(-0.47,\,0.88,\,0.008).
\]
This behavior reflects the chiral structure of the four-lepton operators: left-polarized beams are primarily sensitive to \(O_{\ell\ell}\), whereas right-polarized beams preferentially probe \(O_{ee}\). Consequently, different beam polarizations constrain complementary combinations of squared  Wilson coefficients, effectively rotating the principal axes of the covariance ellipsoid.

Combining the statistically independent beam-polarization datasets further reduces these degeneracies. The combined test statistic at fixed center-of-mass energy is
\begin{equation}
\left(\chi^2_{\rm Pol.\,comb.}\right)_{\alpha\beta}^{f_1f_2}
=
\sum_{P_{\mu^-}}
\left(\chi^2\right)_{\alpha\beta}^{f_1f_2}
(\sqrt{s},P_{\mu^-}),
\end{equation}
for which the least-constrained direction becomes
\begin{align}
&\sqrt{s}:3~{\rm TeV}:\nonumber\\&-0.246\left(C_{\ell\ell}^{e\tau}\right)^2
-0.254\left(C_{ee}^{e\tau}\right)^2
+0.935\left(C_{\ell e}^{e\tau}\right)^2
=\pm2.18,
\end{align}
After combining the data from the different beam polarizations, we observe that the uncertainty along the least-constrained direction is further reduced to \(\pm 2.18\), with corresponding \(\kappa=26.27\) for 3 TeV center-of-mass energy. A similar procedure is then applied to combine the datasets at the different center-of-mass energies, with the least-constrained directions are as,
\begin{align}
&\sqrt{s}=10~{\rm TeV}:\nonumber\\
&-0.237\left(C_{\ell\ell}^{e\tau}\right)^2
-0.238\left(C_{ee}^{e\tau}\right)^2
+0.942\left(C_{\ell e}^{e\tau}\right)^2
=\pm0.111, 
\nonumber\\[1mm]
&\sqrt{s}=14~{\rm TeV}:\nonumber\\
&-0.236\left(C_{\ell\ell}^{e\tau}\right)^2
-0.235\left(C_{ee}^{e\tau}\right)^2
+0.943\left(C_{\ell e}^{e\tau}\right)^2
=\pm0.080.
\end{align}

We observe that higher collision energies shrink the confidence ellipsoid by reducing the uncertainty along its least-constrained direction, while leaving the direction of the least-constrained combination of squared Wilson coefficients largely unchanged.
\subsection{Global uncertainties and correlations}
\label{sec:global}
The covariance analysis presented in the previous subsection characterizes the sensitivity of a given beam-polarization configuration at a fixed center-of-mass energy and the polarization-combined test statistic for each center-of-mass energy.

The measurements performed at different center-of-mass energies are likewise statistically independent. Consequently, combining the dataset from different center-of-mass energies through the global statistics,
\begin{equation}
\left(\chi^2_{\rm global}\right)_{\alpha\beta}^{f_1f_2}
=
\sum_{\sqrt{s}\in\{3,\,10,\,14\}\,{\rm TeV}}
\left(\chi^2_{\rm Pol.\,comb.}\right)_{\alpha\beta}^{f_1f_2}
(\sqrt{s}),
\end{equation}
which combines all beam-polarization configurations and collision energies considered in this analysis.
\begin{figure*}[!ht]
    \centering
    \includegraphics[width=0.3\textwidth]{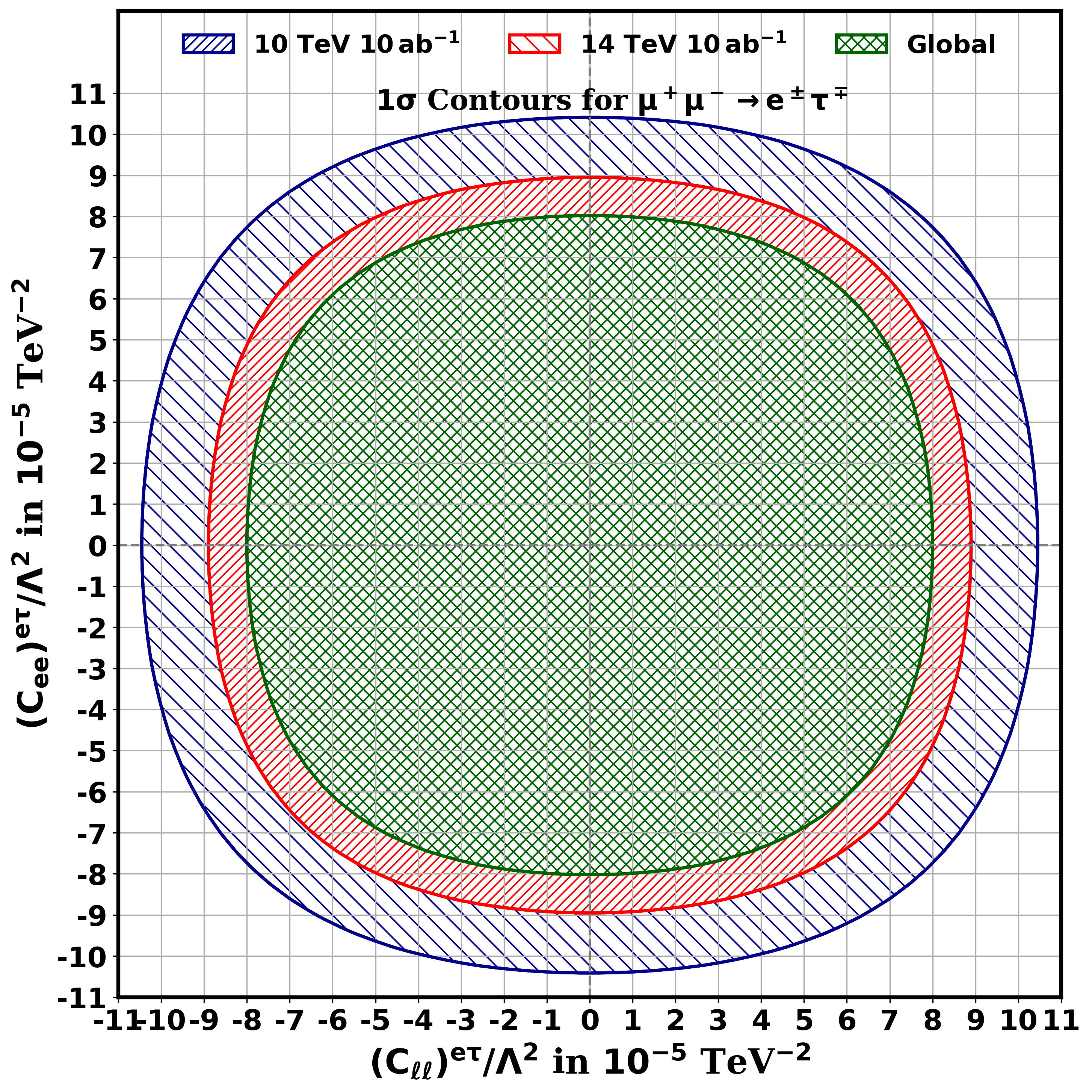}
    \includegraphics[width=0.3\textwidth]{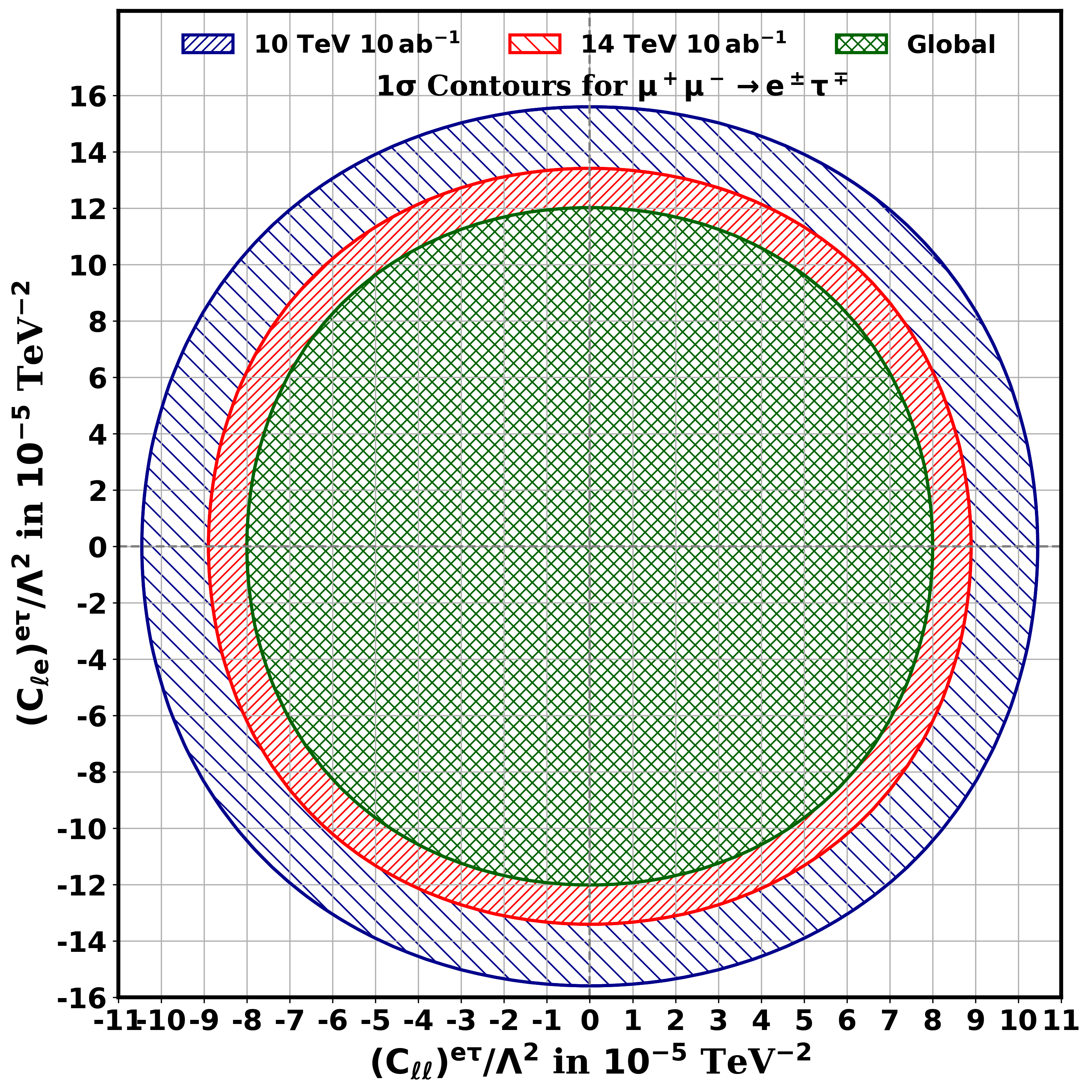}
    \includegraphics[width=0.3\textwidth]{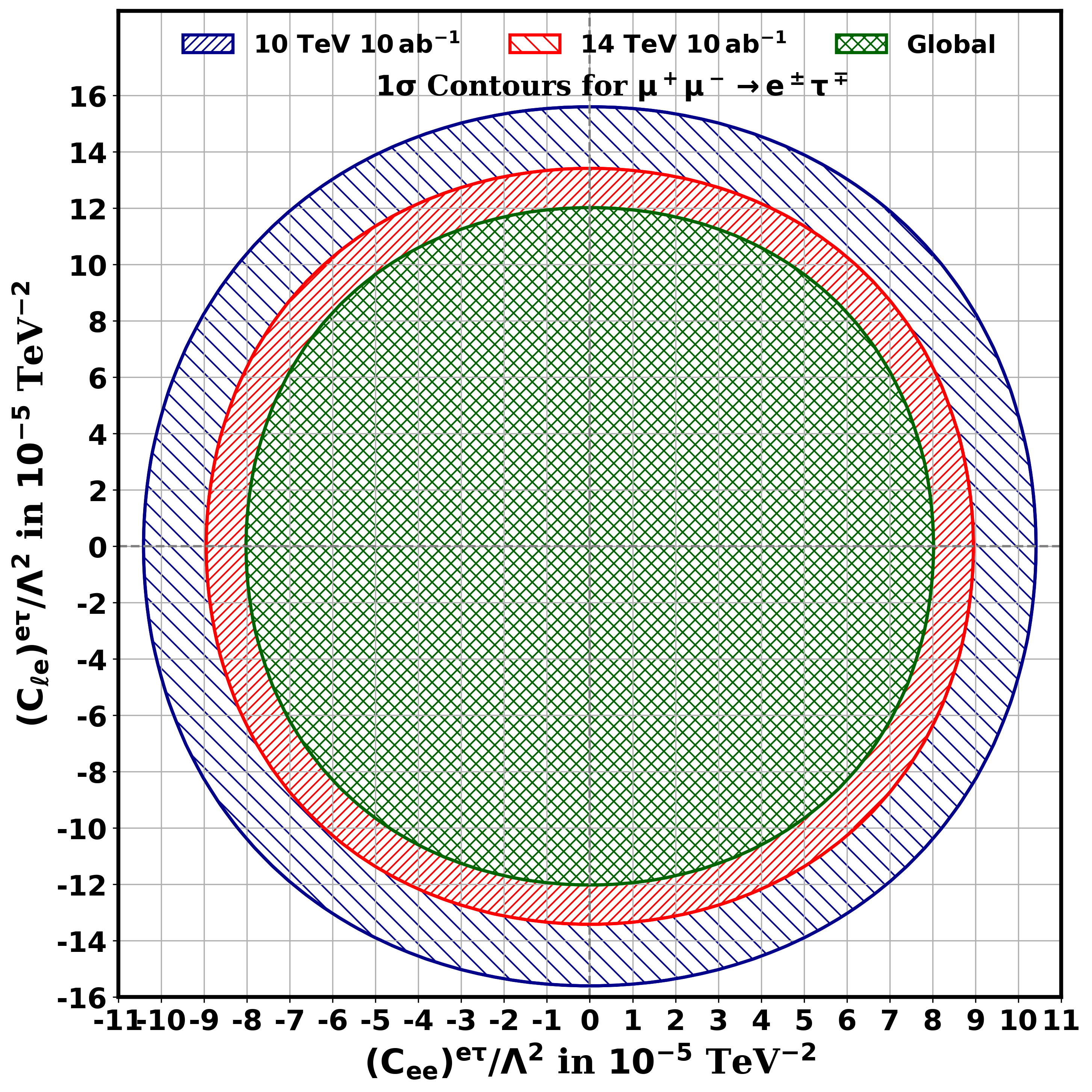}
    \includegraphics[width=0.3\textwidth]{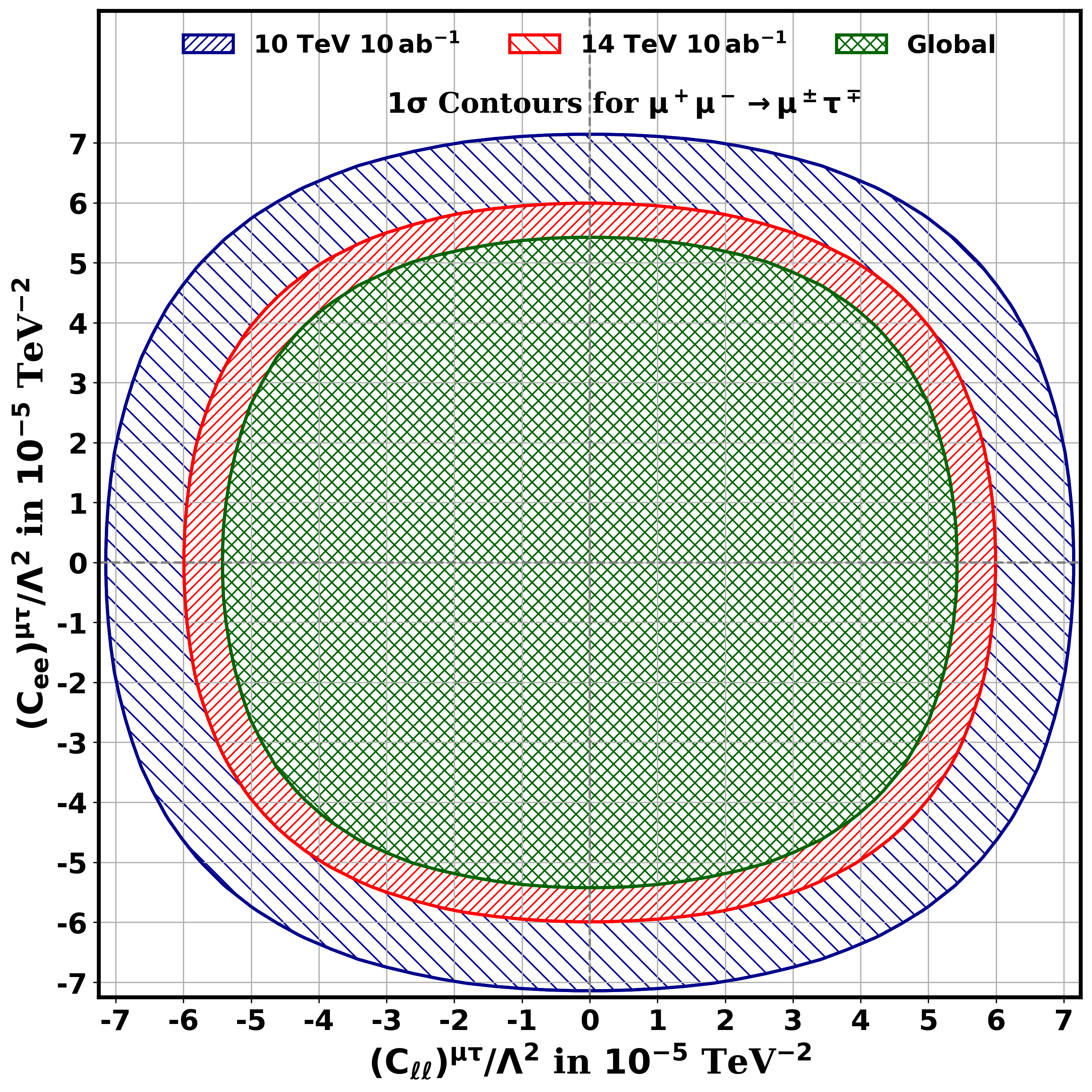}
    \includegraphics[width=0.3\textwidth]{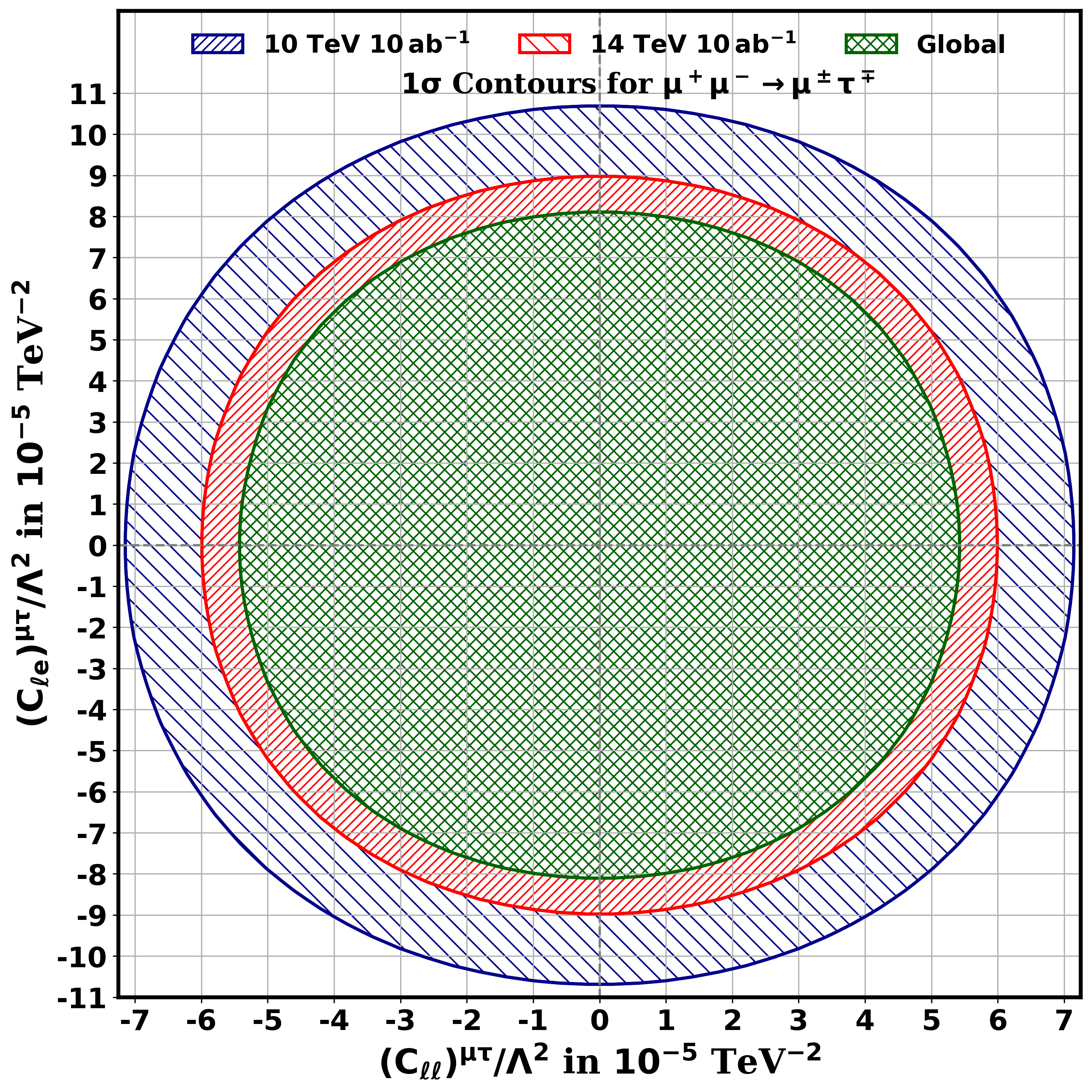}
    \includegraphics[width=0.3\textwidth]{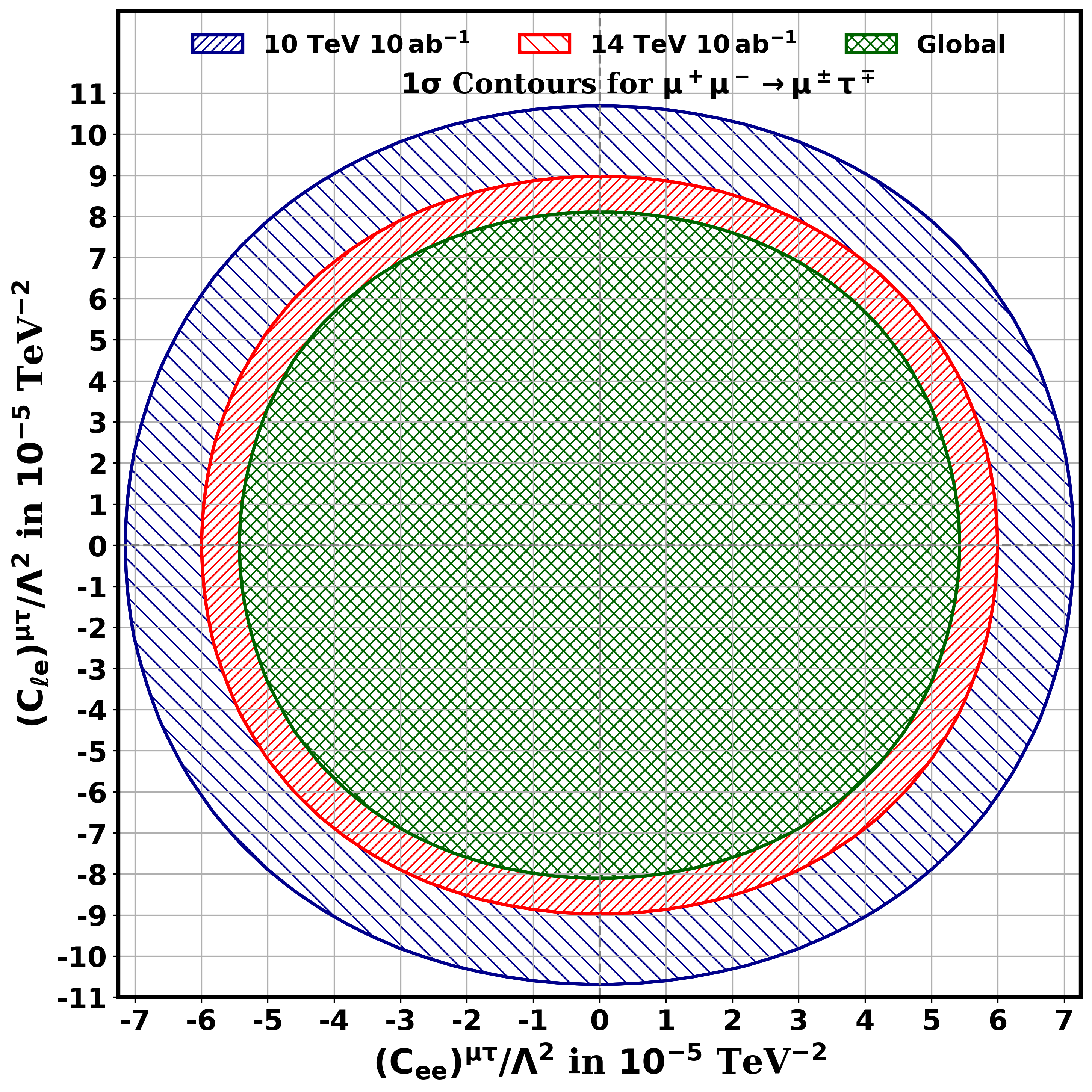}
    \includegraphics[width=0.3\textwidth]{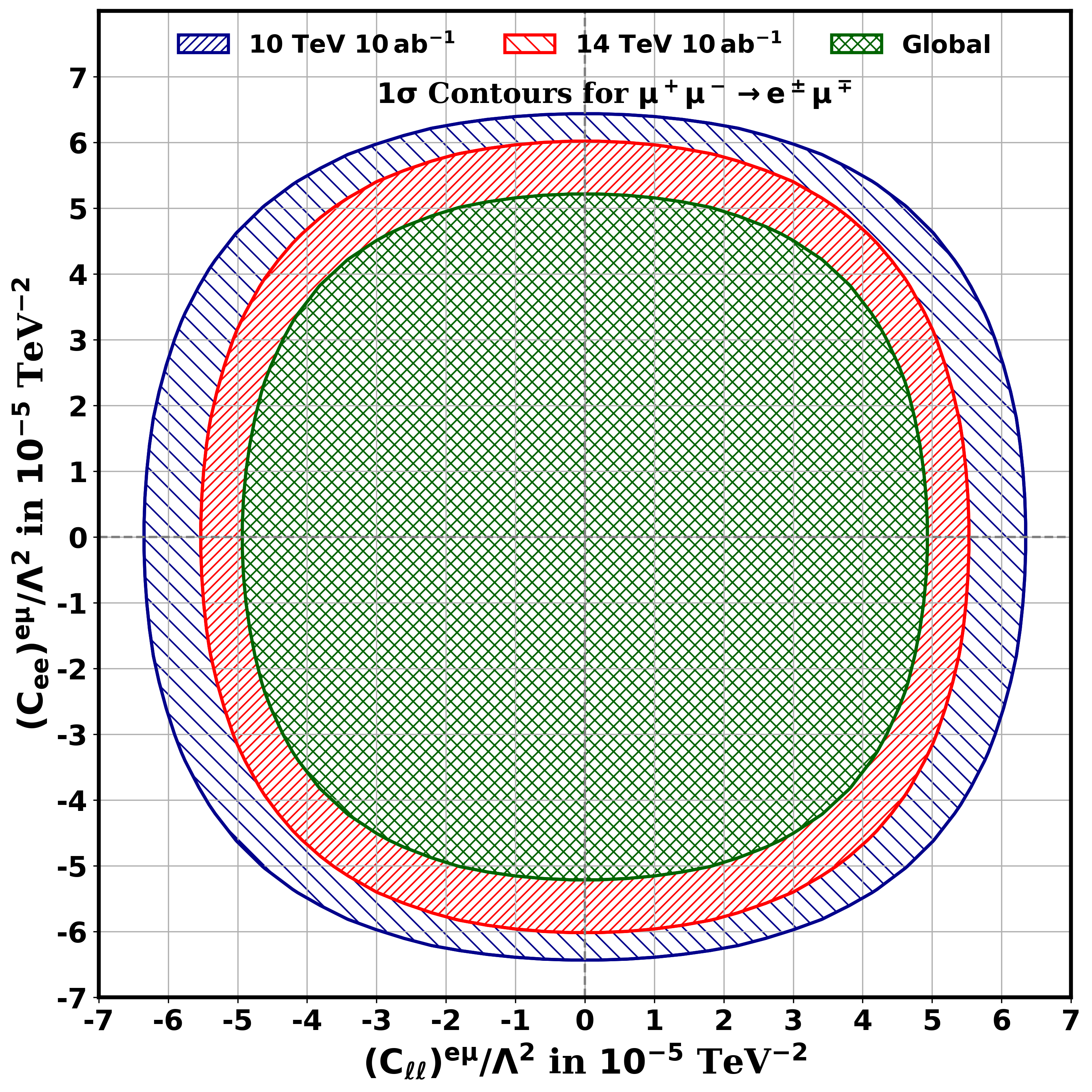}
    \includegraphics[width=0.3\textwidth]{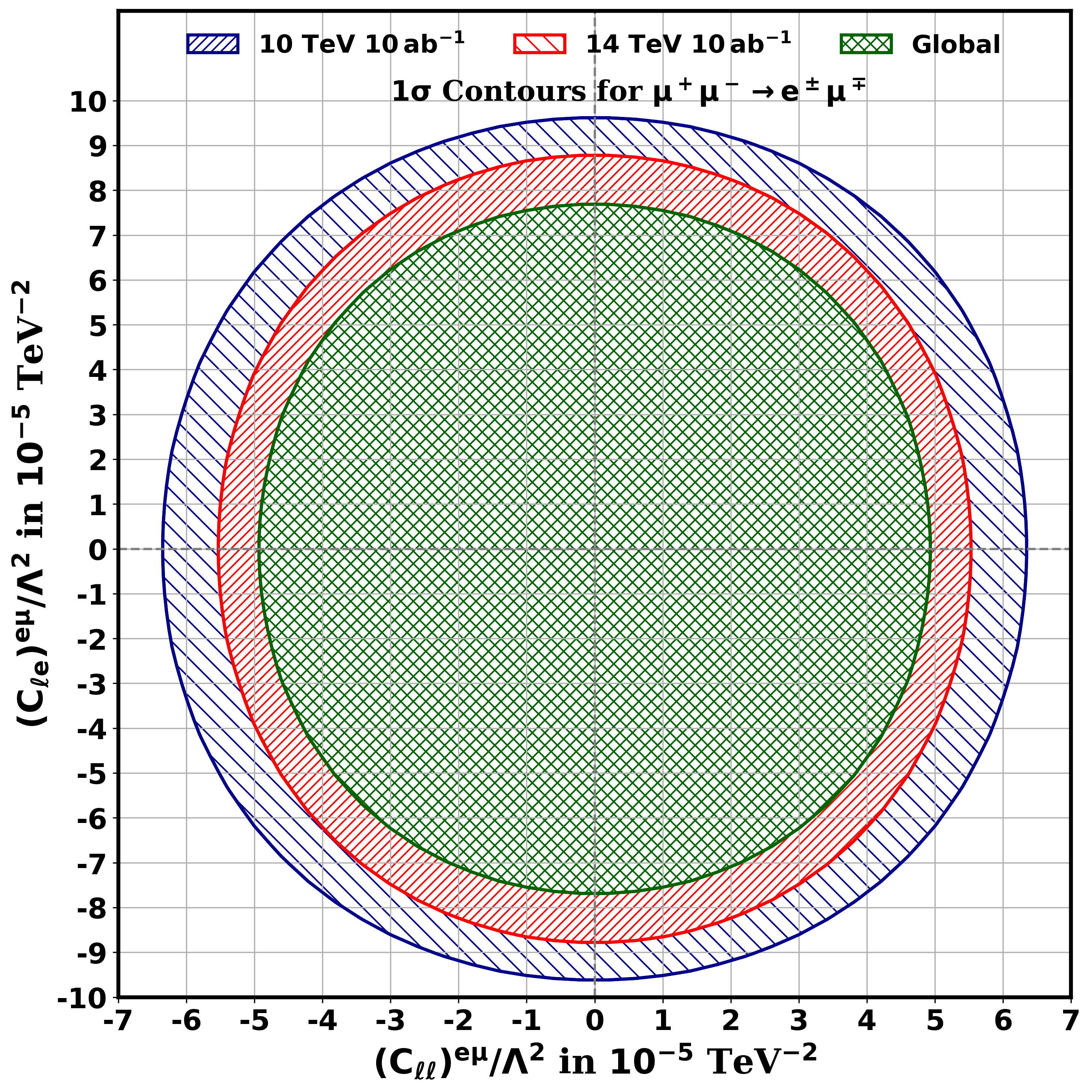}
    \includegraphics[width=0.3\textwidth]{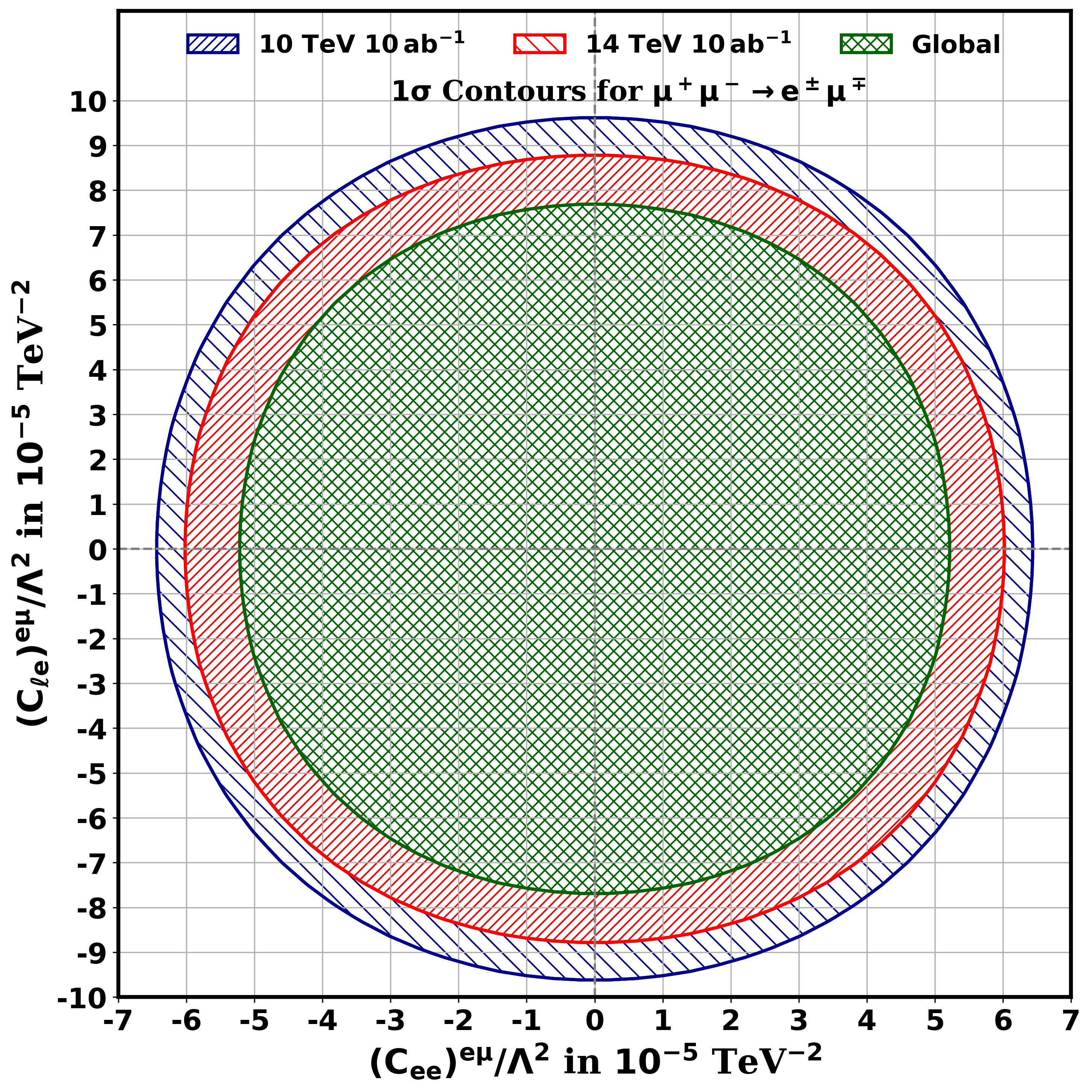}
    \caption{
Projected \(1\sigma\) confidence regions in the \((C_{\ell\ell}^{f_1f_2}/\Lambda^2, \, C_{ee}^{f_1f_2}/\Lambda^2)\), \((C_{\ell\ell}^{f_1f_2}/\Lambda^2,\, C_{\ell e}^{f_1f_2}/\Lambda^2)\), and \((C_{ee}^{f_1f_2}/\Lambda^2,\, C_{\ell e}^{f_1f_2}/\Lambda^2)\) planes for the LFV processes \(\mu^+\mu^-\rightarrow e^\pm\tau^\mp\), \(\mu^+\mu^-\rightarrow\mu^\pm\tau^\mp\), and \(\mu^+\mu^-\rightarrow e^\pm\mu^\mp\). The blue and red contours correspond to the analyses obtained after combining all beam-polarization configurations at \(\sqrt{s}=10\) and \(14~\mathrm{TeV}\), respectively, while the green contours represent the global fit obtained by combining the statistically independent datasets at \(\sqrt{s}=3\), \(10\), and \(14~\mathrm{TeV}\). Throughout the analysis, a \(1\%\) systematic uncertainty is assumed in each angular bin, and the third Wilson coefficient is fixed to zero in each panel.}
\label{SigGlobal}
\end{figure*}

The resulting confidence regions are shown in Fig.~\ref{SigGlobal}. For each LFV channel, the blue and red contours correspond to the polarization-combined analyses at
\(\sqrt{s}=10\) and \(14~\mathrm{TeV}\), respectively, while the green contours are obtained from the global combination of the
\(\sqrt{s}=3\), \(10\), and \(14~\mathrm{TeV}\) datasets. Although the overall sizes of the contours differ among the three LFV channels because of their different reconstruction efficiencies and background levels, their orientations remain remarkably similar.

The projected global \(1\sigma\) uncertainties together with the corresponding correlation matrices as summarized in Table~\ref{tab:global_errors_corr}, 
\begin{table*}[!ht]
\centering
\caption{
Global \(1\sigma\) uncertainties and correlation matrices for the SMEFT Wilson coefficients after combining all beam-polarization configurations and the \(\sqrt{s}=3\), \(10\), and \(14~\mathrm{TeV}\) datasets.}
\label{tab:global_errors_corr}
\begin{tabular}{cc c cc c cc}
\\
$
\begin{aligned}
 \varepsilon_{\ell\ell}^{e\tau}&= \pm0.0161\\
 \varepsilon_{ee}^{e\tau}&= \pm0.0161\\
 \varepsilon_{\ell e}^{e\tau}  &= \pm0.0611
\end{aligned}
$
&
$
\begin{pmatrix}
1    & .85  & -.95\\
.85  & 1    & -.95\\
-.95 & -.95 & 1
\end{pmatrix}
$
&&
$
\begin{aligned}
 \varepsilon_{\ell\ell}^{\mu\tau} &= \pm0.0073\\
 \varepsilon_{ee}^{\mu\tau}       &= \pm0.0073\\
 \varepsilon_{\ell e}^{\mu\tau}   &= \pm0.0277
\end{aligned}
$
&
$
\begin{pmatrix}
1    & .85  & -.95\\
.85  & 1    & -.95\\
-.95 & -.95 & 1
\end{pmatrix}
$
&&
$
\begin{aligned}
 \varepsilon_{\ell\ell}^{e\mu} &= \pm0.0067\\
 \varepsilon_{ee}^{e\mu}       &= \pm0.0063\\
 \varepsilon_{\ell e}^{e\mu}   &= \pm0.0255
\end{aligned}
$
&
$
\begin{pmatrix}
1    & .85  & -.95\\
.85  & 1    & -.95\\
-.95 & -.95 & 1
\end{pmatrix}
$
\end{tabular}
\end{table*}
 shows that, despite the different final-state flavor compositions, all three LFV channels exhibit nearly identical correlation matrices, indicating that the correlation pattern is governed primarily by the Lorentz and chirality structure of the underlying four-lepton operators rather than by the final-state flavor. In particular, \(C_{\ell\ell}^2\) and \(C_{ee}^2\) are strongly positively correlated, \(\rho(C_{\ell\ell}^2,C_{ee}^2)\simeq0.85\), while both are strongly anticorrelated with the mixed-chirality coefficient \(C_{\ell e}^2\), with \(\rho\simeq-0.95\). 

Finally, combining the statistically independent datasets collected at
\(\sqrt{s}=3\), \(10\), and \(14~{\rm TeV}\), reduces the corresponding condition number to \(\kappa=20.15\) and yields the global least-constrained direction,
\begin{equation}
-0.236\left(C_{\ell\ell}^{e\tau}\right)^2
-0.236\left(C_{ee}^{e\tau}\right)^2
+0.942\left(C_{\ell e}^{e\tau}\right)^2
=\pm0.065.
\end{equation}
The global least-constrained direction is nearly identical to those obtained at the individual center-of-mass energies after combining all beam-polarization configurations. Similarly, for the other two LFV processes \(\mu \mu \to \mu \tau\) and \(\mu \mu \to \mu e \), the least-constrained directions remain almost similar, and the corresponding global uncertainties are \(\pm0.0294\) and \(\pm0.0264\), respectively.



\begin{table}[!ht]
\centering
\setlength{\tabcolsep}{0.1cm} 
\caption{Projected 1$\sigma$ bounds of all three channels on the individual coupling coefficients from the combined beam polarization analysis (\(0\%,\pm30\%,\pm80\%\)) at \(\Delta\chi_{\text{Pol. comb.}}^2 = 2.3\), for \(\sqrt{s}\) 3, 10 and 14 TeV, and for the global combination \(\Delta\chi_{\text{global}}^2\) of all three center-of-mass energies.}
\label{tab:sensitivities}
\begin{tabular}{ccccc}
\hline
\hline
\multirow{2}{*}{$\sqrt{s}$ (TeV)}
& \multirow{2}{*}{Coupling} & \multicolumn{3}{c}{1$\sigma$ bounds ($10^{-4}$ TeV$^{-2}$)}\\
\cline{3-5}
& & $\mu\mu\to e\tau$ & $\mu\mu\to e\mu$ & $\mu\mu\to \mu\tau$ \\
\hline
\multirow{3}{*}{$\begin{array}{c}
        3\\ \text{(Pol. Combined)}\end{array}$}
& $C_{ll}/{\Lambda^2}$ & 4.133 & 2.237 & 2.622 \\
\cline{2-5}
& $C_{ee}/{\Lambda^2}$ & 4.074 & 2.602 & 2.649 \\
\cline{2-5}
& $C_{le}/{\Lambda^2}$ & 5.948 & 3.401 & 3.808 \\
\hline
\multirow{3}{*}{$\begin{array}{c}
        10\\ \text{(Pol. Combined)}\end{array}$}
& $C_{ll}/{\Lambda^2}$ & 1.045 & 0.636 & 0.715 \\
\cline{2-5}
& $C_{ee}/{\Lambda^2}$ & 1.041 & 0.645 & 0.715 \\
\cline{2-5}
& $C_{le}/{\Lambda^2}$ & 1.560 & 0.963 & 1.069 \\
\hline
\multirow{3}{*}{$\begin{array}{c}
        14\\ \text{(Pol. Combined)}\end{array}$}
& $C_{ll}/{\Lambda^2}$ & 0.890 & 0.554 & 0.600 \\
\cline{2-5}
& $C_{ee}/{\Lambda^2}$ & 0.895 & 0.603 & 0.600 \\
\cline{2-5}
& $C_{le}/{\Lambda^2}$ & 1.341 & 0.879 & 0.898 \\
\hline
\multirow{3}{*}{Global}
& $C_{ll}/{\Lambda^2}$ & 0.801 & 0.494 & 0.542 \\
\cline{2-5} 
& $C_{ee}/{\Lambda^2}$ & 0.803 & 0.523 & 0.542 \\
\cline{2-5}
& $C_{le}/{\Lambda^2}$ & 1.202 & 0.769 & 0.811 \\
\hline
\hline
\end{tabular}
\end{table}

\section{Summary}
\label{sec:sum}
In this work, we have investigated charged LFV processes induced by dimension-six four-lepton operators within the SMEFT at a future high-energy muon collider. Owing to the extreme suppression of charged LFV in the SM, these processes constitute a particularly sensitive probe of new physics. We first translated the existing experimental limits from LFV $\tau$ decay measurements into constraints on the relevant SMEFT Wilson coefficients, and evolved the corresponding parameter space to the multi-TeV scale using the one-loop renormalization-group equations. The small corrections imply that the experimentally allowed low-energy parameter space is mildly modified at future muon collider energies.

Signal events for the LFV processes \(\mu^+\mu^- \rightarrow e^\pm\tau^\mp\), \(\mu^+\mu^- \rightarrow \mu^\pm\tau^\mp\), and \(\mu^+\mu^- \rightarrow e^\pm\mu^\mp\)  and the SM backgrounds both were simulated at \(\sqrt{s}=3\), \(10\), and \(14~\mathrm{TeV}\) and analyzed using a realistic detector-level simulation based on the Delphes muon collider detector card. The characteristic hard kinematic behavior of the contact interactions enables efficient suppression of the SM backgrounds through simple event selections while maintaining high signal efficiencies.

To obtain the most stringent constraints on the SMEFT Wilson coefficients, we employed the OOT on the binned angular distributions of the reconstructed charged leptons. As illustrated by the projected confidence regions in Figs.~\ref{etauConts}, \ref{mtauConts}, and \ref{emuConts}, the projected \(1\sigma\) sensitivities improve upon the existing bounds by reaching the \(\mathcal{O}(10^{-4})~\mathrm{TeV}^{-2}\) level at \(\sqrt{s}=3\)~TeV and are further strengthened to the \(\mathcal{O}(10^{-5})~\mathrm{TeV}^{-2}\) level at \(\sqrt{s}=10\) and \(14\)~TeV, depending on the operator and LFV channel. The projected sensitivities improve with increasing center-of-mass energy and vary slightly across the three LFV processes due to different final-state topologies. Beam polarization helps disentangle the chirality structure of the underlying operators, with left- (right-) polarized muon beams providing the best sensitivity to \(O_{\ell\ell}\) (\(O_{ee}\)), while \(O_{\ell e}\) is largely insensitive to polarization. 

To further investigate the correlations among the fitted squared Wilson coefficients, we performed a covariance matrix analysis. Combining different beam polarizations significantly reduced the uncertainty of the least-constrained direction of the squared Wilson coefficients. The global combination of all center-of-mass energies further reduced the corresponding uncertainties for the three LFV processes to \(\pm0.065,\ \pm0.0294 \,\,\text{and}\,\,\pm0.0264\), leading to substantially stronger constraints on the Wilson coefficients. This is also observed in the projected confidence contours shown in Fig.~\ref{SigGlobal}, which have shrunk, providing more stringent constraints. We have summarized the final results of the bounds on the individual couplings and their uncertainties in Table \ref{tab:sensitivities}.

Overall, our results demonstrate the excellent potential of a future multi-TeV muon collider as a precision probe of charged LFV within the SMEFT framework. With its high center-of-mass energies, beam polarization capabilities, and clean experimental environment, the muon collider is expected to tighten current bounds by orders of magnitude, achieving unprecedented sensitivities of $(0.5-1.2)\times10^{-4}\,\mathrm{TeV}^{-2}$. Crucially, for processes such as $\mu^+\mu^- \rightarrow e^\pm\mu^\mp$, where low-energy observables do not provide tree-level bounds, the muon collider offers a unique and unparalleled environment to establish direct, stringent constraints on these purely leptonic four-fermion operators.

\section*{Acknowledgments}
We thank Debajyoti Choudhury and Rameswar Sahu for fruitful discussions and feedback throughout the work. We acknowledge partial financial support from the ANRF Grant No. CRG/2023/008234 and the University of Delhi project Grant No. IoE/2025-26/12/FRP.

\section*{Data Availability}
There are no publicly available research data or software supporting this manuscript. Requests for further information or data should be sent to the authors.

\newpage
\onecolumngrid
\appendix
\section{Plots for additional processes:}
\label{app-plots}
This appendix details the remaining two-dimensional $1\sigma$ exclusion contours for the LFV channels evaluated in this study. The constraints are derived using the same statistical procedure, benchmark assumptions, and analysis framework established in the main text, thereby providing the full set of exclusion contours for all flavor combinations.

\begin{figure*}[!ht]
    \centering
    \includegraphics[width=0.32\textwidth]{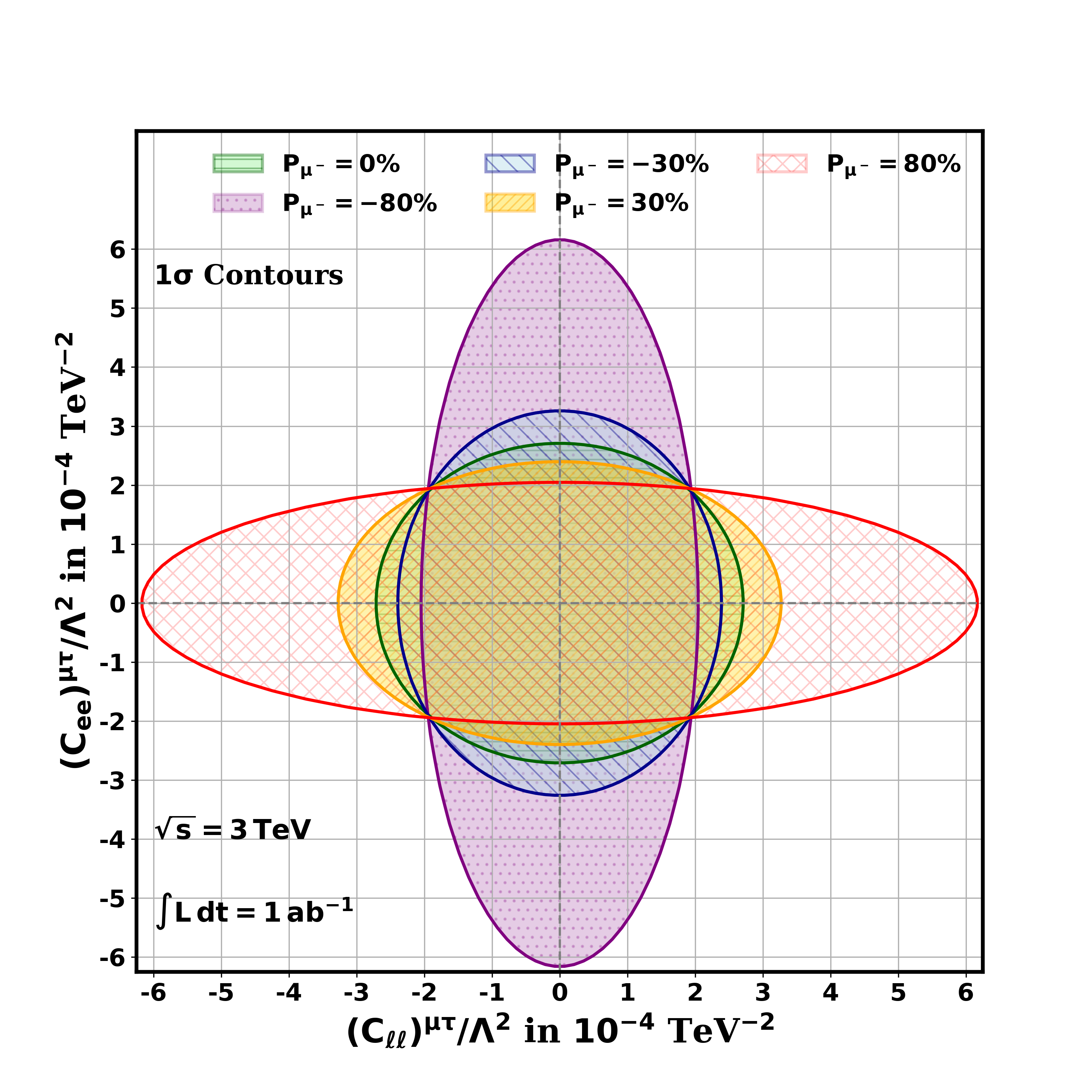}
    \includegraphics[width=0.32\textwidth]{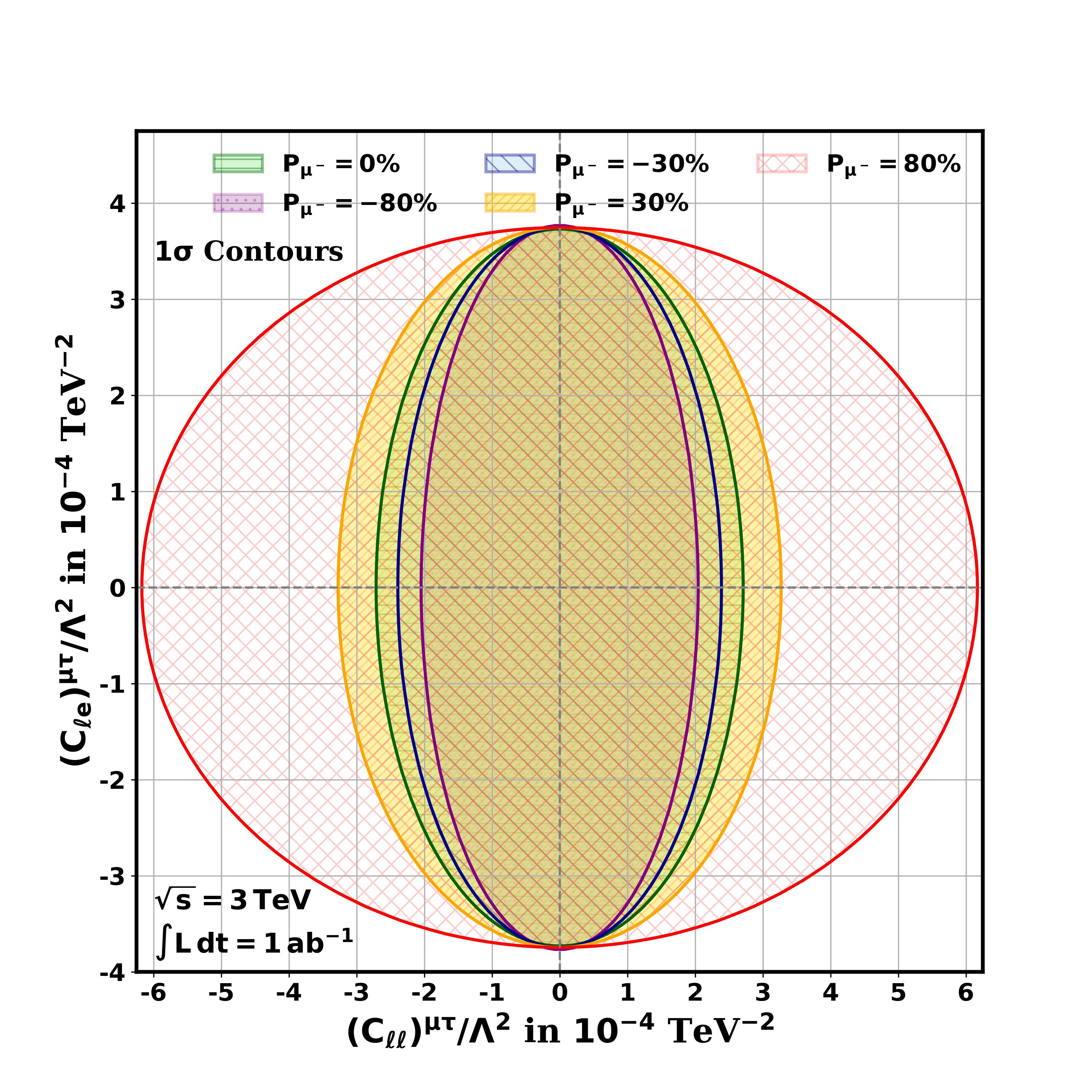}
    \includegraphics[width=0.32\textwidth]{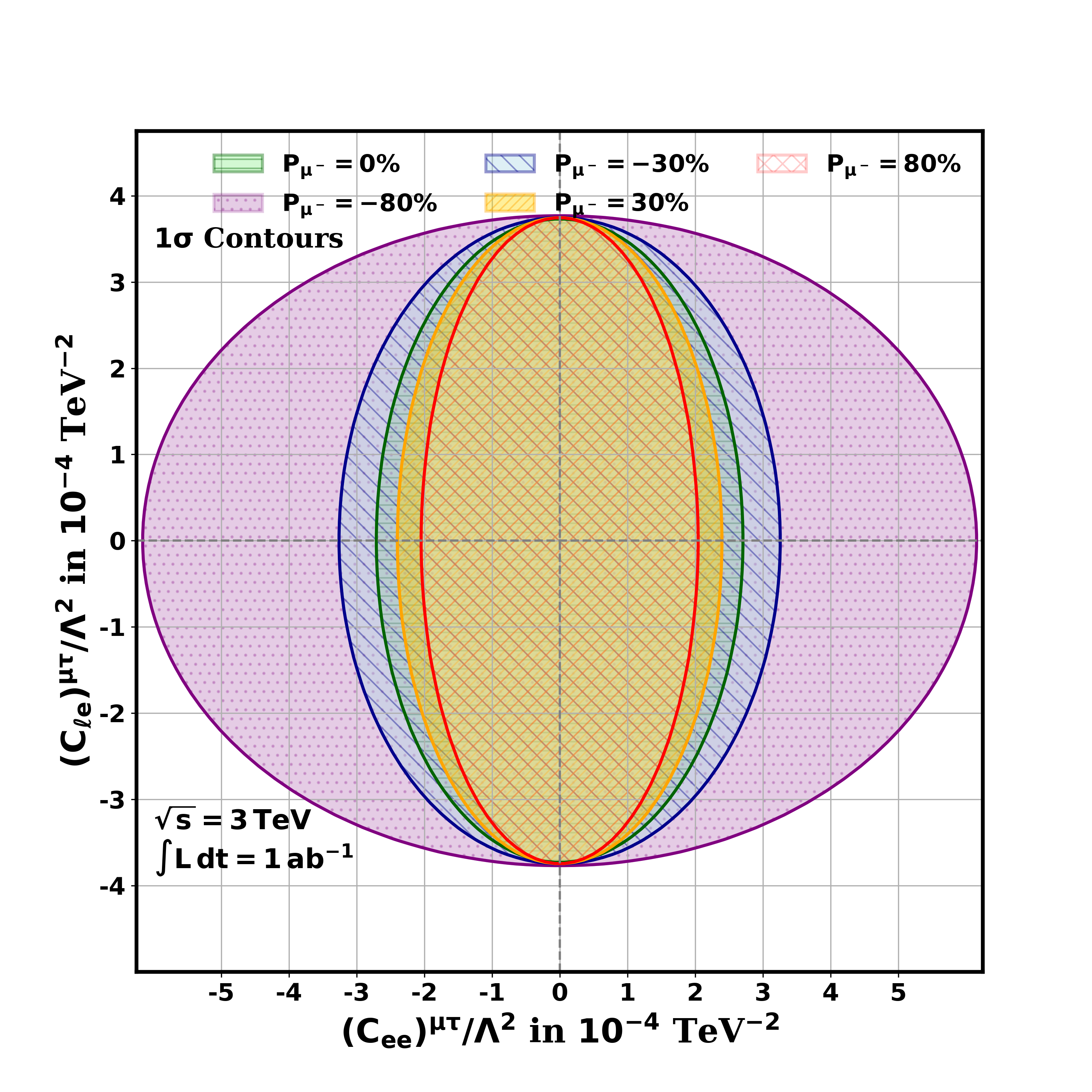}
    \includegraphics[width=0.32\textwidth]{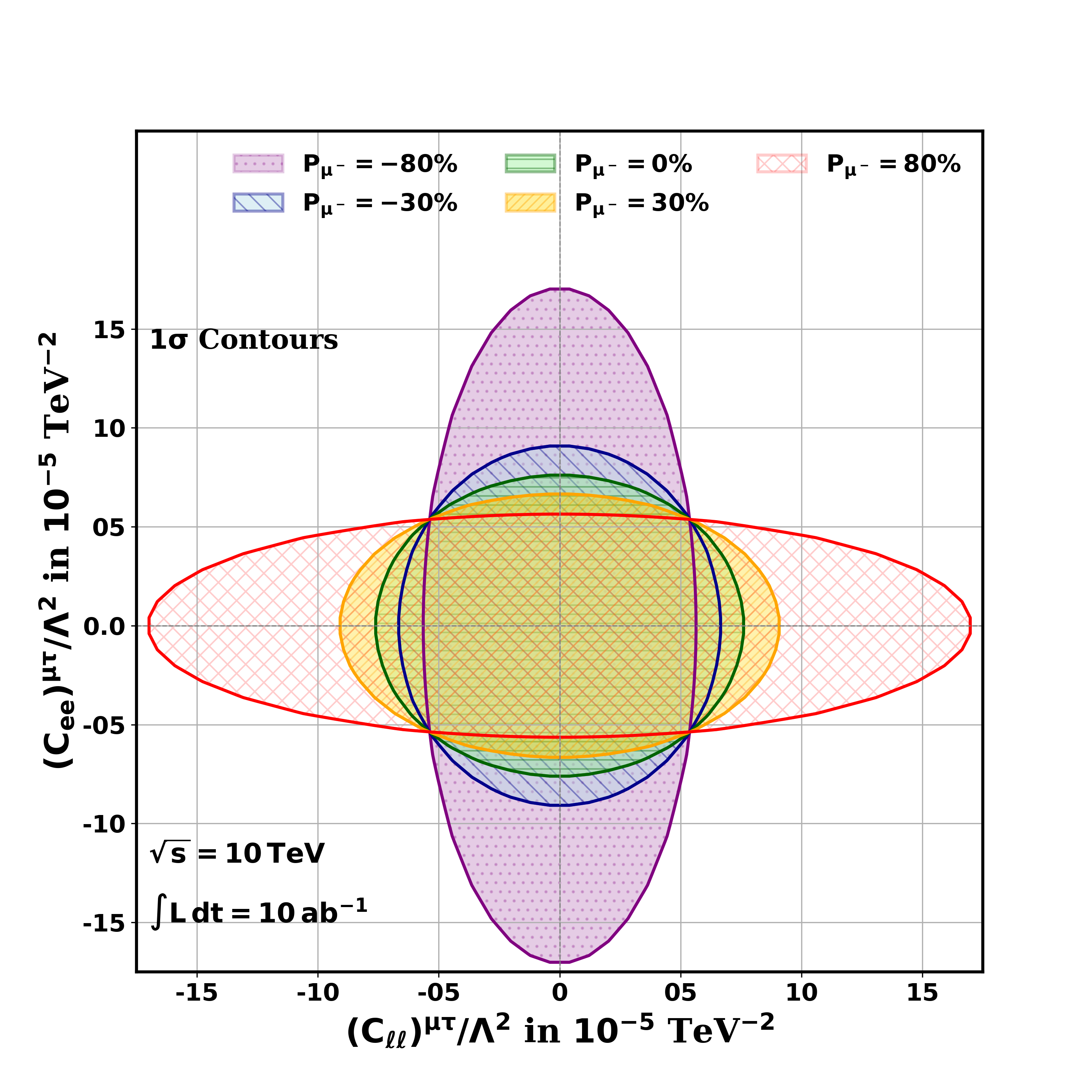}
    \includegraphics[width=0.32\textwidth]{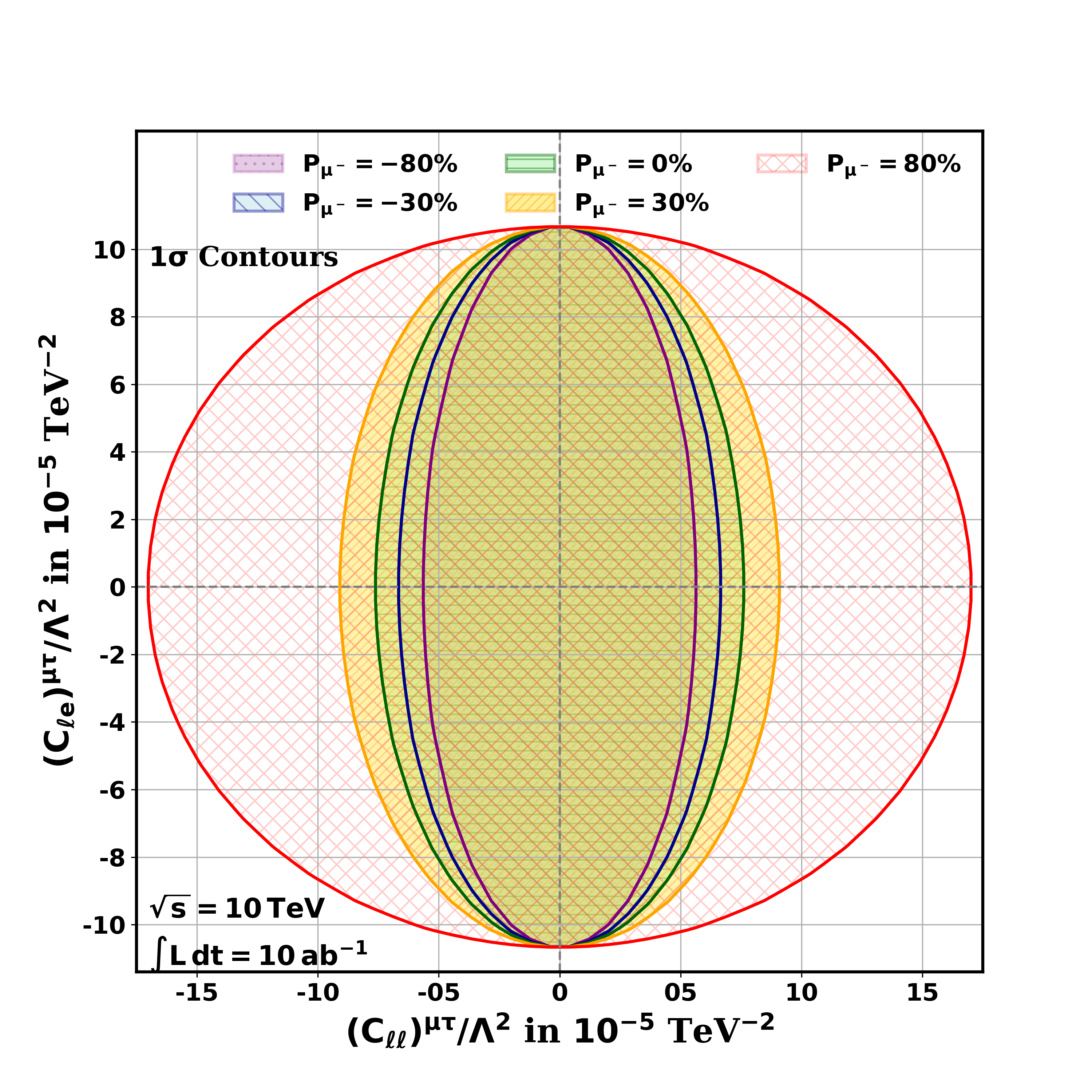}
    \includegraphics[width=0.32\textwidth]{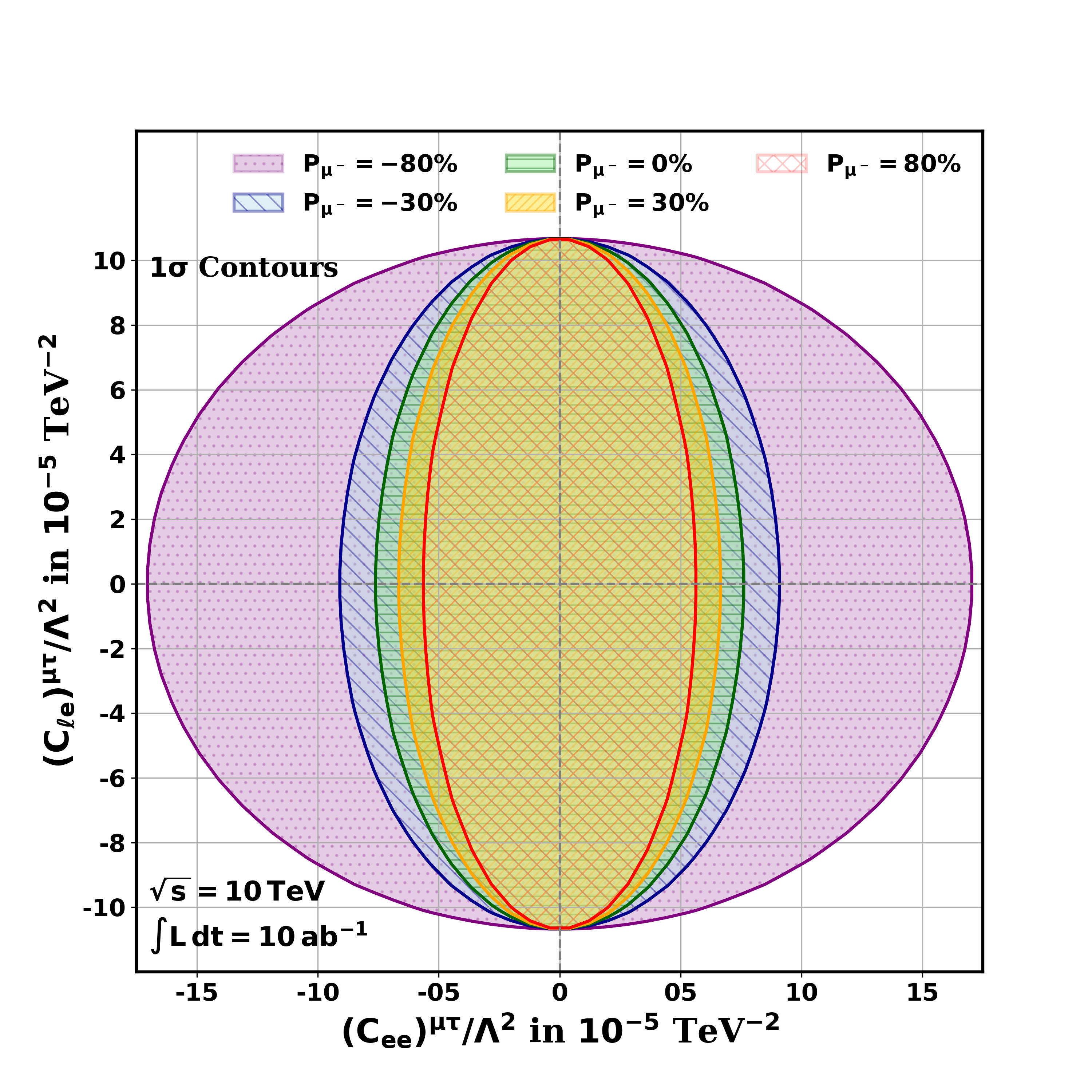}
    \includegraphics[width=0.32\textwidth]{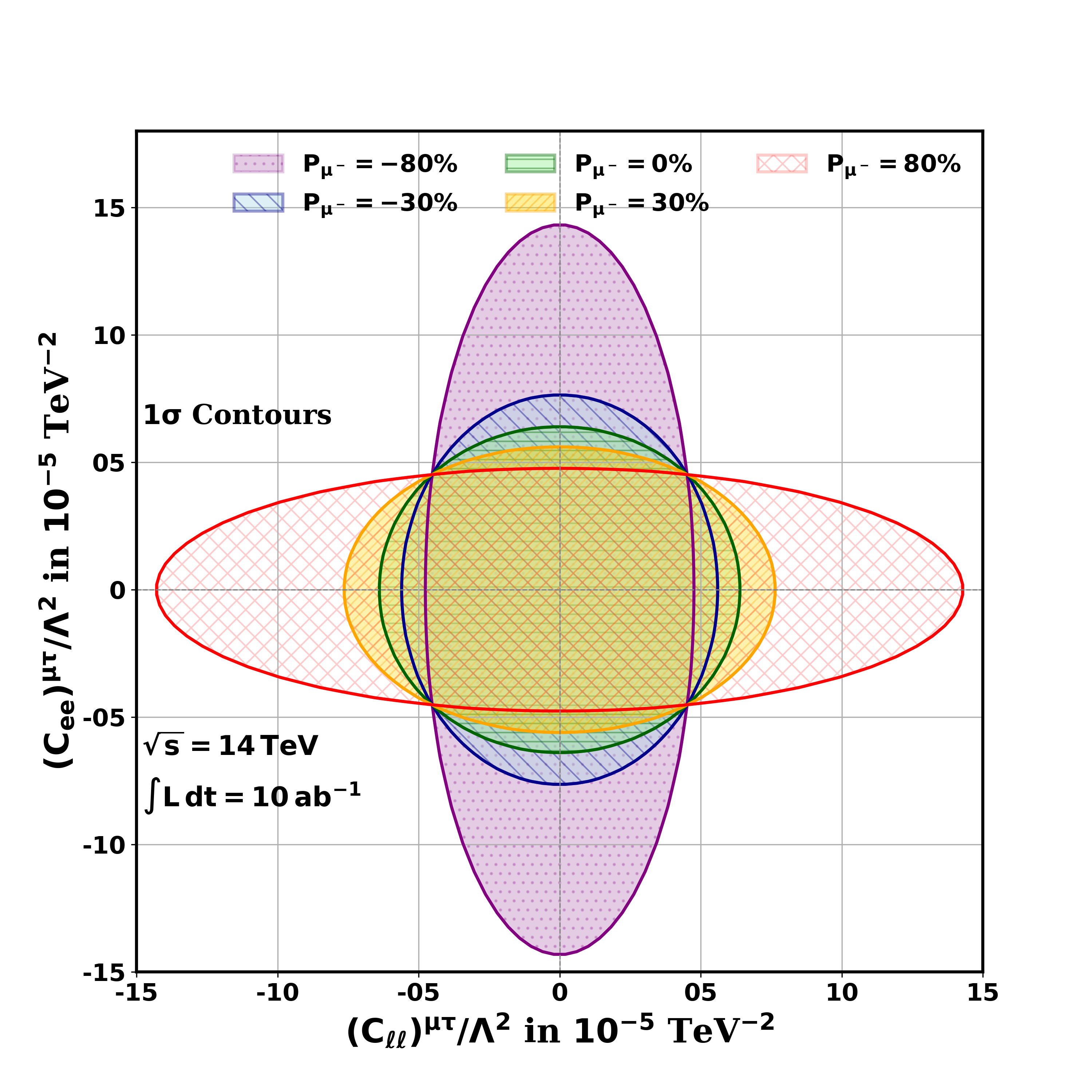}
    \includegraphics[width=0.32\textwidth]{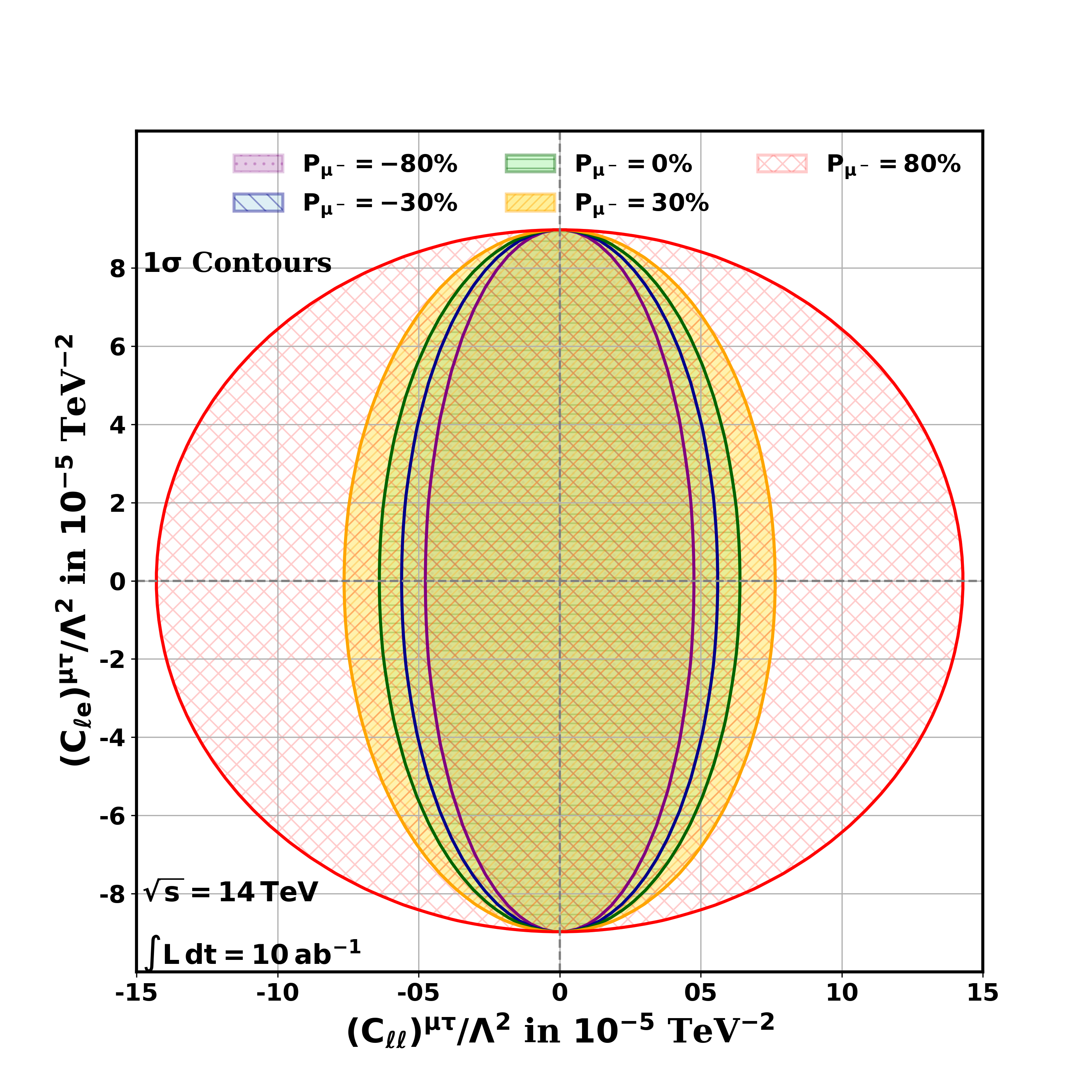}
    \includegraphics[width=0.32\textwidth]{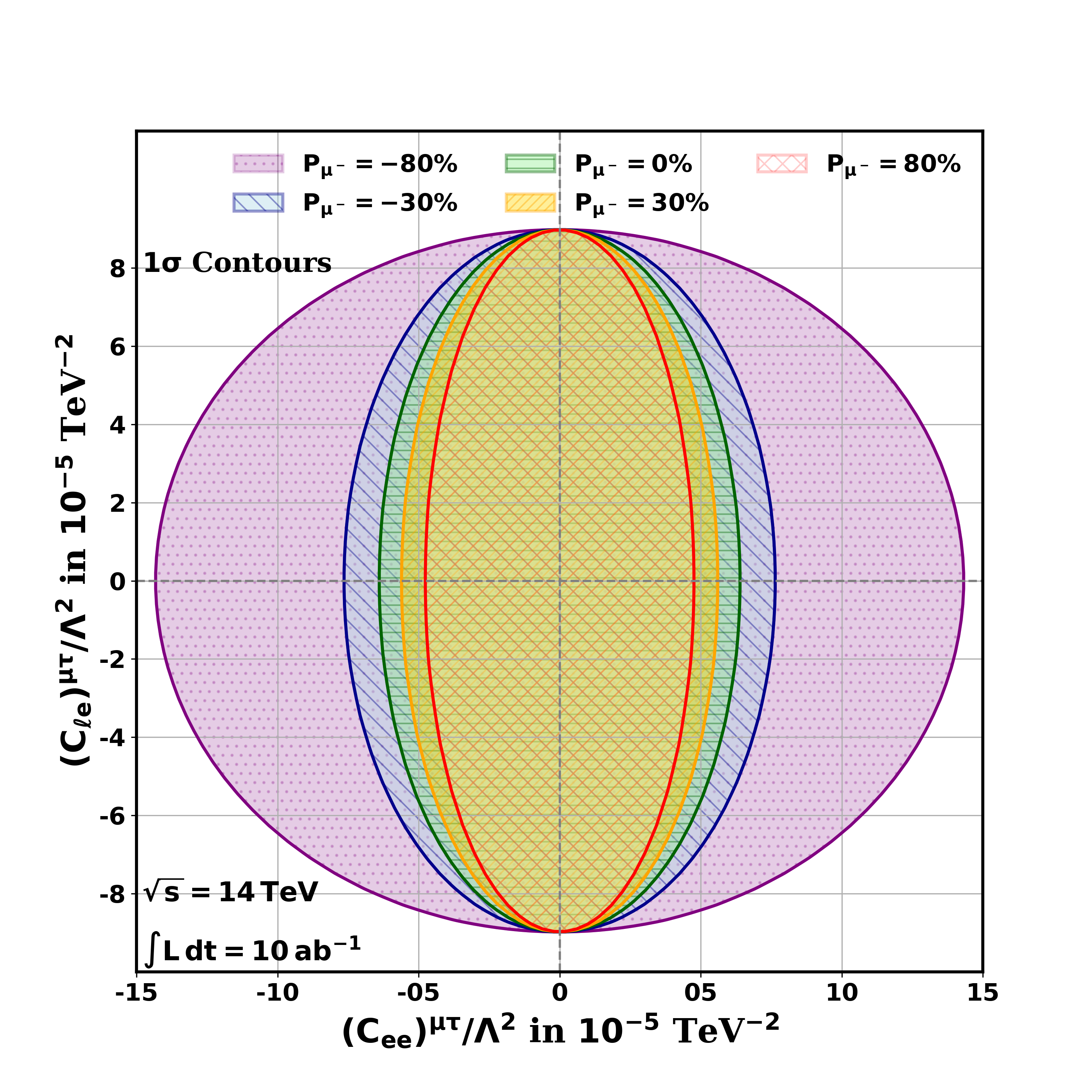}      
    \caption{The shaded \(1\sigma\) confidence contours in the (\(C_{\ell\ell}^{\mu\tau}/\Lambda^2\,-\, C_{ee}^{\mu\tau}/\Lambda^2\)), (\(C_{\ell\ell}^{\mu\tau}/\Lambda^2\,-\, C_{\ell e}^{\mu\tau}/\Lambda^2\)), and (\(C_{ee}^{\mu\tau}/\Lambda^2\,-\, C_{\ell e}^{\mu\tau}/\Lambda^2\)) planes for the LFV channels \(\mu^+\mu^- \to \mu^\pm\tau^\mp\).}
    \label{mtauConts}
\end{figure*}

The exclusion contours for the process $\mu^+\mu^- \to \mu^\pm\tau^\mp$ are shown in Fig.~\ref{mtauConts}. The three panels in each row correspond to the $(C_{\ell\ell}^{\mu\tau}/\Lambda^2,\,C_{ee}^{\mu\tau}/\Lambda^2)$, $(C_{\ell\ell}^{\mu\tau}/\Lambda^2,\,C_{\ell e}^{\mu\tau}/\Lambda^2)$, and $(C_{ee}^{\mu\tau}/\Lambda^2,\,C_{\ell e}^{\mu\tau}/\Lambda^2)$ parameter planes, respectively, for a fixed center-of-mass energy considered and the different rows correspond to different center-of-mass enegies.

\begin{figure*}[!ht]
    \centering
    \includegraphics[width=0.32\textwidth]{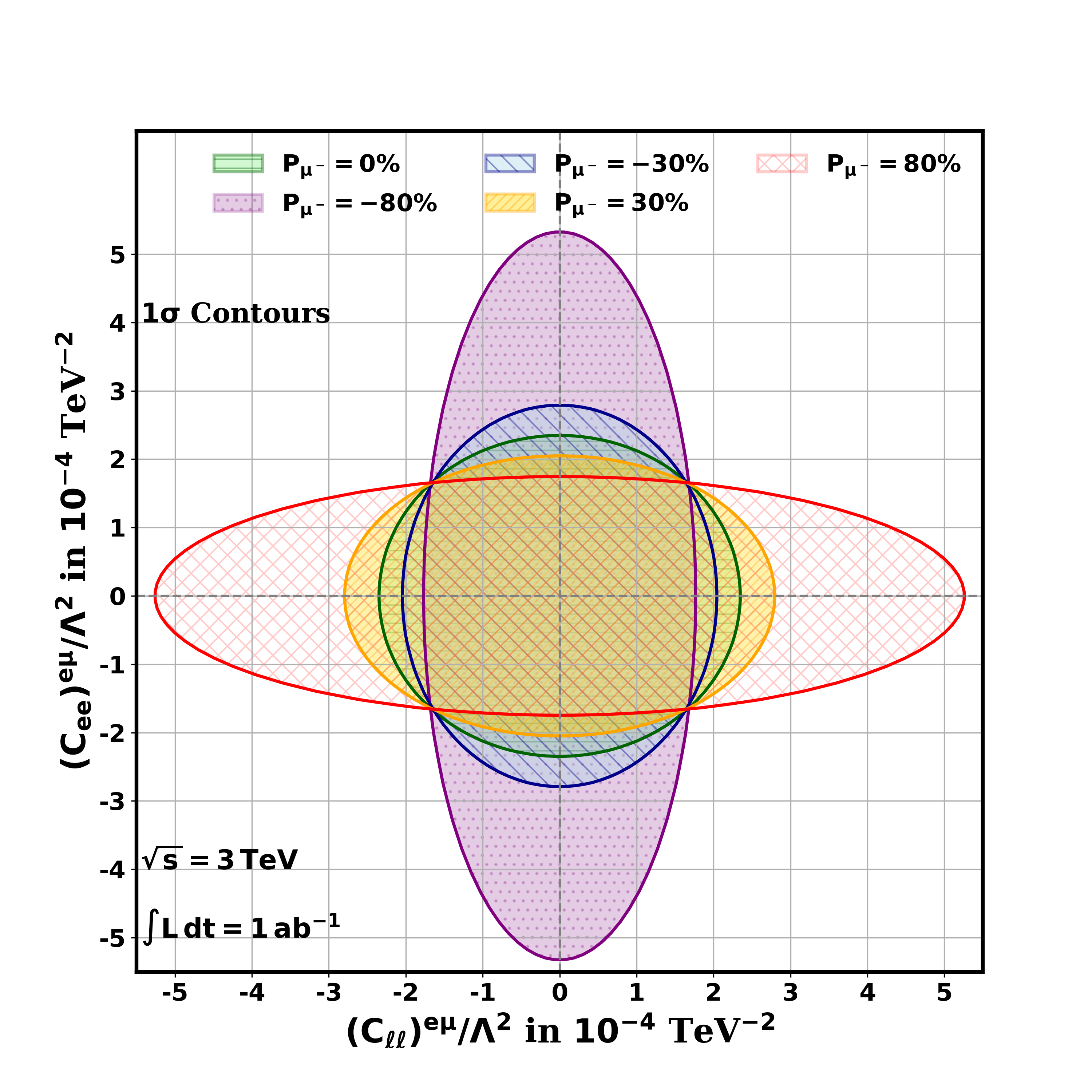}
    \includegraphics[width=0.32\textwidth]{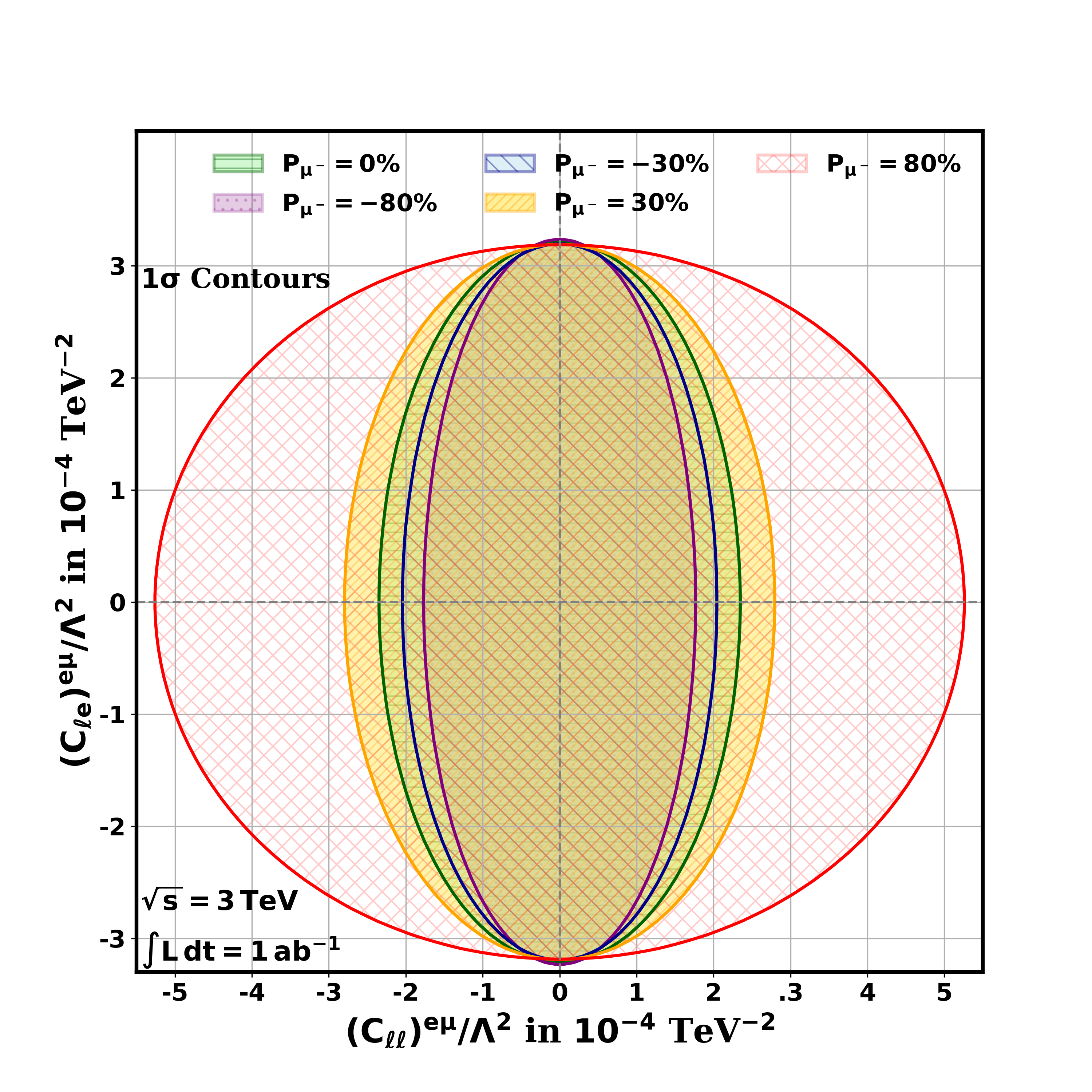}
    \includegraphics[width=0.32\textwidth]{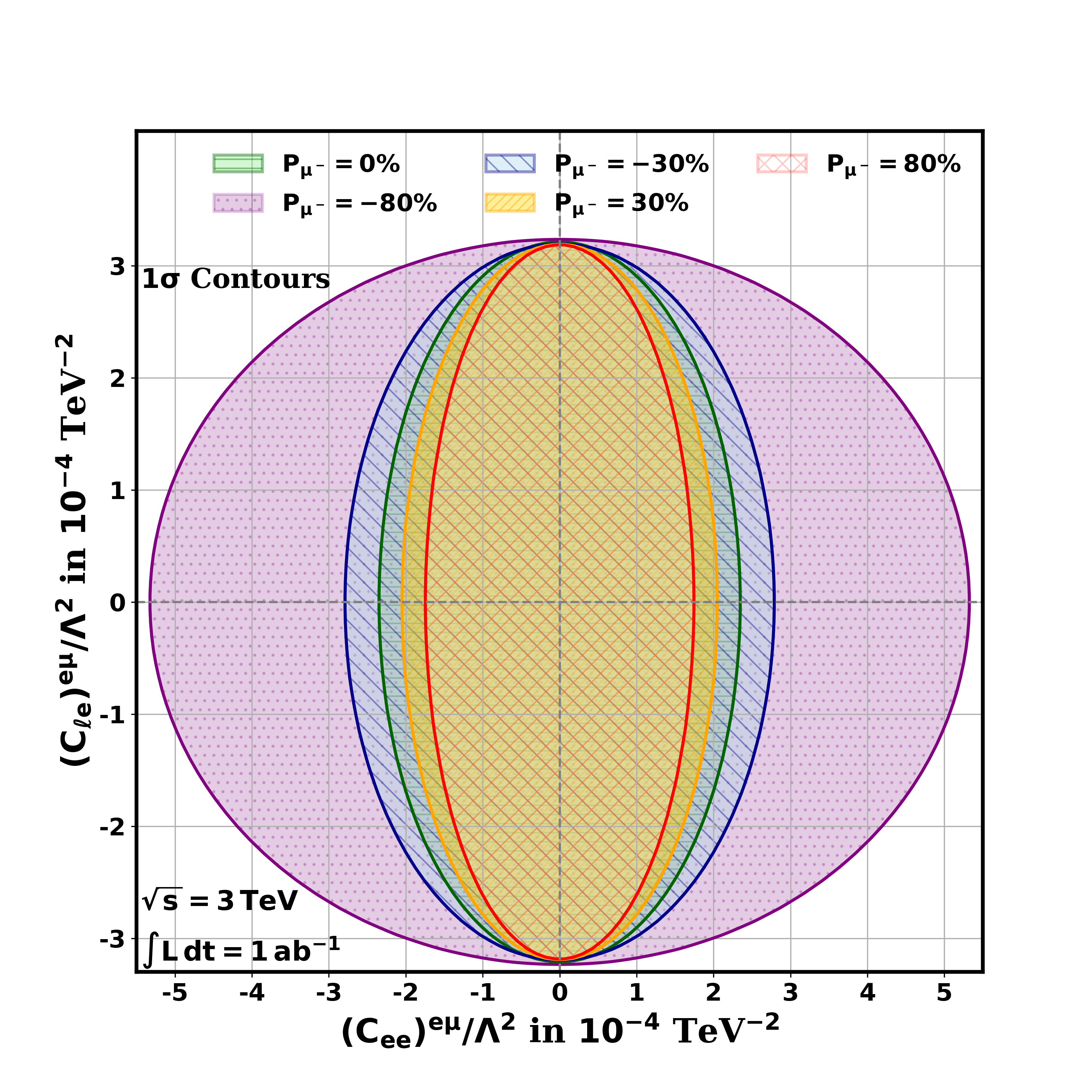}
    \includegraphics[width=0.32\textwidth]{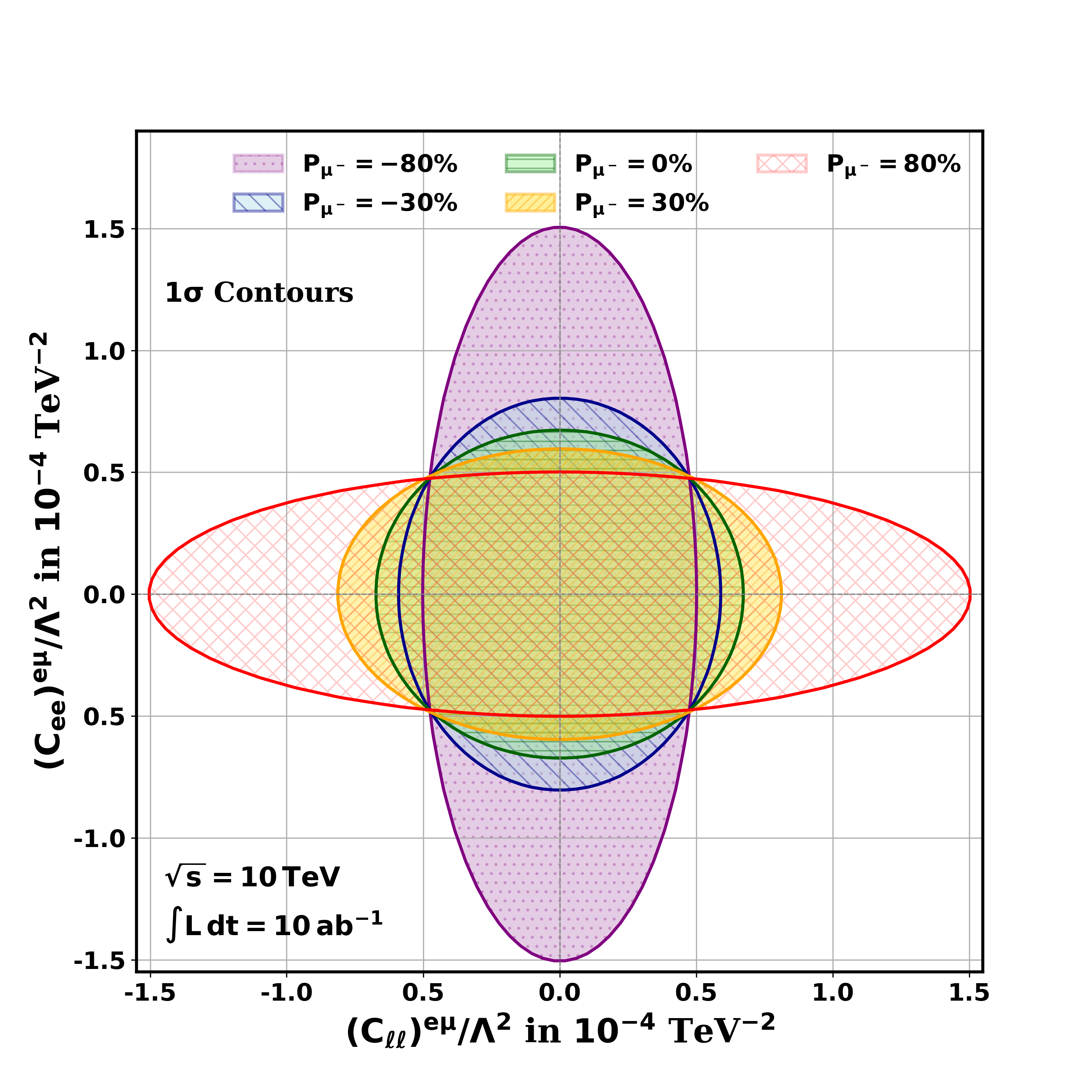}
    \includegraphics[width=0.32\textwidth]{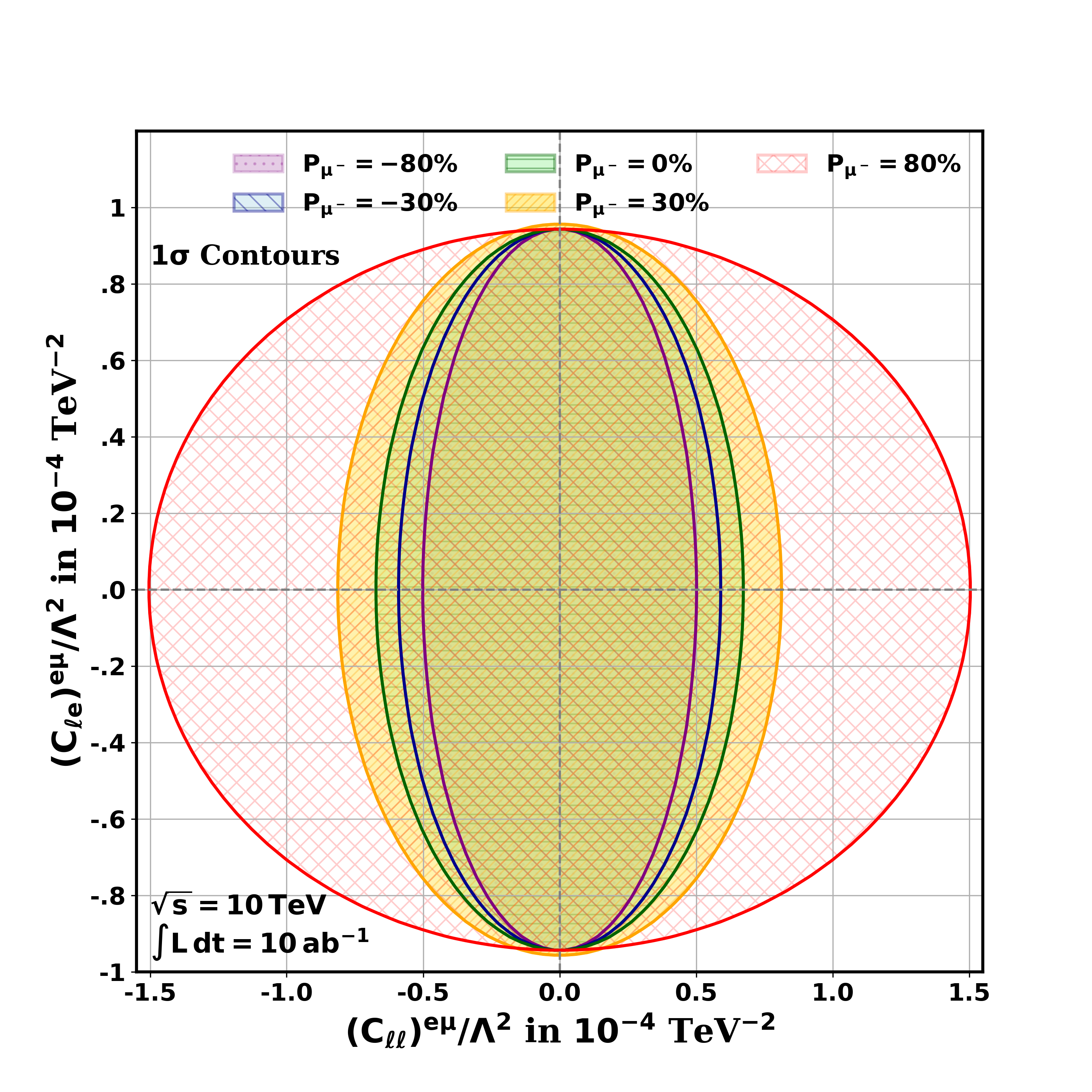}
    \includegraphics[width=0.32\textwidth]{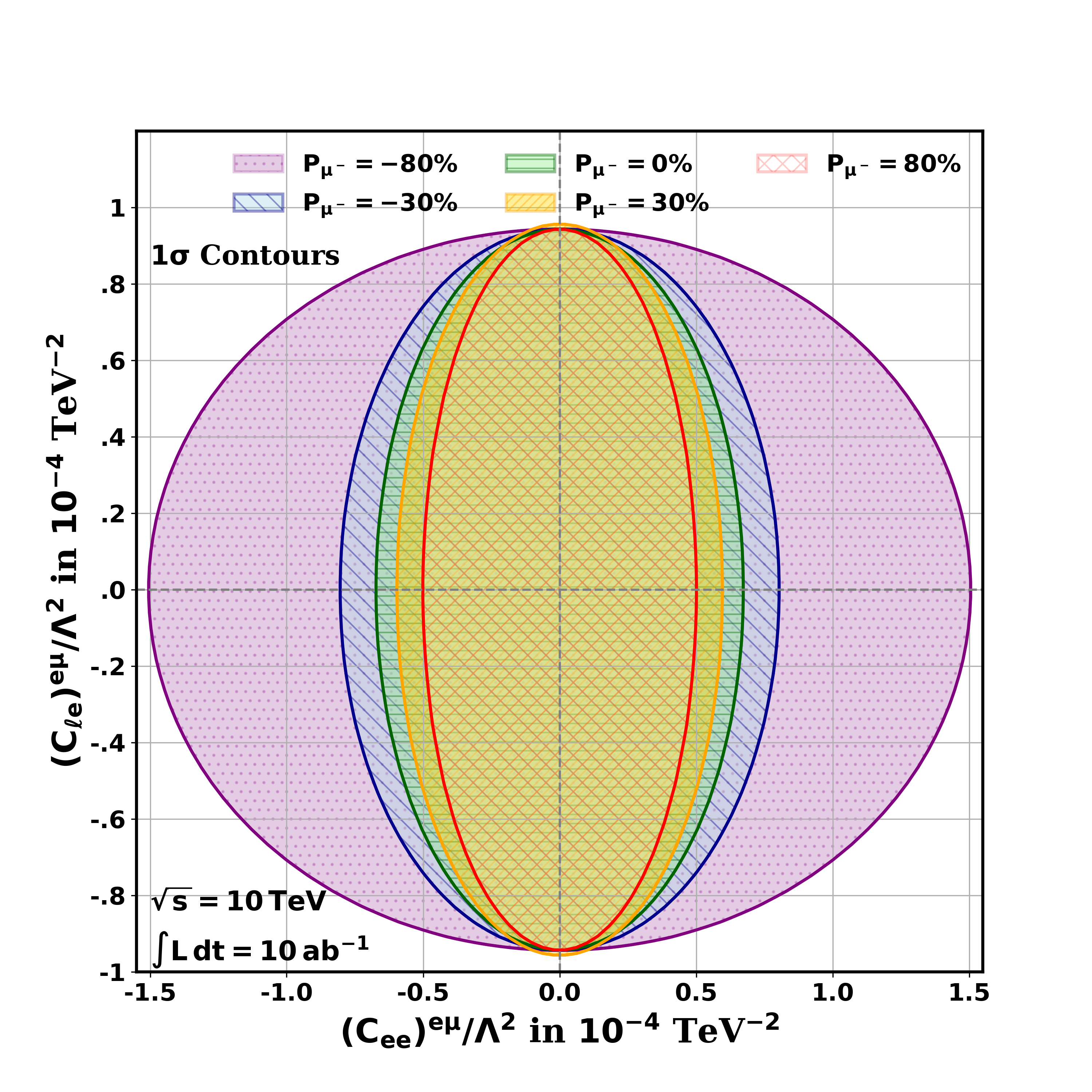}
    \includegraphics[width=0.32\textwidth]{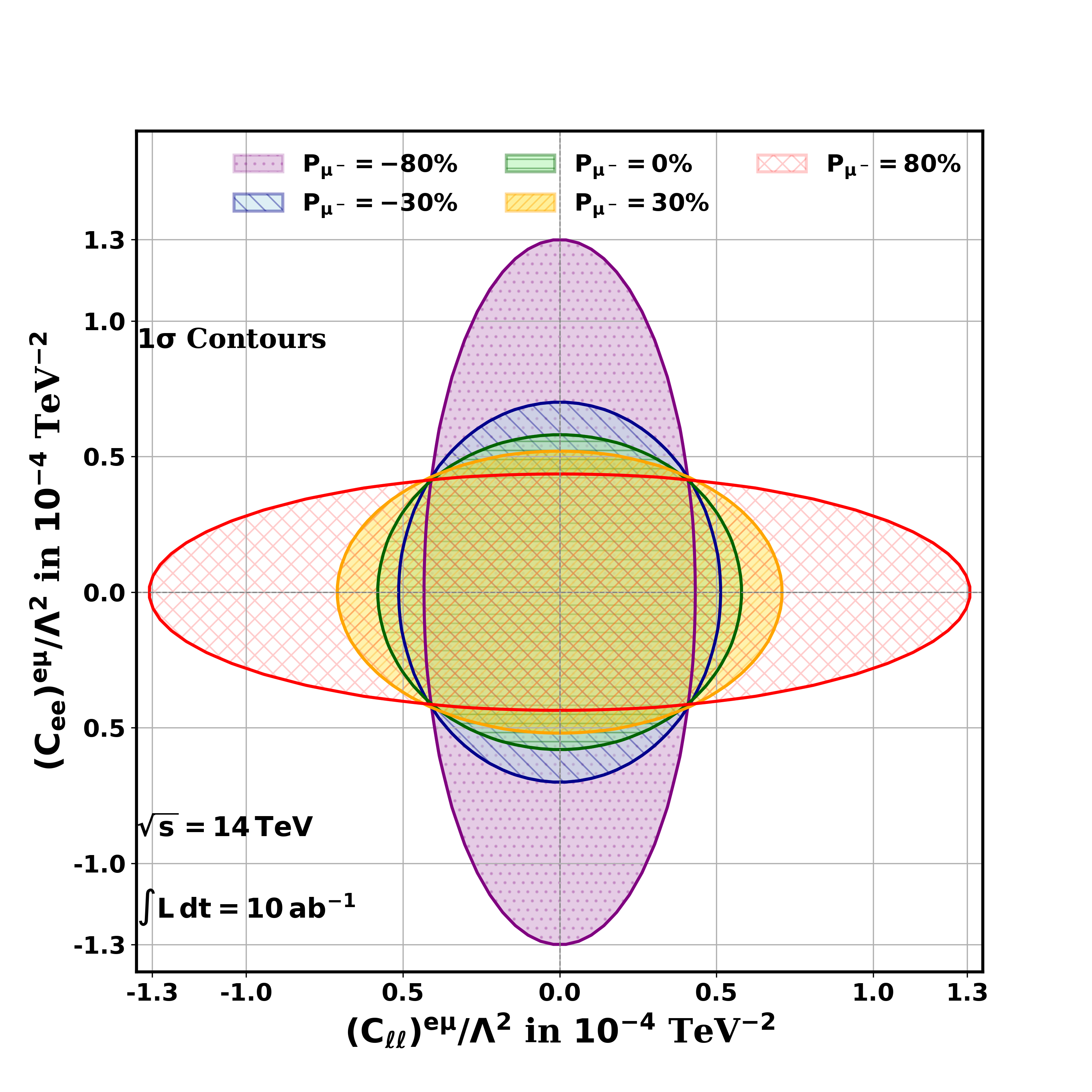}
    \includegraphics[width=0.32\textwidth]{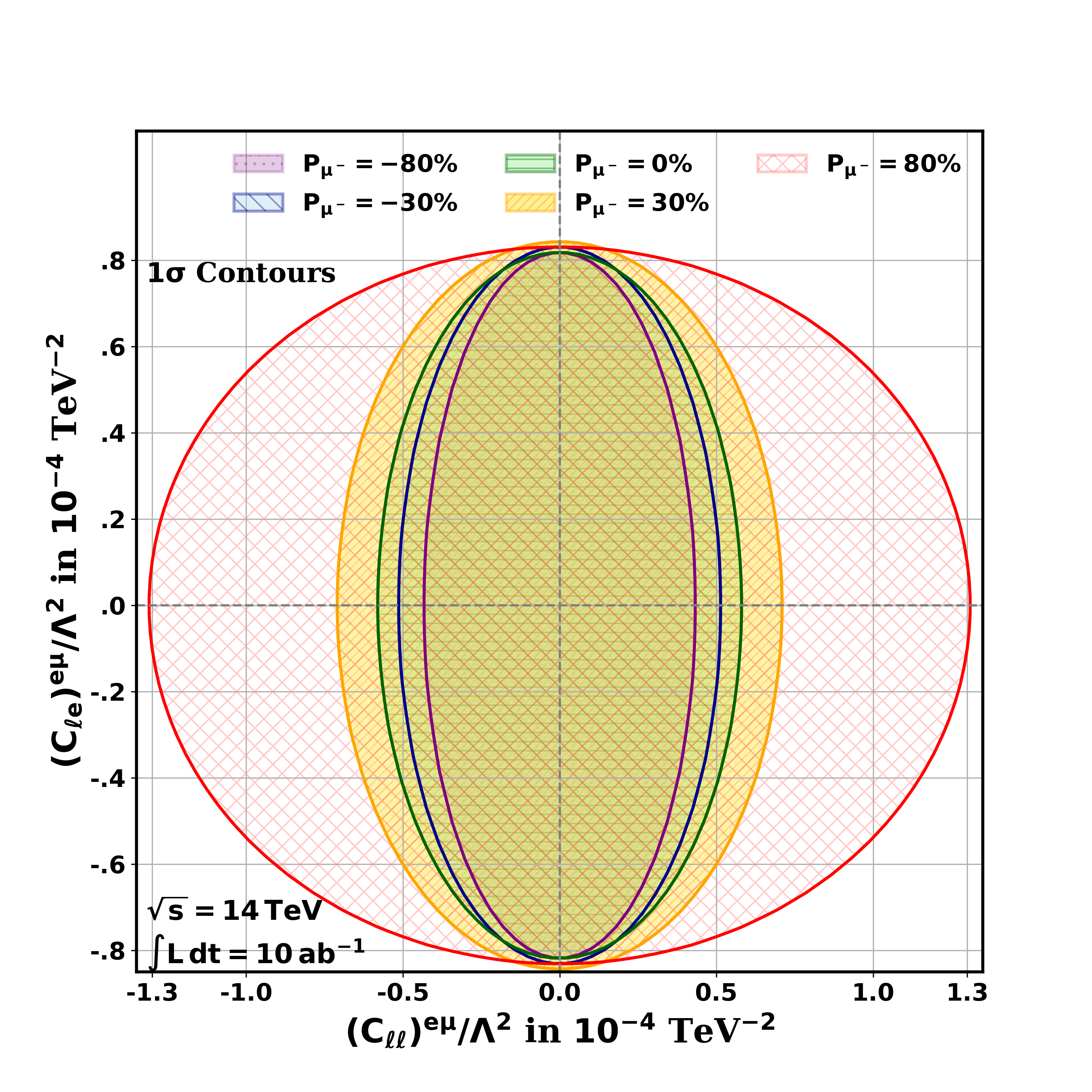}
    \includegraphics[width=0.32\textwidth]{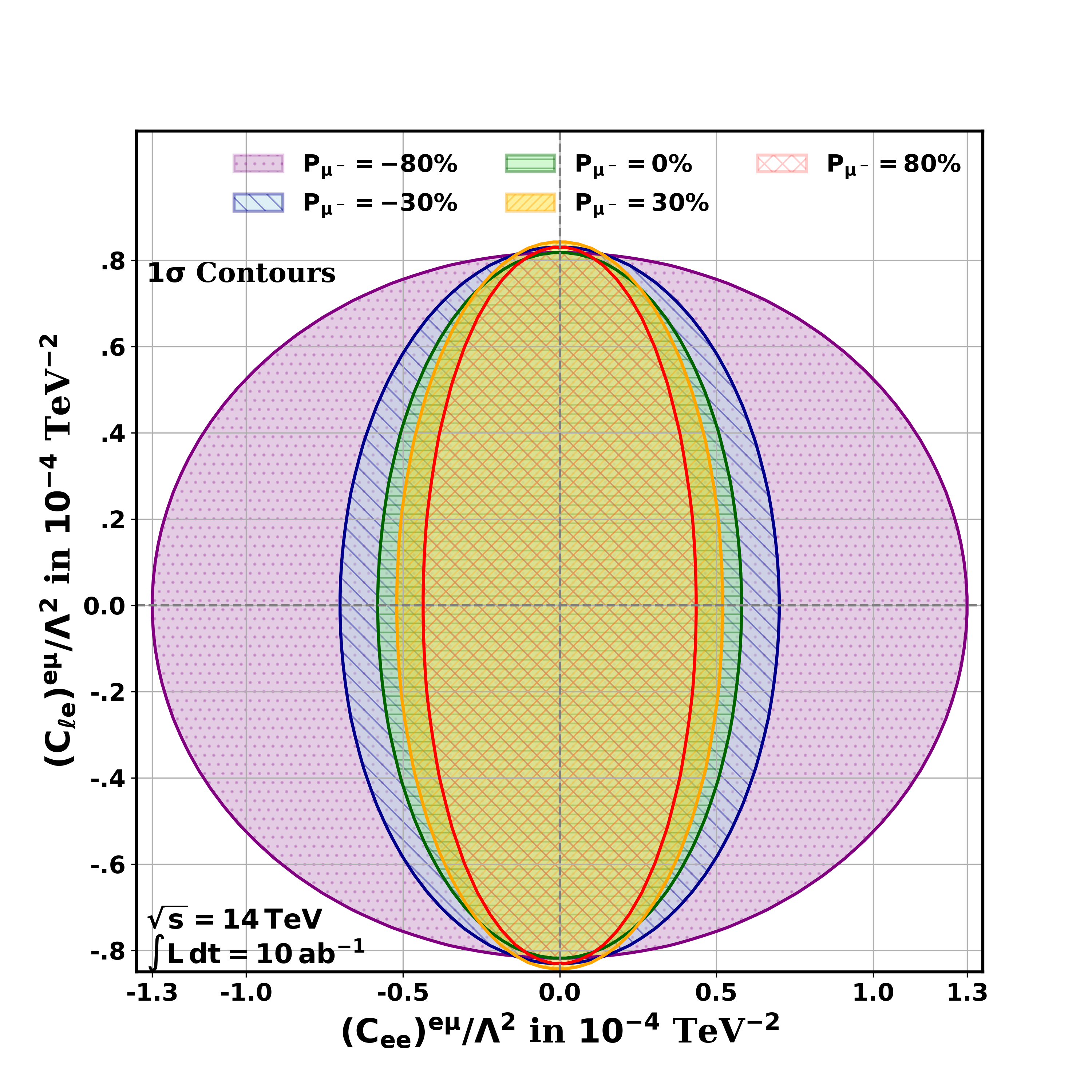}  
    \caption{ The shaded \(1\sigma\) confidence contours in the (\(C_{\ell\ell}^{e\mu}/\Lambda^2\,-\, C_{ee}^{e\mu}/\Lambda^2\)), (\(C_{\ell\ell}^{e\mu}/\Lambda^2\,-\, C_{\ell e}^{e\mu}/\Lambda^2\)), and (\(C_{ee}^{e\mu}/\Lambda^2\,-\, C_{\ell e}^{e\mu}/\Lambda^2\)) planes for the LFV channel \(\mu^+\mu^- \to e^\pm\mu^\mp\).  }
    \label{emuConts}
\end{figure*}

\newpage
For the LFV process $\mu^+\mu^- \to e^\pm\mu^\mp $,the corresponding exclusion contours are given in Fig.~\ref{emuConts}. As in the previous case, the constraints are displayed in the three independent Wilson coefficient planes while fixing the remaining coefficient to its SM value.

\clearpage
\newpage
\twocolumngrid
\bibliography{ref.bib}

\end{document}